\documentclass[twocolumn]{autart}    

\usepackage{graphicx}          

\usepackage{algorithm}
\usepackage{algpseudocode}

\usepackage[outdir=./]{epstopdf} 

\usepackage[OT2,T1]{fontenc}
\usepackage{mleftright} 
\usepackage{widetext}

\usepackage{xcolor}
\newcommand{\ssnote}{\textcolor{black}}

\newcommand{\ssnoteblue}{\textcolor{black}}
\newcommand{\ssnotebluetwo}{\textcolor{black}}

\newcommand{\ssnotebluethree}{\textcolor{black}}

\newcommand{\ssnotebluefour}{\textcolor{black}}

\newcommand{\ssredd}{\textcolor{black}}

\usepackage{amssymb}
\usepackage{mathtools}
\usepackage{bm} 

\newtheorem{theorem}{Theorem}
\newtheorem{lemma}{Lemma}
\newtheorem{proposition}{Proposition}
\newtheorem{corollary}{Corollary}
\newtheorem{assumption}{Assumption}

\newtheorem{definition}{Definition}

\newtheorem{remark}{Remark}

\mathtoolsset{showonlyrefs=true}
\allowdisplaybreaks

\DeclareMathOperator*{\argmin}{arg\,min} 

\newcommand{\comp}{\mathsf{c}}

\DeclareMathOperator{\sat}{sat}

\newcommand{\floor}[1]{\lfloor #1 \rfloor}
\newcommand{\ceil}[1]{\lceil #1 \rceil}

\begin{document}

\begin{frontmatter}

\title{A Framework for Adaptive Stabilisation of Nonlinear Stochastic Systems
\thanksref{footnoteinfo}} 

\thanks[footnoteinfo]{This paper was not presented at any IFAC 
meeting. Corresponding author S.~Siriya.}

\author[Germany]{Seth Siriya}\ead{siriya@irt.uni-hannover.de}, 
\author[Australia]{Jingge Zhu}\ead{jingge.zhu@unimelb.edu.au},
\author[Australia]{Dragan Ne\v{s}i\'{c}}\ead{dnesic@unimelb.edu.au},
\author[Australia]{Ye Pu}\ead{ye.pu@unimelb.edu.au}

\address[Germany]{Leibniz University Hannover, Institute of Automatic Control, 30167 Hannover, Germany}        
\address[Australia]{University of Melbourne, Department of Electrical and Electronic Engineering, Parkville VIC 3010, Australia}        


\begin{abstract}                          
We consider the adaptive control problem for discrete-time, nonlinear stochastic systems with linearly parameterised uncertainty. Assuming access to a parameterised family of controllers that can stabilise the system in a bounded set within an informative region of the state space when the parameter is well-chosen, we propose a certainty equivalence learning-based adaptive control strategy, and subsequently derive stability bounds on the closed-loop system that hold for some probabilities. We then show that if the entire state space is informative, and the family of controllers is globally stabilising with appropriately chosen parameters, high probability stability guarantees can be derived.%
\end{abstract}

\end{frontmatter}

\section{Introduction}
Adaptive control (AC) is concerned with the control of dynamical systems with parameters which are uncertain, online, in a single trajectory.
It is useful in situations where one is not afforded the luxury of running many experiments to collect data offline for the purpose of system identification.
Rigorously characterising the stochastic stability properties of these methods is important for understanding the situations in which an AC strategy can be successfully deployed on a system.
Of great interest are methods that can provably handle nonlinear systems subject to stochastic process noise, since many real-world systems are nonlinear and influenced by factors with random characteristics.

The stability problem in stochastic AC has a rich history, with the first stability results for AC in unstable linear plants obtained back in the 1980s by Goodwin and Caines \cite{goodwin1981discrete}.
Although results for linear systems are well explored, the nonlinear setting continues to attract attention.
In a series of works \cite{guo1997critical,xie1998adaptive,li2013stabilization,liu2019possible}, the fundamental limits of the stabilisation problem for a class of discrete-time systems with \textit{linearly parameterised uncertainties} --- that is, systems where the model uncertainties are formulated as a linear combination of known nonlinear basis functions of the states and controls --- is investigated. These works require that it is possible for controls to completely cancel all system dynamics in a single step, and they show that a certainty equivalence least squares-based AC strategy provably stabilises the system under various assumptions, often involving restrictions on the growth of basis functions. In \cite{boffi2021regret}, stability-based regret bounds were derived for an online gradient-based discrete-time AC method for linearly parameterised systems with \textit{matched uncertainties} --- that is, problems where the controls may not necessarily cancel the entire system dynamics, but just the model uncertainties. 
The AC problem for discrete-time input-constrained linear systems subject to unbounded stochastic disturbances was also recently addressed in \cite{siriya2022learning} and \cite{siriya2023stability} under various assumptions, which can be viewed as a linearly parameterised nonlinear system.
Moreover, the adaptive stabilisation problem for stochastic control-affine SISO systems and stochastic discrete-time parametric strict-feedback systems has also been addressed in \cite{xie1998adaptive} and \cite{wei1999prediction} respectively.
Although all these results are interesting, it is desirable to obtain results which are applicable to other systems with linearly parameterised uncertainty without such structural restrictions.

Some works in the continuous-time, deterministic setting do not impose such structural restrictions. 
In \cite{benosman2014learning}, a framework for \textit{learning-based} AC is proposed which combines an integral input-to-state stabilising controller with an extremum seeking-based parameter update law.
By \textit{learning}, we mean that the parameter estimates converge towards the true parameters \cite{annaswamy2021historical}, and by learning-based we mean that the stability guarantees are reliant on learning.
Moreover, in \cite{karafyllis2018adaptive}, an AC framework is proposed for a class of linearly parameterised systems that combines a dead-beat parameter estimator with
a \textit{certainty equivalence} control law, meaning the controller as chosen as if the estimated parameter is the true parameter. These results provide a \textit{framework} for AC, such that the systems they can be applied to depend on the specifically chosen controller. 
However, stability guarantees require that the chosen controller can globally stabilise the system when given the true parameter, which is restrictive in many real-world situations.

There are some AC works applicable to linearly parameterised systems without requiring the aforementioned structural restrictions and global stabilisability, mainly from the robust adaptive model predictive control (MPC) community (see \cite{adetola2009adaptive}, \cite{adetola2011robust}, \cite{kohler2021robust}, \cite{sasfi2023robust}). However, they typically require the knowledge of a bounded uncertainty set which the true parameter belongs to, and that a feasible solution exists to the robust MPC problem that accounts for all possible parameters in this set. However, one may not always know enough about the system to determine such a set.

Motivated by these aforementioned gaps, we aim to provide a modular framework for the AC of a broad class of discrete-time stochastic nonlinear systems with linearly parameterised uncertainties, which does not require a priori knowledge of a set of parameters which renders the system bounded.
Moreover, we aim to derive stochastic stability guarantees for the closed-loop system which are applicable even when the system is not globally stabilisable.
This guarantee need to be derived under assumptions that can be easily verified based on the system model \ssnotebluefour{offline}. Our contributions are as follows.

Firstly, we provide a framework for certainty equivalence AC in linearly parameterised systems, that combines a parameterised stabilising policy with regularised least squares (RLS) for parameter estimation. The framework is intuitive since the parameterised controller is selected based on the CE principle. 

Secondly, we derive a probabilistic guarantee that the states of the system will remain positively invariant on a subset of the state space for all time, and that a non-asymptotic bounds on the parameter estimation error will holds for sufficiently large time, under some assumptions. These bounds asymptotically converge towards zero, and thus characterise a bound on how fast learning is occurring. The assumptions are related to the instability of the system and the growth rate of the basis functions. They also rely on the assumption that when the parameter estimation error is sufficiently small, the chosen controller parameterised policy will render the states of the system positively invariant in a subset of the state space satisfying a \textit{regional excitation} \ssnotebluefour{condition. This latter condition} ensures the regressor data is informative for learning, \ssnotebluefour{so the parameter estimation error will eventually be small enough for the control policy to enforce invariance}. 

Thirdly, we derive probabilistic stability bounds under the previous assumptions, and by additionally assuming the existence of stochastic Lyapunov function over the positively invariant set for the closed-loop system when the parameter estimate used for control is close to the true parameter.
This guarantee is enabled via a learning-based analysis, where the convergence of the parameter estimates ensures that eventually the controller performs similar to the case when the true parameter is known, and thus is able to stabilise the system. 

\ssnotebluefour{Fourthly}, by strengthening the assumptions to require that the system produces informative regressor data over the entire state space such that \textit{global excitation} holds, and the existence of a global stochastic Lyapunov function when the parameter estimation error is sufficiently small, we derive \textit{high probability stability bounds} --- that is, stability bounds which hold with arbitrary probability less than 1. 

Finally, we \ssnotebluefour{demonstrate our proposed} framework \ssnotebluefour{on the} adaptive control \ssnotebluefour{problem} for a stochastic PWA system, and an input-constrained linear system subject to Gaussian disturbances.
\ssnotebluefour{In particular, we show that with an appropriately selected policy, the probabilistic guarantee for the estimation error and positive invariance holds on both examples. Moreover, the probabilistic stability bounds can be established for the PWA system example, and the stronger high probability stability bound result can be established for the input-constrained linear system example.}

The work is organised as follows. In Section~\ref{ac:sec:control-problem}, we introduce the class of discrete-time stochastic systems and the AC framework considered in this paper, as well as standing assumptions. Section~\ref{ac:sec:control-method-results} contains our main results, where we recall the definition of regional and global excitation, introduce our notion of a stochastic Lyapunov function, and derive the estimation error bound and stability bound results.
Section~\ref{ac:sec:control-examples} contains the PWA system and input-constrained linear systems examples that demonstrate the benefit of our main results. Section~\ref{ac:sec:control-proofs} contains the proof of our results. Our conclusions are presented in Section~\ref{ac:sec:control-conclusion}.

\paragraph*{Notation}
Denote by $\mathbb{R}^{n \times m}$ the set of real matrices of dimension $n \in \mathbb{N}$ and $m \in \mathbb{N}$.
The space $\mathbb{R}^n$ stands for $\mathbb{R}^{n \times 1}$.
For a matrix $A \in \mathbb{R}^{n \times m}$, $\vert A \vert$ denotes its induced $2$-norm, $\vert A \vert_F$ denotes its Frobenius norm, and $B_r(A):=\{ \tilde{A} \in \mathbb{R}^{n \times m} \mid \vert \tilde{A} - A \vert \leq r \}$.
For a vector $v \in \mathbb{R}^n$ and a positive definite matrix $A \in \mathbb{R}^n$, $\vert v \vert_A = \sqrt{v^{\top} A v}$ denotes the weighted Euclidean norm.
Denote the unit sphere embedded in $\mathbb{R}^d$ by $\mathcal{S}^{d-1}$.
Consider the function $f: \mathbb{R}_{\geq 0} \rightarrow \mathbb{R}_{\geq 0}$. 
Given the function $g: \mathbb{R}_{\geq 0} \rightarrow \mathbb{R}_{\geq 0}$, we say $f(r) = O(g(r))$ if $\lim_{r \rightarrow \infty} \frac{f(r)}{g(r)} < \infty$, and $h(r) = o(r)$ if $\lim_{r \rightarrow \infty} \frac{f(r)}{g(r)} = 0$.
Consider a set $S$. The \textit{power set of $S$} is denoted by $2^S$. The \textit{indicator function of $S$} is defined by $\mathbf{1}_{S}(x):=1$ if $x \in S$, and $\mathbf{1}_{S}(x):=0$ otherwise. Consider a collection of subsets $\mathcal{C} \subseteq 2^S$.
The collection $\mathcal{C}\subseteq 2^S$ is called a $\sigma$-\textit{field in} $S$ if it is closed under countable union and intersection, as well as under complement.
$\sigma(C)$ denotes the \textit{$\sigma$-field generated by $\mathcal{C}$}, which is the intersection of all $\sigma$-fields in $S$ containing $\mathcal{C}$.
Consider the matrix subsets $X \subseteq Y \subseteq \mathbb{R}^{n \times m}$.
We say $X$ is an \textit{open subset of $Y$} if for every $x \in X$, there exists $r > 0$ such that \ssnotebluefour{the open ball of radius $r$ centred at $x$ is a subset of} $X$.
$\mathcal{B}(Y)$ denotes the \textit{Borel} $\sigma$-\textit{field of} $Y$, which is defined as $\mathcal{B}(Y):=\sigma(\mathcal{O})$ where $\mathcal{O}\subseteq 2^Y$ is the collection of all open subsets of $Y$.
Consider a measurable space $(S,\mathcal{S})$ and a mapping $f:S\rightarrow \mathbb{R}^{n \times m}$. 
The set $A \subseteq S$ is $\mathcal{S}$-\textit{measurable} if $A \in \mathcal{S}$. 
Given $Y \subseteq T$, $f^{-1}(Y):=\{ s \in S \mid f(s) \in Y \}$. 
We say that $f$ is \textit{$\mathcal{S}$-measurable} if $f^{-1}(X)$ is $\mathcal{S}$-\textit{measurable} for all $X \in \mathcal{B}(\mathbb{R}^{n \times m})$.
Additionally, if $S \subseteq \mathbb{R}^{p \times q}$, we say \textit{$f$ is Borel measurable} if $f$ is $\mathcal{B}(S)$-measurable.    
Consider a probability space $(\Omega,\mathcal{F},\mathbb{P})$.
We call a mapping $X:\Omega \rightarrow \mathbb{R}^{n \times m}$ a \textit{random variable} if $X$ is measurable, and additionally a \textit{random vector} if $m=1$, and a \textit{scalar random variable} if $n=m=1$.
We define a \textit{random sequence} as a sequence $\{X(i)\}_{i \in \mathcal{I}}$ of random variables $X(i):\Omega \rightarrow \mathbb{R}^{n \times m}$ over $i$ in an index set $\mathcal{I} \subseteq \mathbb{N}$. 
Consider a collection of random variables $X_1,X_2,\hdots$ taking values in $\mathcal{X}_1, \mathcal{X}_2,\hdots$ and a predicate $Q:\mathcal{X}_1 \times \mathcal{X}_2 \times \cdots \rightarrow \{ \mathrm{true}, \mathrm{false} \}$. With an abuse of notation, $\{Q(X_1,X_2,\hdots) \}$ stands for $\left \{ \omega \in \Omega \mid Q(X_1(\omega),X_2(\omega),\hdots) \right \}$, and $\mathbb{P}\left ( Q(X_1,X_2,\hdots) \right)$ stands for $\mathbb{P} \left ( \{Q(X_1,X_2,\hdots) \} \right )$. 
\ssnotebluetwo{The function $h: \mathbb{R}_{\geq 0} \rightarrow \mathbb{R}_{\geq 0}$ is said to be \textit{$n$-order sub-exponential} ($n$-SE) if  $\ln( h(r^n) ) = o (r)$.    The function $\alpha : \mathbb{R}_{\geq 0} \rightarrow \mathbb{R}_{\geq 0}$ is in class $\mathcal{K}$ if it is continuous, strictly increasing, and $\alpha(0) = 0$.
The function $\alpha : \mathbb{R}_{\geq 0} \rightarrow \mathbb{R}_{\geq 0}$ is in class $\mathcal{K}_{\infty}$ if $\alpha(\cdot) \in \mathcal{K}$ and unbounded, i.e., $\lim_{s \rightarrow \infty} \alpha(s) = \infty$. 
The function $\alpha : \mathbb{R}_{\geq 0} \rightarrow \mathbb{R}_{\geq 0}$ is in class $\mathcal{K}_{\infty}^{n\textnormal{-SE}}$ if $\alpha \in \mathcal{K}_{\infty}$ and $\alpha$ is $n$-SE.
A function $h:\mathbb{R}_{\geq 0} \rightarrow \mathbb{R}_{\geq 0}$ is said to be \textit{asymptotically polynomially bounded (APB)} if there exists $k \in \mathbb{N}$ such that $h(r) = O(r^k)$.}
\ssnotebluetwo{The function $\eta: \mathbb{N}_0 \rightarrow \mathbb{R}_{\geq 0}$ is in class $\mathcal{L}$ if it is non-increasing and $\lim_{t \rightarrow \infty} \eta(t) = 0$. The function $\beta : \mathbb{R}_{\geq 0} \times \mathbb{N}_0 \rightarrow \mathbb{R}_{\geq 0}$ is in class $\mathcal{K}\mathcal{L}$ if $\beta(\cdot,t)\in \mathcal{K}$ for any $t \in \mathbb{N}_0$ and $\beta(r,\cdot) \in \mathcal{L}$ for any $r \geq 0$.}
\section{Problem Setup and Framework} \label{ac:sec:control-problem}

\ssnotebluetwo{We start by describing the stochastic system and adaptive control framework considered in this work in Section~\ref{ac:sec:system}. Afterwards, in Section~\ref{ac:sec:standing-assumptions}, we make various assumptions that ensure stochastic properties of interest for our system are well-defined, and allow us to derive estimation error and stability bounds in our main results.}

\subsection{Discrete-Time Stochastic System and Adaptive Control Framework} \label{ac:sec:system}

\ssnoteblue{Let $(\Omega,\mathcal{F},\mathbb{P})$ be the probability space on which all of our random variables are defined, and denote expectation and variance by $\mathbb{E}$ and $\mathbb{V}\mathrm{ar}$ respectively. 
} Consider the discrete-time, stochastic, nonlinear system
\begin{align}
    X(t+1) =& f(X(t),U(t)) + \theta_*^{\top} \psi(X(t),U(t)) \\
    &\quad+ W(t+1), \ t \in \mathbb{N}_0,  \label{ac:eqn:system-dynamics}
\end{align}
with $X(0)=x_0$. Here, \ssnoteblue{the \textit{process noise} $\{ W(t) \}_{t \in \mathbb{N}}$ is a random sequence taking values in  $\mathbb{W} \ssnotebluetwo{\subseteq \mathbb{R}^n}$, and}
$\{X(t)\}_{t \in \mathbb{N}_0}$ and $\{U(t)\}_{t \in \mathbb{N}_0}$ are the sequence of \textit{states} and \textit{controls} taking values in $\mathbb{X} \subseteq \mathbb{R}^n$ and $\mathbb{U} \subseteq \mathbb{R}^m$ respectively. 
Moreover, $f:\mathbb{X} \times \mathbb{U} \rightarrow \ssnoteblue{\mathbb{R}^n}$ is a known \textit{nominal system model}, $\psi:\mathbb{X}\times\mathbb{U}\rightarrow \mathbb{R}^d$ is a known vector of \textit{basis functions}, $\theta_* \in \mathbb{R}^{d \times n}$ is the \textit{true unknown system parameter}, and $x_0 \in \mathbb{X}$ is the \textit{initial state}. 
For convenience, we also define $g(x,u,w):=f(x,u)+\theta_*^{\top}\psi(x,u)+w$, \ssnoteblue{and we only consider $g$ with the codomain $\mathbb{X}$ to ensure solutions are well-defined for all time.}

\ssnotebluetwo{Our adaptive control framework for stabilising \eqref{ac:eqn:system-dynamics} is summarised in Algorithm~\ref{ac:alg:framework}.}
\begin{algorithm}[H]
\caption{Adaptive Control Framework}
\begin{algorithmic}
\State Hello
\State \textbf{Inputs:} \ssnotebluethree{$f$}, \ssnotebluethree{$\psi$}, $\alpha$, \ssnotebluethree{$\mu_s$}, $\vartheta_0 \in \mathbb{R}^{d \times n}$, $\gamma > 0$
\State Measure $X(0)$
\For{$t = 0,1,\hdots$}
\State Compute estimate $\hat{\theta}(t-1)$ following \eqref{ac:eqn:estimator}
\State Sample $S(t)\stackrel{\textnormal{i.i.d.}}{\sim} \mu_s$
\State Compute control $U(t)$ following \eqref{ac:eqn:controller}
\State Apply $U(t)$ to \eqref{ac:eqn:system-dynamics}
\State Measure state $X(t+1)$ from \eqref{ac:eqn:system-dynamics}
\EndFor
\end{algorithmic}\label{ac:alg:framework}
\end{algorithm}
\ssnotebluetwo{We now describe our framework in further detail.} The control sequence is generated via
\begin{align}
    &U(t) = \alpha(X(t),S(t), \hat{\theta}(t-1) ), \ t \in \mathbb{N}_0. \label{ac:eqn:controller}
\end{align}
where $\alpha: \mathbb{X} \times \mathbb{S} \times \mathbb{R}^{d \times n} \rightarrow \mathbb{U}$ is a \textit{known control policy} that designs that control input $U(t)$ according to the state $X(t)$, an injected noise term $S(t)$, and a parameter estimate $\hat{\theta}(t-1)$. 
\ssnotebluetwo{Intuitively, we can view $\alpha$ as a policy we could apply to \textit{stabilise} the system when the true parameter $\theta_*$ is used instead of a parameter estimate. In this sense, we can view strategies following the framework outlined in Algorithm~\ref{ac:alg:framework} as a certainty equivalence adaptive control law. This will be formalised in Section~\ref{ac:sec:control-method-results}.}
The \ssnoteblue{random} sequence \ssnoteblue{of \textit{exploratory noise}} $\{S(t)\}_{t \in \mathbb{N}}$ \ssnoteblue{is sampled i.i.d. from a distribution $\mu_s$ with the support $\mathbb{S} \subseteq \mathbb{R}^q$}, and is injected to help facilitate convergence of the estimate $\hat{\theta}(t)$. 
\ssnoteblue{The sequence of \textit{parameter estimates} $\{\hat{\theta}(t)\}_{t \in \mathbb{N}}$ taking values in $\mathbb{R}^{d \times n}$} are obtained via regularised least squares (RLS) estimation:
\begin{align}
    \hat{\theta}(t) = \begin{cases}
        \vartheta_0, \ t  \in \{ -1,0 \}, \\
        \argmin_{\theta \in \mathbb{R}^{d \times n}} \sum_{s=1}^t \big \vert X(s) \ssnotebluetwo{- \theta^{\top} Z(s)} \\
        \ssnotebluetwo{- f(X(s-1), U(s-1)) \big \vert^2 }+ \gamma \vert \theta \vert_F^2, \ t \in \mathbb{N},
    \end{cases} \label{ac:eqn:estimator}
\end{align}
where $\{ Z(t) \}_{t \in \mathbb{N}}$ is sequence of \textit{regressors} taking values in $\mathbb{R}^{d}$ and satisfying $Z(t) = \psi(X(t-1),U(t-1))$. \ssnoteblue{Moreover,} $\gamma > 0$ is a user-chosen regularisation parameter, and $\vartheta_0 \in \mathbb{R}^{d \times n}$ is an initial deterministic parameter estimate. Note that the RLS estimates also satisfy 
\begin{align}
    &\hat{\theta}(t) = G(t)^{-1} \sum_{s=1}^t Z(s) \left ( X(s) \right.  \left. - f(X(s-1),U(s-1)) \right )^{\top}  \label{ac:eqn:estimator-equivalent}
\end{align}
for $t \in \mathbb{N}$, where $G(t) = \sum_{i=1}^t Z(i)Z(i)^{\top} + \gamma I$ is known as the \textit{regularised Gramian}.

\ssnoteblue{Note that $\{ X(t) \}_{t \in \mathbb{N}_0}$, $\{ U(t) \}_{t \in \mathbb{N}_0}$, $\{Z(t)\}_{t \in \mathbb{N}}$ and $\{ \hat{\theta}(t) \}_{t \in \mathbb{N}}$ are defined via construction through 
\eqref{ac:eqn:system-dynamics}-\eqref{ac:eqn:controller}, and depend on the 
noise sequences $\{W(t)\}_{t \in \mathbb{N}}$ and $\{S(t)\}_{t \in \mathbb{N}_0}$. 
Since the former are sequences of mappings from $\Omega$ to $\mathbb{X}$, $\mathbb{U}$ and $\mathbb{R}^{n \times d}$ respectively, they can \textit{informally} be viewed as random sequences. However it is unclear from our description so far whether they \textit{formally} 
satisfy the required definition, and hence whether \ssnotebluetwo{all stochastic properties of interest are well-defined}. We address this concern in Section~\ref{ac:sec:standing-assumptions} 
by imposing assumptions that ensure such requirements are satisfied.}

\subsection{Standing Assumptions} \label{ac:sec:standing-assumptions} 

Firstly, we make the following regularity assumption on $f$, $\psi$ and $\alpha$. It helps ensure that all stochastic properties of interest for our system \ssnotebluetwo{are well-defined}.
\begin{assumption} (Measurability of system mappings) \label{ac:assump:measurable}
    The nominal system model $f$,
    basis functions $\psi$, 
    and policy $\alpha$, are all Borel measurable.
\end{assumption}

Next, we recall the definition of sub-Gaussian random variables, which are random variables whose distributions have tails that decay at least as fast as the Gaussian distribution.
\begin{definition} \label{ac:def:sub-gaussian} (sub-Gaussian random variable)
    Given a sub sigma-field $\mathcal{G} \subseteq \mathcal{F}$ and scalar random variable $X$, we say $X\mid \mathcal{G}$ is $\sigma_x^2$-sub-Gaussian if for all $\lambda \in \mathbb{R}$, $\mathbb{E}[\exp(\lambda X) \mid \mathcal{G}] \leq \exp(\lambda^2 \sigma_x^2 / 2)$\ssnotebluetwo{\footnote{\ssnotebluetwo{Unless otherwise stated, inequalities involving random variables should be interpreted \textit{surely}, i.e. given two random variables $X$ and $Y$, $X \leq Y$ stands for $X(\omega) \leq Y(\omega)$ for all $\omega \in \Omega$.}}}. If $X$ is an $\mathbb{R}^d$-valued random vector, we say $X\mid \mathcal{G}$ is $\sigma_x^2$-sub-Gaussian if $\zeta^{\top}X \mid \mathcal{G}$ is $\sigma_x^2$-sub-Gaussian for all $\zeta \in \mathcal{S}^{d-1}$. For both cases, we say $X$ is $\sigma_x^2$-sub-Gaussian when $\mathcal{G}$ is chosen as the trivial sigma-field $\{\emptyset, \Omega \}$.
\end{definition}

We make the following assumption on $\{W(t)\}_{t \in \mathbb{N}}$ over the probability space. It is related to independence, and ensures the tails of its distribution decays sufficiently fast for parameter estimation.

\begin{assumption} \label{ac:assump:process-noise}
    (Independent and sub-Gaussian process noise)
    \begin{enumerate}
        \item The process noise sequence $\{W(t)\}_{t \in \mathbb{N}}$ is i.i.d. with distribution denoted by $\mu_w$.
        \item The sequences $\{W(t)\}_{t \in \mathbb{N}}$ and $\{S(t)\}_{t \in \mathbb{N}_0}$ are mutually independent.
        \item $W(t)$ is zero-mean $\sigma_w^2$-sub-Gaussian for any $t \in \mathbb{N}$, i.e. $\mathbb{E}[\exp(\gamma \zeta^{\top}W(t))] \leq \exp(\gamma^2 \sigma_w^2 / 2)$ for any $\gamma \in \mathbb{R}$ and $\zeta \in \mathcal{S}^{n-1}$.
    \end{enumerate}
\end{assumption}

Let $\phi(t,\xi,\{u(i)\}_{i = 0}^{t-1},\{w(i)\}_{i = 1}^t)$ denote the \ssnotebluetwo{\textit{deterministic} state solution at time $t \in \mathbb{N}_0$ to the problem described in \eqref{ac:eqn:system-dynamics}} given the initial state $\xi \in \mathbb{X}$, deterministic control input sequence $\{u(i)\}_{i=0}^{t-1} \in \mathbb{U}^t$ and deterministic process noise sequence $\{w(i)\}_{i = 1}^t \in \mathbb{W}^t$. We make the following assumption, which bounds the reachable states of this deterministic solution as a function of energy-like estimates of the control inputs and process noise. It naturally restricts the instability of the system which helps establish the convergence of parameter estimates.
, as required for estimation in \cite{sysidpaper}.
The continuous-time analogue of this property is equivalent to a concept called \textit{forward completeness} \cite{angeli1999forward}.
\begin{assumption} \label{ac:assump:forward-complete}
    (Sub-exponential input-to-state bound on reachable states)
    There exist functions $\chi_1,\chi_3,\sigma_2 \in \mathcal{K}_{\infty}^{1\textnormal{-SE}}$, $\chi_4 \in \mathcal{K}_{\infty}^{2\textnormal{-SE}}$, $\chi_2,\sigma_1 \in \mathcal{K}_{\infty}$ and a constant $c_1 \geq 0$ such that
    \begin{align}
        &\left\vert \phi(t,\xi,\{u(i)\}_{i = 0}^{t-1}, \{w(i)\}_{i=1}^t) \right\vert \leq \chi_1(t) + \chi_2(\left \vert \xi \right \vert )  \\
        &\quad + \chi_3 \left( \sum_{i=0}^{t-1}  \sigma_1 \left ( \left \vert u(i) \right \vert \right ) \right )  + \chi_4 \left( \sum_{i=1}^t  \sigma_2 \left ( \left \vert w(i) \right \vert \right ) \right ) + c_1  \label{ac:eqn:forward-complete}
    \end{align}
    for all $t \in \mathbb{N}$, $\xi \in \mathbb{X}$, $\{u(i)\}_{i=0}^{t-1} \in \mathbb{U}^t$ and $\{w(i)\}_{i = 1}^t \in \mathbb{W}^t$.
\end{assumption}

We assume the controls generated by the policy $\alpha$ satisfy a magnitude constraint. Such constraints are commonly satisfied in strategies like model predictive control (MPC), and are natural in most real-world applications due to actuator saturation. They are useful for restricting the growth of the states of the system during instability, as required for estimation in \cite{sysidpaper}.

\begin{assumption} \label{ac:assump:constrained-controls}
    (Constrained controls)
    There exists $u_{\textnormal{max}}\geq 0$ such that for any $x \in \mathbb{X}$, $s \in \mathbb{S}$ and $\vartheta \in \mathbb{R}^{d \times n}$, $\left \vert \alpha(x,s,\vartheta) \right \vert \leq u_{\textnormal{max}}$.
\end{assumption}

We assume that the magnitude of the basis functions $\psi$ grow polynomially as a function of the states and controls, which aids in bounding the regressor sequence, as required for estimation in \cite{sysidpaper}.

\begin{assumption} \label{ac:assump:poly-feature-map} \ssnoteblue{
    (APB basis functions)
    There exists an APB function $\chi_5:\mathbb{R}_{\geq 0} \rightarrow \mathbb{R}_{\geq 0}$ such that $\left \vert \psi(x,u) \right \vert \leq \chi_5 \left ( \left \vert \ssnotebluetwo{[x^{\top} \ y^{\top}]^{\top}} \right \vert \right )$.}
\end{assumption}

\section{Main Results} \label{ac:sec:control-method-results}

\ssnotebluetwo{We now provide the main theoretical results of this work. In Section~\ref{ac:sec:stability-regional} we define the \textit{regional excitation} and \textit{global excitation} conditions, alongside the concept of a \textit{robustly positive invariant} (RPI) set. Under the assumption that regional excitation holds and an RPI set exists, we provide bounds on the estimaton error for a class of systems controlled by our adaptive framework. We then define the concept of stochastic Lyapunov functions and global stochastic Lyapunov functions, and under the assumption that the former exists, we provide probabilistic stability bounds for the adaptively controlled system.
Subsequently, in Section~\ref{ac:sec:stability-global}, we provide high probability stability bounds for the system under the assumption of global excitation and the existence of a global stochastic Lyapunov function.}

\subsection{Stability and Estimation with Regional Control and Excitation} \label{ac:sec:stability-regional}

For any Borel measurable function $h:\mathbb{S} \times \mathbb{W} \rightarrow \mathbb{R}$, we denote $\mathbf{E}\left [ h(S,W) \right ] := \int_{\mathbb{W}} \int_{\mathbb{S}} h(s',w') d\mu_s(s') d\mu_w(w')$ and $\mathbf{V}\mathrm{ar}(h(S,W)):=\\\mathbf{E}\left [ \left ( h(S,W) - \mathbf{E} \left[ h(S,W) \right] \right )^2 \right ]$.
\ssnotebluefour{They can be respectively interpreted as the expectation and variance of $h(S,W)$ viewing $(S,W)$ as random variables sampled from the product distribution $ \mu_s \times \mu_w$ over $\mathbb{S}\times\mathbb{W}$}
\ssnotebluetwo{Moreover, given a predicate\footnote{\ssnotebluefour{Recall that a \textit{predicate} is just a mapping from some objects to the truth-values $\mathrm{true}$ and $\mathrm{false}$. For example, given $a \in \mathbb{R}$, ``$x \leq a$'' is a predicate with $x$ as the free variable, which takes the value $\mathrm{true}$ if $x \leq a$ holds, and $\mathrm{false}$ if $x > a$ holds.}} $Q:\mathbb{S}\times \mathbb{W} \rightarrow \{ \mathrm{true},\mathrm{false} \}$, we define $\mathbf{P}(Q(S,W)):=\mathbf{E} [ \mathbf{1}_{\{ (s,w) \in \mathbb{S} \times \mathbb{W} \mid Q(s,w) \}}(S,W) ]$}
\ssnotebluetwo{, which can be interpreted as the probability of $Q(S,W)$ holding.}
\ssnotebluefour{This notation will be used to emphasise that some important objects can be constructed as a simple integral with respect to only $\mu_s \times \mu_w$, which is useful from a verification standpoint, in contrast to $\mathbb{E}$, $\mathbb{V}\mathrm{ar}$ and $\mathbb{P}$.}
We now define regional \ssnotebluetwo{and global excitation} as follows, which we recall from \cite{sysidpaper}.

\begin{definition} \label{ac:def:excitation}
    (Regional \ssnotebluetwo{and global} excitation) 
    The feature map $\psi$ and controller family $\alpha$ are said to be \textit{regionally excited} with noise distribution $\mu_s$ and $\mu_w$ over $\mathcal{X} \subseteq \mathbb{X}$ with 
    \ssnotebluetwo{finite constants $c_{\textnormal{PE}},p_{\textnormal{PE}} > 0$} if for all $\vartheta \in \mathbb{R}^{d \times n}$, $x \in \mathcal{X}$ and $\zeta \in \mathcal{S}^{d-1}$,
    \begin{align}
        \ssnotebluetwo{\mathbf{P}\left (  \left \vert \zeta^{\top} \psi \left ( x + W , \alpha(x+W,S,\vartheta) \right ) \right \vert^2 \geq c_{\textnormal{PE}} \right ) \geq p_{\textnormal{PE}}.} \label{ac:eqn:excitation}
    \end{align}
    For convenience, we say $(\psi,\alpha,\mu_w,\mu_s)$ is \ssnotebluetwo{\textit{$(\mathcal{X},c_{\textnormal{PE}},p_{\textnormal{PE}})$-regionally excited}}.
    \ssnotebluetwo{Moreover, we say $(\psi,\alpha,\mu_w,\mu_s)$ is \textit{$(c_{\textnormal{PE}},p_{\textnormal{PE}})$-globally excited} if $(\psi,\alpha,\mu_w,\mu_s)$ is $(\mathbb{X},c_{\textnormal{PE}},p_{\textnormal{PE}})$-regionally excited}
\end{definition}

Recall that from the perspective of LS estimation, regional excitation ensures that for trajectories evolving through \ssnotebluetwo{$\mathcal{X}$}, the corresponding regressors will have a non-degenerate distribution and will therefore be \textit{informative} for estimation, by which we mean that the \ssnotebluetwo{minimum eigenvalue of the} Gramian $G(t)$ will be lower bounded with high probability by a linearly growing function, which is known as \textit{persistency of excitation} (PE). 
\ssnotebluetwo{Global excitation in Definition~\ref{ac:def:excitation} is a strengthening of regional excitation so that it holds over the entire state space $\mathbb{X}$. In the general context of nonlinear system identification, such a condition may be restrictive, but stronger statements on the estimation error can be made when the global excitation condition holds.}

Alongside being used to establish PE, recall that regional excitation is useful since it is a data-independent condition characterised entirely by the basis functions $\psi$, control policy $\alpha$, and the distribution of the process and exploratory noises $\mu_w$, $\mu_s$ entering the system. All of these objects are assumed to be known \textit{a priori}, and therefore regional excitation in Definition~\ref{ac:def:excitation} can be readily checked without knowing the entirety of the system dynamics. 
\ssnotebluetwo{There are many different ways to verify regional excitation, one being to bound the first and second moments of $\left \vert \zeta^{\top} \psi \left ( x + W , \alpha(x+W,S,\vartheta) \right ) \right \vert^2 $ from below and above respectively, as in Lemma~\ref{ac:lemma:regional-excitation-sufficient}. This result is the same as Lemma~1 from \cite{sysidpaper}.}

\begin{lemma} \label{ac:lemma:regional-excitation-sufficient}
    \ssnotebluetwo{Consider feature map $\psi$ and controller family $\alpha$ satisfying Assumption~\ref{ac:assump:measurable}, and noise distributions $\mu_w,\mu_s$ satisfying Assumption~\ref{ac:assump:process-noise}. Suppose there exists a subset $\mathcal{X} \subseteq \mathbb{X}$ and constants $c_{\textnormal{PE1}}>0$ and $c_{\textnormal{PE2}}\geq 0$ such that for all $\vartheta \in \mathbb{R}^{d \times n}$, $x \in \mathcal{X}$ and $\zeta \in \mathcal{S}^{d-1}$,
    \begin{align}
        &\mathbf{E} \left[ \left \vert \zeta^{\top} \psi \left ( x + W , \alpha(x+W,S,\vartheta) \right ) \right \vert \right]  \geq c_{\textnormal{PE1}}, \\
        &\mathbf{V}\mathrm{ar}\left ( \left \vert \zeta^{\top} \psi \left ( x + W , \alpha(x+W,S,\vartheta) \right ) \right \vert \right ) \leq c_{\textnormal{PE2}}.
    \end{align}
    Then, $(\psi,\alpha,\mu_s,\mu_w)$ is $(\mathcal{X},c_{\textnormal{PE}},p_{\textnormal{PE}})$\textit{-regionally excited} with 
    \begin{align}
        p_{\textnormal{PE}} & := \frac{1}{4} \left( \frac{ c_{\textnormal{PE}2}}{c_{\textnormal{PE}1}^2} + 1 \right)^{-1}, \quad c_{\textnormal{PE}} := \frac{1}{4} c_{\textnormal{PE}1}^2.
    \end{align}}
\end{lemma}

Throughout \ssnotebluefour{this section}, we assume that regional excitation holds.
\begin{assumption} \label{ac:assump:regional-excitation}
    $(\psi,\alpha,\mu_w,\mu_s)$ is \ssnotebluetwo{$(\mathcal{X}_{\textnormal{PE}},c_{\textnormal{PE}},p_{\textnormal{PE}})$\textit{-regionally excited}} for some set $\mathcal{X}_{\textnormal{PE}} \subseteq \mathbb{X}$ and constants \ssnotebluetwo{$c_{\textnormal{PE}},p_{\textnormal{PE}} > 0$}.
\end{assumption}

We define the \textit{\ssnoteblue{one-step predicted} reachable set map} as 
\begin{align}
    \Gamma(\mathcal{X}) := \left \{ g(x,u,0) \mid x \in \mathcal{X}, \left \vert u \right \vert \leq u_{\textnormal{max}} \right \}, \ \mathcal{X} \subseteq \mathbb{X}.
\end{align}
which is the \ssnoteblue{set of} reachable states from $\mathcal{X} \subseteq \mathbb{X}$ \ssnoteblue{when the} control input \ssnoteblue{magnitude is bounded by} $u_{\textnormal{max}}$ and the process noise \ssnotebluetwo{is set to zero}.

\ssnotebluetwo{Next, we define the concept of a robustly positive invariant (RPI) set.}
\begin{definition} \label{ac:def:rpi} 
    \ssnotebluetwo{The set $\mathcal{X} \subseteq \mathbb{X}$ is said to be a robustly positive invariant (RPI) set with error constant $\bar{\vartheta}>0$ for the} 
    system $g$ under the influence of the policy $\alpha$ if \ssnotebluetwo{for all $x \in \mathcal{X}$, $w \in \mathbb{W}$, $s \in \mathbb{S}$, and $\vartheta \in B_{\bar{\vartheta}}(\theta_*)$, it holds that    $g(x,\alpha(x,s,\vartheta),w) \in \mathcal{X}$}.
    For convenience, we say that $(g,\alpha)$ is \textit{$(\ssnotebluetwo{\mathcal{X}},\bar{\vartheta},\ssnotebluetwo{\mu_w,\mu_s})$-RPI}.
\end{definition}
\ssnotebluetwo{Intuitively, a set $\mathcal{X}$ is RPI if starting from any state $x$ inside $\mathcal{X}$, regardless of value of the process noise $w \in \mathbb{W}$ and injected noise $s \in \mathbb{S}$, if the parameter $\vartheta$ used for control is sufficiently close to the true parameter $\theta_*$ such that $\vert \vartheta - \theta_* \vert \leq \bar{\vartheta}$, then the evolved state $g(x,\alpha(x,s,\vartheta),w)$ remains inside $\mathcal{X}$. Throughout Section~\ref{ac:sec:control-method-results}, we assume the existence of such a set $\mathcal{X}_{\textnormal{RPI}}$ and an associated error constant $\bar{\vartheta}$. 
\begin{assumption} \label{ac:assump:rpi}
    There exists a subset $\mathcal{X}_{\textnormal{RPI}} \subseteq \mathbb{X}$ and error bound
    $\bar{\vartheta} > 0$ such that $(g,\alpha,\mu_w,\mu_s)$ is $(\mathcal{X}_{\textnormal{RPI}},\bar{\vartheta},\ssnotebluetwo{\mu_w,\mu_s})$-RPI \ssnotebluetwo{and $\Gamma({\mathcal{X}_{\textnormal{RPI}}}) \subseteq \mathcal{X_{\textnormal{PE}}}$}.
\end{assumption}
Note that in Assumption~\ref{ac:assump:rpi}, we also require that $\Gamma(\mathcal{X}_{\textnormal{RPI}})\subseteq \mathcal{X}_{\textnormal{PE}}$. This connects the RPI set $\mathcal{X}_{\textnormal{RPI}}$ to the region of excitation $\mathcal{X}_{\textnormal{PE}}$, allowing us to establish PE when the state trajectory $X(t)$ evolves inside $\mathcal{X}_{\textnormal{RPI}}$.
}

\begin{remark}
\ssnotebluefour{In many adaptive control works based on robust control techniques like robust adaptive MPC, knowledge of a bounded parameter set with nonempty interior containing $\theta_*$ is needed for controller design, such that the robust control algorithm can render some set robustly positive invariant accounting for all possible parameters in the bounded set (e.g. \cite{adetola2009adaptive,adetola2011robust,kohler2021robust,sasfi2023robust}). This is \textit{not} what is required in this work. Even though $\bar{\vartheta}$ is required in Assumption~\ref{ac:assump:rpi} for the estimation and stability results, it is not required to implement the framework described in Algorithm~\ref{ac:alg:framework}.}
\end{remark}

Before providing our main \ssnotebluetwo{results}, we introduce a few objects required to state our result. We define the \textit{process noise bound} $\overline{w}:\mathbb{N}_0 \times (0,1) \rightarrow \mathbb{R}_{\geq 0}$, \textit{state bound} $\overline{x}:\mathbb{N} \times (0,1) \times \mathbb{X}\rightarrow \mathbb{R}_{\geq 0}$ and \textit{regressor bound} $\overline{z}:\mathbb{N} \times (0,1) \times \mathbb{X}\rightarrow \mathbb{R}_{\geq 0}$ as
\begin{align}
    &\overline{w}(t,\delta) := \begin{cases}
        \sigma_w \sqrt{\ssnotebluetwo{2 n \ln \left( \frac{\ssnotebluetwo{n}\pi^2t^2}{3\delta} \right )} }, \quad t \in \mathbb{N}, \\
        0, \quad t = 0,
    \end{cases} \label{ac:eqn:high-prob-noise-bound} \\
    &\overline{x}(t,\delta,x_0) := \chi_1(t) + \chi_2\left (\left \vert x_0 \right \vert \right ) + \chi_3 \left ( t \sigma_1 \left ( u_{\textnormal{max}} \right ) \right )  + \chi_4 \left ( t \sigma_2 \left ( \overline{w}(t,\delta) \right ) \right ) + c_1, \label{ac:def:high-prob-state-bound} \\
    &\overline{z}(t,\delta,x_0) := \chi_5 \left ( \ssnotebluetwo {\sqrt{\overline{x}(t-1,\delta,x_0)^2 + u_{\textnormal{max}}^2} } \right ), \label{ac:def:high-prob-regressor-bound}
\end{align}
\ssnoteblue{with $\chi_1$-$\chi_4$ and $c_1$ from Assumption~\ref{ac:assump:forward-complete},} and \ssnotebluetwo{and $\chi_5$ from Assumption~\ref{ac:assump:poly-feature-map}.}
Under Assumptions~\ref{ac:assump:process-noise}-\ref{ac:assump:poly-feature-map}, these objects provide high probability upper bounds on the magnitude of $\{W(t)\}_{t \in \mathbb{N}}$, $\{X(t)\}_{t \in \mathbb{N}_0}$ and $\{Z(t)\}_{t \in \mathbb{N}}$ respectively. 

Next, we introduce objects that provide probabilistic upper and lower semi-definite bounds on the regularised Gramian $G(t)$. We define the \textit{Gramian upper bound} as
\begin{align}
    \beta_{\textnormal{max}}(t,\delta,x_0) &:= \sum_{i=1}^t\overline{z}^2(i,\delta,x_0) + \gamma, \label{ac:def:gram-max-bound}
\end{align}
\ssnotebluetwo{with $\gamma$ from \eqref{ac:eqn:estimator}. It} is a high probability upper bound on the maximum eigenvalue of $G(t)$, \ssnotebluetwo{which is established in the proof of Theorem~\ref{ac:theorem:rls-error-rpi}.
}

\ssnotebluetwo{Next, we define the \textit{contained time bound} as
\begin{align}
    T_{\textnormal{contained}}(\delta,x_0) := \sup \left \{ T \in \mathbb{N} \mid B_{\overline{x}(\ssnotebluetwo{T},\delta/3,x_0) }(0) \cap \mathbb{X} \subseteq \mathcal{X}_{\textnormal{RPI}} \right \}.   \label{ac:eqn:def-T-contained}
\end{align}
$T_{\textnormal{contained}}(\delta,x_0)$ can be interpreted as a high probability lower bound on the horizon over which the state $X(t)$ remains inside $\mathcal{X}_{\textnormal{RPI}}$.
}
Moreover, define the \textit{burn-in time bound} as
\begin{align}
        &T_{\textnormal{burn-in}}(\delta,x_0) := \inf \bigg \{  T \in \mathbb{N} \;\biggm\vert\; t \geq \frac{2}{(1 - \ln(2))  p_{\textnormal{PE}}}  \\
        & \quad \times \Bigg ( d \ln \mleft (1 + \frac{16 \sum_{i=1}^t \overline{z}^2 \mleft (i, \ssnoteblue{\delta/3}, x_0 \mright ) }{ c_{\textnormal{PE}3} p_{\textnormal{PE}} (t-1) } \mright ) \\
        & \quad + \ln \mleft ( \frac{\pi^2 (t - T + 1)^2 }{\ssnoteblue{2\delta}} \mright ) \Bigg )  + 1 \text{ for all } t \geq T \bigg \}, \label{eqn:theorem-rls-error-exciting-T-burn-in}
\end{align}
Intuitively, $T_{\textnormal{burn-in}}(\delta,x_0)$ can be viewed as a high probability upper bound on the time it takes for PE to start holding in the sense that $\frac{\ssnotebluetwo{c_{\textnormal{PE}}}p_{\textnormal{PE}}}{4} (t-1) + \gamma$ is a lower bound on $\lambda_{\textnormal{min}}(G(t))$, assuming that the predicted state $g(X(t-1),U(t-1),0)$ remains inside $\mathcal{X}_{\textnormal{PE}}$, \ssnotebluetwo{which notably occurs if $X(t-1)$ is inside $\mathcal{X}_{\textnormal{RPI}}$ due to Assumption~\ref{ac:assump:rpi}}. 

\ssnotebluetwo{
We also define $T_{\textnormal{converge}}$ as
\begin{align}
    T_{\textnormal{converge}}(\delta,x_0) := \max\left(T_{\textnormal{burn-in}}(\delta,x_0), \right. \\
    \left. \inf \left \{ T \in \mathbb{N} \mid e(t,\delta,x_0) \leq \bar{\vartheta}, \ \forall t \geq T  \right \}\right). \label{ac:eqn:def-T-converge}
\end{align}
$T_{\textnormal{converge}}(\delta,x_0)$ can be interpreted as a high probability upper bound on the time the parameter estimate $\hat{\theta}(t)$ converges into a $\bar{\vartheta}$-ball around the true parameter $\theta_*$, assuming that the predicted state $g(X(t-1),U(t-1),0)$ remains inside $\mathcal{X}_{\textnormal{PE}}$.
}

We are now ready to state Theorem~\ref{ac:theorem:rls-error-rpi}, which provides probabilistic guarantees on the estimation error $\vert \hat{\theta}_t - \theta_* \vert$, and the invariance of the state $X_t$ within the set $\mathcal{X}_{\text{RPI}}$, uniformly over all time.
\begin{theorem} \label{ac:theorem:rls-error-rpi}
    (Regional estimation error bound and positive invariance guarantee)
    Suppose Assumptions~\ref{ac:assump:measurable}, \ref{ac:assump:process-noise}, \ref{ac:assump:forward-complete}, \ref{ac:assump:constrained-controls}, \ref{ac:assump:poly-feature-map}, \ref{ac:assump:regional-excitation} and \ref{ac:assump:rpi} are satisfied. 
    Then, for any $x_0 \in \mathcal{X}_{\textnormal{RPI}}$ and $\delta \in (0,1)$, if 
    \begin{align}
        \ssnotebluetwo{T_{\textnormal{converge}}(\delta,x_0)} + 1 \leq T_{\textnormal{contained}}(\delta,x_0), \label{ac:eqn:theorem-rls-error-rpi-condition}
    \end{align}
    then
    \begin{align}
        \mathbb{P}  \Big(& \big \vert \hat{\theta}(t) - \theta_* \big \vert \leq e(t,\delta,x_0), \ \forall t \geq T_{\textnormal{burn-in}}(\delta,x_0) \\ 
        &\textnormal{and} \ X(t) \in \mathcal{X}_{\textnormal{RPI}}, \ \forall t \geq 0  \Big) \geq 1 - \delta, \label{ac:eqn:theorem-rls-error-rpi-main}
    \end{align}
    where the error bound $e(t,\delta,x_0)$ is defined as 
    \begin{align}
        &e(t,\delta,x_0) := \frac{1}{ \sqrt{ \ssnoteblue{\frac{c_{\textnormal{PE}}p_{\textnormal{PE}}}{4} (t-1) + \gamma }} } \Big( \gamma^{1/2} \mleft \vert \theta_* \mright \vert_F + \label{ac:eqn:theorem-rls-error-exciting-e} \\
        & \ \sigma_w  \sqrt {  2 n   \mleft (  \ln \mleft ( \ssnoteblue{3}n / \delta \mright )        + \mleft ( d/2 \mright ) \ln \mleft (  \beta_{\textnormal{max}}\mleft ( t,\ssnoteblue{\delta/3},x_0 \mright ) \gamma^{-1}  \mright )   \mright )   }  \bigg) 
    \end{align}
    and $\beta_{\textnormal{max}}$ is defined in \eqref{ac:def:gram-max-bound}.
    Moreover, for any $x_0 \in \mathcal{X}_{\textnormal{RPI}}$ and $\delta \in (0,1)$,
    \begin{enumerate}
        \item $\lim_{t \rightarrow \infty} e(t,\delta,x_0) = 0$;
        \item $T_{\textnormal{burn-in}}(\delta,x_0) < \infty$;
        \item $T_{\textnormal{converge}}(\delta,x_0) < \infty$.
    \end{enumerate}
\end{theorem}
Theorem~\ref{ac:theorem:rls-error-rpi} provides 1) an upper bound $e(t,\delta,x_0)$ on the estimation error $\left \vert \hat{\theta}(t) - \theta_* \right \vert$ that holds \ssnoteblue{uniformly} \ssnotebluetwo{over all $t \geq T_{\textnormal{burn-in}}(\delta,x_0)$}, \ssnotebluetwo{and 2) a guarantee of the invariance of $X(t)$ inside $\mathcal{X}_{\textnormal{RPI}}$ for all $t \in \mathbb{N}_0$,} \ssnotebluetwo{that are satisfied with probability $1-\delta$.} \ssnotebluetwo{These guarantees hold} whenever condition \eqref{ac:eqn:theorem-rls-error-rpi-condition} is satisfied, for any $\delta \in (0,1)$ and $x_0 \in \mathbb{X}$. 
Condition \eqref{ac:eqn:theorem-rls-error-rpi-condition} ensures that \ssnotebluetwo{with high probability, the estimate $\hat{\theta}(t-1)$ will converge into a $\bar{\vartheta}$-ball around $\theta_*$ before $X(t)$ escapes $\mathcal{X}_{\textnormal{RPI}}$. 
Then, since $\Gamma(\mathcal{X}_{\textnormal{RPI}}) \subseteq \mathcal{X}_{\textnormal{PE}}$, by making use of Assumption~\ref{ac:assump:rpi} which guarantees RPI when the estimation error is smaller than $\bar{\vartheta}$, it can be \textit{inductively} established for subsequent time steps that PE holds, ensuring the estimation error remains smaller than $\bar{\vartheta}$, so the state remains inside $\mathcal{X}_{\textnormal{RPI}}$, and so on, producing the invariance portion of the statement in \eqref{ac:eqn:theorem-rls-error-rpi-main}. 
}
\ssnotebluetwo{Simultaneously, the error bound portion of \eqref{ac:eqn:theorem-rls-error-rpi-main} holds for infinite time since the state trajectory is guaranteed to remain inside $\mathcal{X}_{\textnormal{RPI}}$ with high probability, ensuring PE holds (see Remark~\ref{ac:remark:comparing-ac-sysid} for a more detailed discussion). We also observe that the error bound is asymptotically convergent to zero in the sense that $e(t,\delta,x_0) \rightarrow \infty$ as $t \rightarrow \infty$, which is particularly insightful because it means that with probability at least $1- \delta$, the the parameter estimate $\hat{\theta}(t)$ will eventually enter and remain inside any arbitrarily small ball around $\theta_*$.}

\begin{remark} \label{ac:remark:comparing-ac-sysid}
    \ssnotebluetwo{Theorem~\ref{ac:theorem:rls-error-rpi} bears similarities with Theorem~1 from \cite{sysidpaper}. The formula for the error bound $e(t,\delta,x_0)$ on the estimation error $\vert \hat{\theta}(t) - \theta_* \vert$ is the same in both results, which means that the bound in this work is similarly smaller in systems experiencing a large degree of excitation and less instability. 
    Moreover, both guarantees require some kind of condition to be verified for a given $\delta$ and $x_0$. The key difference between the two is that the error bound in Theorem~\ref{ac:theorem:rls-error-rpi} holds for infinite time, whereas the one in \cite{sysidpaper} holds only over a finite interval. This occurs because in Theorem~\ref{ac:theorem:rls-error-rpi}, the RPI property of the controlled system in Assumption~\ref{ac:assump:rpi} combined with condition \eqref{ac:eqn:theorem-rls-error-rpi-condition} ensures the system remains inside $\mathcal{X}_{\textnormal{RPI}}$ for \textit{all time}, which guarantees PE holds over an infinite time interval. On the other hand, there is no RPI assumption in Theorem~1 from \cite{sysidpaper}, such that we are unable to characterise a probabilistic event where the state trajectory is remains inside a region guaranteeing PE for all time. Instead, Theorem~1 from \cite{sysidpaper} involves verifying a condition that enables the construction of a probabilistic event where the lower bound $\frac{\ssnotebluetwo{c_{\textnormal{PE}}}p_{\textnormal{PE}}}{4} (t-1) + \gamma$ on $\lambda_{\textnormal{min}}(G(t))$ starts holding before $X(t)$ escapes the exciting region, allowing us to derive an error bound over a finite time interval.}
\end{remark}

\ssnotebluetwo{Just like Theorem~1 from \cite{sysidpaper}}, \ssnoteblue{\eqref{ac:eqn:theorem-rls-error-rpi-condition} is verifiable offline, meaning that it does not rely on any online data collection.}
\ssnotebluetwo{Moreover, although $T_{\textnormal{converge}}(\delta,x_0)$ (and $T_{\textnormal{burn-in}}(\delta,x_0)$) are finite for any $\delta\in (0,1)$ and $x_0 \in \mathcal{X}_{\textnormal{RPI}}$, without choosing a particular system, we cannot say whether or not the condition will be satisfied. }
\ssnotebluetwo{However, qualitatively, it is more likely to hold in systems with a greater degree of excitation, and with less instability. This is because $T_{\textnormal{burn-in}}(\delta,x_0)$ is smaller and $e(t,\delta,x_0)$ decays faster when $\overline{z}(t,\delta/3,x_0)$ grows slowly, and $\ssnotebluetwo{c_{\textnormal{PE}}}$ and $p_{\textnormal{PE}}$ are larger, resulting in a smaller value for $T_{\textnormal{converge}}(\delta,x_0)$. Moreover, $T_{\textnormal{contained}}(\delta/3,x_0)$ is larger when $\overline{z}(t,\delta/3,x_0)$ grows slowly. On the other hand, it may not hold in systems with a smaller RPI set, small degree of excitation, and greater instability. Of particular note is that $T_{\textnormal{converge}}(\delta,x_0)$ is smaller when $\bar{\vartheta}$ is larger, and $T_{\textnormal{contained}}(\delta,x_0)$ is larger when $\mathcal{X}_{\textnormal{RPI}}$ is larger. However, $\bar{\vartheta}$ and $\mathcal{X}_{\textnormal{RPI}}$ are intrinsically linked, and a tradeoff between them is likely to exist in specific problems.}

\ssnotebluetwo{We now progress towards stating our main stability result with regional control and excitation. Before doing so, we introduce some concepts.}
Given a Borel measurable function $V:\mathcal{X} \rightarrow \mathbb{R}_{\geq 0}$ with $\mathcal{X} \subseteq \mathbb{X}$, we define 
\begin{align}
    \Delta V(x,\tilde{\vartheta}) := \mathbf{E} [ V(g(x,W,\alpha(x,S,\tilde{\vartheta} + \theta_*) ] - V(x) \label{ac:eqn:def-lyapunov-difference}
\end{align}
for $x \in \mathcal{X}$, $\tilde{\vartheta} \in \mathbb{R}^{d \times n}$. Note that $\tilde{\vartheta}$ in \eqref{ac:eqn:def-lyapunov-difference} does not represent the parameter estimate, but the estimation error.

\ssnotebluetwo{Next, we introduce the notion of a stochastic Lyapunov function.}
\begin{definition} \label{ac:def:lyapunov} (Stochastic Lyapunov function)
    The Borel measurable function $V: \mathcal{X}_{\textnormal{RPI}} \rightarrow \mathbb{R}_{\geq 0}$ over the domain $\mathcal{X}_{\textnormal{RPI}} \subseteq \mathbb{X}$ is said to be a stochastic Lyapunov function with the estimation error bound $\bar{\vartheta}$, for the system $g$ under the influence of the policy $\alpha$ and the noise distributions $\mu_w$, $\mu_s$, if 1) \textit{$(g,\alpha)$ is $(\mathcal{X}_{\textnormal{RPI}},\bar{\vartheta},\ssnotebluetwo{\mu_w,\mu_s})$-RPI}, and 2) there exist $\alpha_1, \alpha_2, \alpha_3 \in \mathcal{K}_{\infty}$, $\sigma_3 \in \mathcal{K}$, and $\tilde{d} \geq 0$, such that 
    \begin{align}
        &\alpha_1(\vert x \vert) \leq V(x) \leq \alpha_2(\vert x \vert ), \label{ac:eqn:prop-agip-conj-full-pd}  \\
        & \Delta V(x,\tilde{\vartheta}) \leq - \alpha_3(\vert x \vert ) + \tilde{d} + \sigma_3(\vert \tilde{\vartheta} \vert), \label{ac:eqn:prop-agip-conj-full-dissip}
    \end{align}
    are satisfied for all $x \in \mathcal{X}_{\textnormal{RPI}}$, and $\tilde{\vartheta} \in B_{\bar{\vartheta}}(0)$, and $\alpha_3 \circ \alpha_2^{-1}$ is lower bounded by some convex function in $\mathcal{K}_{\infty}$. For convenience, we say $(g,\alpha,\mu_w,\mu_s)$ is \textit{$(V,\mathcal{X}_{\textnormal{RPI}},\bar{\vartheta})$-stochastic Lyapunov}. \ssnotebluetwo{If $(g,\alpha,\mu_w,\mu_s)$ is \textit{$(V,\mathbb{X},\bar{\vartheta})$-stochastic Lyapunov}, we say that $(g,\alpha,\mu_w,\mu_s)$ is \textit{$(V,\bar{\vartheta})$-global stochastic Lyapunov}.}
\end{definition}
\ssnotebluetwo{Intuitively, the existence of the stochastic Lyapunov function in Definition~\ref{ac:def:lyapunov} implies that if the parameter $\vartheta$ used for control is sufficiently close to the true parameter $\theta_*$ such that $\vert \tilde{\vartheta} \vert = \vert \vartheta - \theta_* \vert \leq \bar{\vartheta}$, then from any initial state $x$ inside an RPI set $\mathcal{X}_{\textnormal{RPI}}$, we \textit{expect} the closed-loop trajectory to decrease in magnitude, up to some value that monotonically depends on $\tilde{d}$ and $\vert \tilde{\vartheta} \vert$. 
On the other hand, a \textit{global} stochastic Lyapunov function means that from any state $x$ over the entire state space $\mathbb{X}$, this decreasing property is expected.
The definition of the stochastic Lyapunov function in Definition~\ref{ac:def:lyapunov} is similar to existing definitions for stochastic Lyapunov functions in nonlinear stochastic stability theory, in particular the stochastic input-to-state Lyapunov function from \cite{mcallister2022stochastic}. The main difference is that the dissipation inequality \eqref{ac:eqn:prop-agip-conj-full-dissip} includes an additional determinisitc additive component $\sigma_3(\vert \tilde{\vartheta} \vert)$ representing the influence of the estimation error. 
Another key difference is the requirement that $\alpha_3 \circ \alpha_2^{-1}$ is lower bounded by some convex function in $\mathcal{K}_{\infty}$, which is trivially satisfied when $\mathcal{X}_{\textnormal{RPI}}$ is bounded as in \cite{mcallister2022stochastic}.
We assume the existence of a stochastic Lyapunov function for the remainder of this section.}
\begin{assumption} \label{ac:assump:lyapunov}
    There exist a Borel measurable function $V:\mathcal{X}_{\textnormal{RPI}} \rightarrow \mathbb{R}_{\geq 0}$ over $\mathcal{X}_{\textnormal{RPI}} \subseteq \mathbb{X}$ satisfying $\Gamma(\mathcal{X}_{\textnormal{RPI}}) \subseteq \mathcal{X}_{\textnormal{PE}}$, and a constant $\bar{\vartheta} > 0$, such that 
    $(g,\alpha,\mu_w,\mu_s)$ is $(V,\mathcal{X}_{\textnormal{RPI}},\bar{\vartheta})$-stochastic Lyapunov.
\end{assumption}

\ssnotebluetwo{We now our main stability result in this section.}
\begin{theorem} \label{ac:theorem:regional-stability} (Probabilistic regional stability bound)
    Suppose Assumptions~\ref{ac:assump:measurable}, \ref{ac:assump:process-noise}, \ref{ac:assump:forward-complete}, \ref{ac:assump:constrained-controls}, \ref{ac:assump:poly-feature-map}, \ref{ac:assump:regional-excitation}, and \ref{ac:assump:lyapunov} are satisfied. 
    Then, for all \ssnotebluetwo{$\delta \in (0,1)$}, there exists $c_2 \geq 0$ such that for any initial state $x_0 \in \mathcal{X}_{\textnormal{RPI}}$, if $(\delta,x_0)$ satisfy 
    \begin{align}
        \ssnotebluetwo{T_{\textnormal{converge}}(\delta/2,x_0)} + 1 \leq T_{\textnormal{contained}}(\ssnotebluethree{\delta/2},x_0), \label{ac:eqn:theorem-rls-error-rpi-condition-mod}
    \end{align}
    then there exists $\eta \in \mathcal{L}$ such that
    \begin{align}
        \mathbb{P}( \vert X(t) \vert \leq \eta(t) + c_2 ) \geq 1 - \ssnotebluethree{\delta} \label{ac:eqn:theorem-regional-stability-main}
    \end{align}
    for all $t \in \mathbb{N}_0$.
\end{theorem}
\ssnotebluetwo{Theorem~\ref{ac:theorem:regional-stability} says that, for any chosen failure probability $\delta \in (0,1)$, there is some constant $c_2 \geq 0$ such that for any initial state $x_0 \in \mathcal{X}_{\textnormal{RPI}}$, if $\delta$ and $x_0$ satisfy condition \eqref{ac:eqn:theorem-rls-error-rpi-condition-mod}, there exists some function $\eta \in \mathcal{L}$ such that magnitude of $X(t)$ will be less than $\eta(t) + c_2$ with probability at least $1-\delta$. We can interpret this magnitude bound as a \textit{stability bound} since it exhibits the properties of both \textit{boundedness} and \textit{convergence} --- boundedness in the sense that it is bounded from above by $\eta(0) + c_2$ over all time (with $\eta(0)$ dependent on $x_0$), and convergent in the sense that it converges to $c_2$ regardless of $x_0$. 
Intuitively, we can view $\eta(t)$ as a \textit{transient} component, and $c_2$ as a \textit{steady-state} component.
This result requires condition \eqref{ac:eqn:theorem-rls-error-rpi-condition-mod} to hold since it is essentially established by relying on the convergence of the estimate $\hat{\theta}(t)$ into a $\bar{\vartheta}$-ball around $\theta_*$, then using the existence of a Lyapunov function when the estimation error is sufficiently small in Assumption~\ref{ac:assump:lyapunov} to derive the stability bound. We require \eqref{ac:eqn:theorem-rls-error-rpi-condition-mod} rather than \eqref{ac:eqn:theorem-rls-error-rpi-condition} due to the use of the union bound in the derivation of Theorem~\ref{ac:theorem:regional-stability}.}

\begin{remark} \label{ac:remark:construct-stability-bounds}
\ssnotebluethree{
Although Theorem~\ref{ac:theorem:regional-stability} only talks about the existence of transient component $\eta$ and steady-state component $c_2$, they can be explicitly constructed. Given $\delta \in (0,1)$, $c_2$ can be constructed via
\begin{align}
    c_2 = \alpha_1^{-1} \circ \frac{2}{\delta} \tilde{\gamma}(2 \tilde{d}) \label{ac:eqn:transient-construction}
\end{align}
where $\alpha_1 \in \mathcal{K}_{\infty}$ and $\tilde{d}$ are part of Definition~\ref{ac:def:lyapunov}, and $\tilde{\gamma}\in \mathcal{K}_{\infty}$ is constructed as
\begin{align}
    &\tilde{\gamma}(r):=2\max ( \alpha_v^{-1}(r), r ).
\end{align}
Here, $\alpha_v$ is any convex function in $\mathcal{K}_{\infty}$ that lower bounds $\alpha_3 \circ \alpha_2^{-1}$, whose existence is required in Definition~\ref{ac:def:lyapunov}.
On the other hand, given $x_0 \in \mathcal{X}_{\textnormal{RPI}}$ and $\delta \in (0,1)$, $\eta$ can be constructed via
\begin{align}
    \eta(t) := \max_{t' \geq t} \tilde{\eta}(t') \label{ac:eqn:steady-state-construction}
\end{align}
where $\tilde{\eta}$ is $x_0$- and $\delta$-dependent and in turn constructed via \eqref{eqn:eta-tilde}. In turn, it relies on
\begin{align}
    \lambda_1(r) &= r - \alpha_v(r) + \alpha_v(r/2), \\
    \lambda(r) &= \frac{1}{2}(r + \max_{r' \in [0,1]} \lambda_1(r')), \\
    \beta_1(r,t) &= \alpha_1^{-1} \circ \frac{2}{\delta} \lambda^t \circ \alpha_2(r), \\
    \beta_2(r,t)&=\alpha_1^{-1} \circ \frac{2}{\delta} \lambda^{t_0 + \ceil{t/2}} \circ \tilde{\gamma}(2\sigma_3(r) ), \\
    \eta_2(t)&=\alpha_1^{-1} \circ \frac{2}{\delta} \lambda^{t_0 + \ceil{t/2}} \circ \tilde{\gamma}( 2\tilde{d} ), \\
    \gamma_3(r) &= \alpha_1^{-1} \circ \frac{2}{\delta}\tilde{\gamma}( 2\sigma_3( r  ) ).\\
    \
\end{align}
Here, $\alpha_1,\alpha_2,\alpha_3 \in \mathcal{K}_{\infty}$ and $\sigma_3 \in \mathcal{K}$ are all from Definition~\ref{ac:def:lyapunov}, $\overline{x}$ is from \eqref{ac:def:high-prob-state-bound}, and $e$ is from \eqref{ac:eqn:theorem-rls-error-rpi-main}.
\ssnotebluetwo{Note that the steady-state component $c_2$ does \textit{not} depend on the estimation error bound $e(t,\delta/2,x_0)$, only on $\tilde{d}$. Instead, this information is captured entirely in the transient component $\eta(t).$}
}

\end{remark}
\begin{widetext}
\begin{align}
    \tilde{\eta}(t) &= \begin{cases} \overline{x}(T_{\textnormal{converge}}(\delta/2,x_0)+1,\delta/6,x_0), \quad 0 \leq t \leq T_{\textnormal{converge}}(\delta/2,x_0), \\
        \max \bigg( \beta_1\left ( \overline{x}\left (T_{\textnormal{converge}}\left (\delta/2,x_0\right )+1,\delta/6,x_0\right ), t - \left (T_{\textnormal{converge}}\left (\delta/2,x_0\right )+1\right )\right ), \\
        \quad \qquad \eta_2\left ( t - \left (T_{\textnormal{converge}}\left (\delta/2,x_0\right )+1\right )\right )  \\
        \quad \qquad + \beta_2(  \max_{T_{\textnormal{converge}}\left (\delta/2,x_0\right ) \leq i \leq T_{\textnormal{converge}}\left (\delta/2,x_0\right ) + \lfloor \left (t - \left (T_{\textnormal{converge}}\left (\delta/2,x_0\right ) + 1\right )\right )/2 \rfloor} e\left (i,\delta/2,x_0\right ), \\
        \quad \qquad \quad \qquad t - (T_{\textnormal{converge}}(\delta/2,x_0)+1) ), \\
        \quad \qquad \gamma_3\left ( \max_{T_{\textnormal{converge}}\left (\delta/2,x_0\right )+1 + \floor{\left (t-\left (T_{\textnormal{converge}}\left (\delta/2,x_0\right )+1\right )\right )/2}\leq i \leq t-1} e\left (i,\delta/2,x_0\right ) \right ) \bigg), \\
        \quad \qquad \quad t \geq T_{\textnormal{converge}}(\delta/2,x_0) + 1,
        \end{cases} \label{eqn:eta-tilde}
\end{align}
\end{widetext}

\subsection{Stability with Global Control and Excitation} \label{ac:sec:stability-global}

\ssnotebluetwo{We now assume global excitation and the existence of a global stochastic Lyapunov function, and establish Corollary~\ref{ac:corollary:global-stability} under these assumptions.}
\begin{assumption} \label{ac:assump:global-excitation}
    $(\psi,\alpha,\mu_w,\mu_s)$ is \ssnotebluetwo{$(c_{\textnormal{PE}},p_{\textnormal{PE}})$\textit{-globally excited}} for some constants \ssnotebluetwo{$c_{\textnormal{PE}},p_{\textnormal{PE}} > 0$}.
\end{assumption}

\begin{assumption} \label{ac:assump:global-lyapunov}
    There exist a Borel measurable function $V:\mathbb{X} \rightarrow \mathbb{R}_{\geq 0}$, and a constant $\bar{\vartheta} > 0$ such that 
    $(g,\alpha,\mu_w,\mu_s)$ is $(V,\bar{\vartheta})$-global stochastic Lyapunov.
\end{assumption}

\begin{corollary} \label{ac:corollary:global-stability} (High probability global stability bound)
    Suppose Assumptions~\ref{ac:assump:measurable}, \ref{ac:assump:process-noise}, \ref{ac:assump:forward-complete}, \ref{ac:assump:constrained-controls}, \ref{ac:assump:poly-feature-map}, \ref{ac:assump:global-excitation}, and \ref{ac:assump:global-lyapunov} are satisfied. 
    Then, for all $\delta \in (0,\ssnotebluethree{1})$, there exists $c_2 \geq 0$ such that for all $x_0 \in \mathbb{X}$, there exists $\eta \in \mathcal{L}$ such that
    \begin{align}
        \mathbb{P}( \vert X(t) \vert \leq \eta(t) + c_2 ) \geq 1 - \ssnotebluethree{\delta}. \label{ac:en:corollary-global-stability-main}
    \end{align}
    for all $t \in \mathbb{N}_0$.
\end{corollary}

\ssnotebluetwo{Corollary~\ref{ac:corollary:global-stability} is similar to Theorem~\ref{ac:theorem:regional-stability}, but considers the specific case where global excitation holds and a global stochastic Lyapunov function exists.
The key difference between the two is that under these strengthened assumptions, condition \eqref{ac:eqn:theorem-rls-error-rpi-condition} no longer needs to be manually verified, such that the statement in \eqref{ac:en:corollary-global-stability-main} can be established for any $x_0$ and $\delta$, resulting in a high probability stability bound.} \ssnotebluefour{The reason for this is explained in Remark~\ref{ac:remark:global-stability}. Moreover, the transient component $\eta(t)$ and steady-state component $c_2$ of the bound in \eqref{ac:en:corollary-global-stability-main} can also be constructed via \eqref{ac:eqn:transient-construction} and \eqref{ac:eqn:steady-state-construction} from Remark~\ref{ac:remark:construct-stability-bounds}.}

\begin{remark} \label{ac:remark:global-stability}
    \ssnotebluefour{Corollary~\ref{ac:corollary:global-stability} follows from Theorem~\ref{ac:theorem:regional-stability} since under}
    Assumptions~\ref{ac:assump:global-excitation} and \ref{ac:assump:global-lyapunov}, Assumption~\ref{ac:assump:regional-excitation} holds with $\mathcal{X}_{\textnormal{PE}}=\mathbb{X}$ and Assumption~\ref{ac:assump:lyapunov} holds with $\mathcal{X}_{\textnormal{RPI}} = \mathbb{X}$, and moreover $T_{\textnormal{contained}}(\delta,x_0) = \infty$, such that condition \eqref{ac:eqn:theorem-rls-error-rpi-condition} is automatically verified. 
\end{remark}

\section{Examples} \label{ac:sec:control-examples}
In Example~1, we consider a piecewise affine (PWA) system with unknown parameters that is only regionally controllable and excited. We propose an adaptive control strategy based on the framework in Algorithm~\ref{ac:alg:framework}, and subsequently show that the assumptions required to establish Theorems~\ref{ac:theorem:rls-error-rpi} and ~\ref{ac:theorem:regional-stability} both hold. This provides probabilistic guarantees on the positive invariance of the system on the excited and controllable region, alongside non-asymptotic bounds on the estimation error and stability bounds.

In Example~2, we propose an adaptive control strategy for input-constrained stochastic linear systems based on the framework described in Algorithm~\ref{ac:alg:framework}, and subsequently show how Corollary~\ref{ac:corollary:global-stability} can be used to establish high probability stability guarantees for the closed-loop system.
\subsection{Example 1: Piecewise-Affine System}

We consider stochastic systems of the form
\begin{align}
    X(t+1) = X(t) + 0.1 \mathbf{1}_{ \{ \vert X(t) \vert \leq \bar{x} \} } U(t) + W(t), \ t \in \mathbb{N}, \\
    X(0) = x_0 \in \mathbb{R}, \qquad  \label{ac:eqn:example-1-system}
\end{align}
where $\{X(t)\}_{t \in \mathbb{N}_0}$, $\{U(t)\}_{t \in \mathbb{N}_0}$ and $\{W(t)\}_{t \in \mathbb{N}}$ are random sequences taking values in $\mathbb{R}$, and $x_0$ is the initial state. 
Here, $\bar{x} > 0$ determines region where \eqref{ac:eqn:example-1-system} is controllable.
We assume $\{W(t)\}_{t \in \mathbb{N}}$ is sampled i.i.d. from $\mathrm{Uniform}(-\bar{w},\bar{w})$ with $\bar{w} > 0$.
The nominal system $f$, true unknown parameter $\theta_*$ and basis function $\psi$ in \eqref{ac:eqn:system-dynamics} then correspond to
\begin{align}
    f(x,u) = 0, \quad \theta_* = \begin{bmatrix} 1 \\ 0.1 \end{bmatrix}, \quad \psi(x,u)= \begin{bmatrix} x \\ \mathbf{1}_{ \{ \tilde{x} \mid \vert \tilde{x} \vert \leq \bar{x} \} } u \end{bmatrix}, \label{ac:eqn:example-1-model}
\end{align}
and $\mu_w$ is the probability measure associated with $\mathrm{Uniform}(-\bar{w},\bar{w})$, such that $\mathbb{W} = [-\bar{w},\bar{w}]$.

We now describe the adaptive control strategy based on the framework in Algorithm~\ref{ac:alg:framework}.
Before doing so, given $r>0$ and $x \in \mathbb{R}^n$, let us define the radial saturation function as $\sat_r(x):=x$ if $\vert x \vert \leq r$ and $\sat_r(x) = \frac{x}{\vert x \vert}r$ if $\vert x \vert \leq r$, and let $M^{\dagger}$ denote the pseudo-inverse of the matrix $M$. 
Now, let $\bar{u}_1, \bar{u}_2>0$ be constants, and let $\bar{u}_{\textnormal{max}} = \bar{u}_1 + \bar{u}_2$. Moreover, let $\mu_s$ be the probability measure associated with $\mathrm{Uniform}(-\bar{s},\bar{s})$ where $\bar{s}>0$, such that $\mathbb{S} = [-\bar{s},\bar{s}]$. The family of control policies $\alpha$ chosen for our framework is then given by
\begin{align}
    \alpha(x,s,\vartheta):= \mathrm{sat}_{\bar{u}_1}(-\vartheta_2^{\dagger}\vartheta_1 x) + s
\end{align}
for $s \in \mathbb{S}$ and $\vartheta = \begin{bmatrix} \vartheta_1 & \vartheta_2  \end{bmatrix}^{\top} \in \mathbb{R}^2$.

Next, we verify the assumptions required to establish Theorems~\ref{ac:theorem:rls-error-rpi} and~\ref{ac:theorem:regional-stability}. In order to do so, we assume that the objects in the problem setup satisfy the following:
\begin{align}
    &\bar{x}\geq\bar{w}+0.1(u_{\textnormal{max}}+\bar{u}_1), \label{ac:eqn:ac-example-regional-18} \\
    &0.1\bar{u}_1 > 0.1 \bar{s} + \bar{w}. \label{ac:eqn:ac-example-regional-19}
\end{align} 
We start by verifying Assumption~\ref{ac:assump:measurable}. We know $f(x,u)=0$ is Borel measurable since constants are Borel measurable, $\psi(x,u)= \begin{bmatrix} x & \mathbf{1}_{ \{ \tilde{x} \mid \vert \tilde{x} \vert \leq \bar{x} \} }(x) u \end{bmatrix}^{\top}$ is Borel measurable since the indicator function of a Borel measurable set is also Borel measurable, and Borel measurability is preserved under multiplication and coordinate functions. Moreover, $\alpha(x,s,\vartheta)= \mathrm{sat}_{\bar{u}_1}(-\vartheta_2^{\dagger}\vartheta_1 x) + s$ is Borel measurable since $\mathrm{sat}_{\bar{u}_1}(\cdot)$ and the pseudo-inverse $(\cdot)^{\dagger}$ of a matrix are Borel measurable, and Borel measurability is preserved under multiplication, addition, and composition (see \cite{kallenberg1997foundations} for these standard results).

Next, Assumption~\ref{ac:assump:process-noise} holds since $\{W(t)\}_{t \in \mathbb{N}}$ is i.i.d., $\{W(t)\}_{t \in \mathbb{N}}$ and $\{S(t)\}_{t \in \mathbb{N}_0}$ are mutually independent, and $W(t)$ is zero-mean and $\bar{w}^2$-sub-Gaussian for all $t \in \mathbb{N}$ (since it has a bounded support).

We now verify Assumption~\ref{ac:assump:forward-complete}. Firstly, note that for any initial state $\xi \in \mathbb{X}$, time $t \in \mathbb{N}_0$, and deterministic input and noise sequences $\{u(i)\}_{i=0}^{\tau-1} \in \mathbb{U}^t$ and $\{w(i)\}_{i=1}^t \in \mathbb{W}^t$, the deterministic trajectory $\phi$ corresponding to the system \eqref{ac:eqn:example-1-system} satisfies
\begin{align}
    &\mleft \vert \phi(t,\xi, \{ u(i) \}_{i=0}^{t-1}, \{ w(i) \}_{i=1}^t ) \mright \vert \\
    & \quad \leq \mleft \vert \xi \mright \vert + 0.1 \mleft ( \sum_{i=0}^{t-1} \mleft \vert u(i) \mright \vert \mright ) + \sum_{i=1}^t \mleft \vert w(i) \mright \vert
\end{align}
Thus, Assumption~\ref{ac:assump:forward-complete} is satisfied with $\chi_2(\cdot) = \mathrm{Id}(\cdot)$, $\chi_3(\cdot)=0.1 \mathrm{Id}(\cdot)$, $\chi_4(\cdot) =\sigma_1(\cdot)=\sigma_2(\cdot)=\mathrm{Id}(\cdot)$, and with arbitrary $\chi_1 \in \mathcal{K}_{\infty}^{\textnormal{1-SE}}$.

Next, Assumption~\ref{ac:assump:constrained-controls} holds since for all $x \in \mathbb{X}$, $s \in \mathbb{S}$ and $\vartheta \in \mathbb{R}$,
\begin{align}
    \vert \alpha(x,s,\vartheta) \vert  \leq \vert \mathrm{sat}_{\bar{u}_1}(-\vartheta^{\dagger} x ) + s \vert \leq \bar{u}_1 + \bar{u}_2 = u_{\textnormal{max}}.
\end{align}

Note that Assumption~\ref{ac:assump:poly-feature-map} is satisfied with $\chi_4(\cdot) = \mathrm{Id}(\cdot)$, since
\begin{align}
    \vert \psi(x,u) \vert = \vert \mathbf{1}_{ \{ \tilde{x} \mid \vert \tilde{x} \vert \leq \bar{x} \} } ( x ) u \vert \leq \vert u \vert. 
\end{align}

We now verify Assumption~\ref{ac:assump:regional-excitation}. \ssnotebluefour{The steps we take bear similarities to Example~1 from \cite{sysidpaper}, where regional excitation was also established by making use of Lemma~\ref{ac:lemma:regional-excitation-sufficient}. There are some minor differences, primarily due to the assumption that the process noise distribution is bounded, and that the control inputs applied do not consist solely of injected noise, but also included a component computed from past data. This required us to make use of some steps inspired by Example~2 in \cite{sysidpaper}, which also included such a component.}
Let $\mathcal{X}_{\textnormal{PE}} := [-(\bar{x} - \bar{w}), \bar{x} - \bar{w}]$.
Suppose $x \in \mathcal{X}_{\textnormal{PE}}$, $\vartheta = \begin{bmatrix} \vartheta_1 & \vartheta_2 \end{bmatrix}^{\top} \in \mathbb{R}^2$, and $\zeta \in \mathcal{S}^1$. 
Then, 
\begin{align}
    & \mathbf{E} \mleft [ \mleft \vert \zeta^{\top} \psi(x+W, \alpha(x+W,S,\vartheta) \mright \vert \mright ] \\
    &= \mathbf{E} \mleft [ \mleft \vert \zeta_1(x+W) + \zeta_2\mathbf{1}_{\{ \tilde{x} \mid \vert \tilde{x} \vert \leq \bar{x} \}}(x + W) \mright. \mright. \\
    &\quad \quad  \mleft. \mleft. \cdot  (\mathrm{sat}_{\bar{u}_1}(-\vartheta_2^{\dagger}\vartheta_1^{\kappa}(x+W)) + S) \mright \vert \mright ] \\
    &= \mathbf{E} \mleft [ \mleft \vert \zeta_1(x+W) + \zeta_2  (\mathrm{sat}_{\bar{u}_1}(-\vartheta_2^{\dagger}\vartheta_1^{\kappa}(x+W)) + S) \mright \vert \mright ] \quad  \label{ac:eqn:ac-example-regional-12}
\end{align}
Moreover,
\begin{align}
    &\mathbf{E} \mleft [ \mleft \vert \zeta_1(x+W) + \zeta_2 (\mathrm{sat}_{\bar{u}_1}(-\vartheta_2^{\dagger}\vartheta_1^{\kappa}(x+W)) + S) \mright \vert \mright ] \\
    &= \mathbf{E} \mleft [ \mathbf{E} \mleft [ \mleft \vert \zeta_1(x+W) \mright. \mright. \mright. \\
    &\quad \mleft. \mleft. \mleft. + \zeta_2 (\mathrm{sat}_{\bar{u}_1}(-\vartheta_2^{\dagger}\vartheta_1^{\kappa}(x+W)) + S) \mright \vert \mid S \mright ] \mright ] \label{ac:eqn:ac-example-regional-4} \\
    & \geq \mathbf{E} \Big [ \Big \vert \mathbf{E} \Big[ \zeta_1(x+W) \\
    & \quad \mleft. \mleft. \mleft. + \zeta_2 (\mathrm{sat}_{\bar{u}_1}(-\vartheta_2^{\dagger}\vartheta_1^{\kappa}(x+W)) + S) \mid S  \mright ] \mright \vert\mright ] \\
    &= \mathbf{E} \mleft [ \mleft \vert \zeta_1 x + \zeta_2 \mathbf{E} \mleft [ \mathrm{sat}_{\bar{u}_1}(-\vartheta_2^{\dagger}\vartheta_1^{\kappa}(x+W)) \mright ] \mright. \mright. \\
    & \quad + \zeta_2 S \Big \vert\Big ] \label{ac:eqn:ac-example-regional-5} \\
    &\geq \vert \zeta_2 \vert \mathbf{E}\mleft[ \mleft \vert S \mright \vert \mright] = \vert \zeta_2 \vert \frac{\bar{u}_2}{2} \label{ac:eqn:ac-example-regional-6}
\end{align}
where \eqref{ac:eqn:ac-example-regional-4} follows from the tower property, \eqref{ac:eqn:ac-example-regional-5} follows from the independence of $W$ and $S$, and \eqref{ac:eqn:ac-example-regional-6} follows from the optimality property of medians and $ \mathbf{E}\mleft[ \mleft \vert S \mright \vert \mright] = \frac{\bar{u}_2}{2}$. 
Similarly, we have
\begin{align}
    &\mathbf{E} \mleft [ \mleft \vert \zeta_1(x+W) + \zeta_2 (\mathrm{sat}_{\bar{u}_1}(-\vartheta_2^{\dagger}\vartheta_1^{\kappa}(x+W)) + S) \mright \vert \mright ] \\
    &= \mathbf{E} \Big [ \mathbf{E} \Big [ \Big \vert \zeta_1(x+W) \\
    &\quad \mleft. \mleft. \mleft. + \zeta_2 (\mathrm{sat}_{\bar{u}_1}(-\vartheta_2^{\dagger}\vartheta_1^{\kappa}(x+W)) + S) \mright \vert \mid W \mright ] \mright ] \label{ac:eqn:ac-example-regional-7} \\
    & \geq \mathbf{E} \Big [ \Big \vert \mathbf{E} \Big [ \zeta_1(x+W) +    \\
    & \quad \mleft. \mleft. \mleft. \zeta_2 (\mathrm{sat}_{\bar{u}_1}(-\vartheta_2^{\dagger}\vartheta_1^{\kappa}(x+W)) + S) \mid W  \mright ] \mright \vert\mright ] \\
    &= \mathbf{E} \mleft [ \mleft \vert \zeta_1(x+W) + \zeta_2 \mathrm{sat}_{\bar{u}_1}(-\vartheta_2^{\dagger}\vartheta_1^{\kappa}(x+W)) \mright \vert\mright ] \label{ac:eqn:ac-example-regional-8} \\
    &\geq \mleft \vert \zeta_1 \mright \vert \mathbf{E} \mleft [ \mleft \vert (x+W) \mright \vert\mright ] \\
    & \quad - \mleft \vert \zeta_2 \mright \vert \mathbf{E} \mleft [ \mleft \vert \mathrm{sat}_{\bar{u}_1}(-\vartheta_2^{\dagger}\vartheta_1^{\kappa}(x+W)) \mright \vert\mright ] \label{ac:eqn:ac-example-regional-9} \\
    &\geq \mleft \vert \zeta_1 \mright \vert \mathbf{E} \mleft [ \mleft \vert W \mright \vert\mright ] - \mleft \vert \zeta_2 \mright \vert \overline{u}_1 \label{ac:eqn:ac-example-regional-10} \\
    &= \vert \zeta_1 \vert \frac{\bar{w}}{2} - \mleft \vert \zeta_2 \mright \vert \overline{u}_1 \label{ac:eqn:ac-example-regional-11}
\end{align}
where \eqref{ac:eqn:ac-example-regional-7} follows from the tower property, \eqref{ac:eqn:ac-example-regional-8} follows from the independence of $W$ and $S$, \eqref{ac:eqn:ac-example-regional-9} follows from the reverse triangle inequality, \eqref{ac:eqn:ac-example-regional-10} follows from the optimality property of medians and the fact that $\mathrm{sat}_{\bar{u}_1}$ is upper bounded by $\overline{u}_1$, and \eqref{ac:eqn:ac-example-regional-11} follows from $\mathbf{E}\mleft [ \mleft \vert W \mright \vert \mright ] = \frac{\bar{w}}{2}$.
Note that if $\vert \zeta_2 \vert \leq \frac{\bar{w}}{ 4 \bar{u}_1 + 2 \bar{w} }$, then $\vert \zeta_1 \vert \geq 1 - \frac{\bar{w}}{ 4 \bar{u}_1 + 2 \bar{w} }$, and so $\vert \zeta_1 \vert \frac{\bar{w}}{2} - \mleft \vert \zeta_2 \mright \vert \overline{u}_1 \geq \mleft ( 1 - \frac{\bar{w}}{ 4 \bar{u}_1 + 2 \bar{w} } \mright ) \frac{\bar{w}}{2} - \frac{\bar{w} \bar{u}_1}{ 4 \bar{u}_1 + 2 \bar{w} } = \frac{\bar{w}}{4}$. 
Moreover, if $\vert \zeta_2 \vert > \frac{\bar{w}}{ 4 \bar{u}_1 + 2 \bar{w} }$, then $\vert \zeta_2 \vert \frac{\overline{u}_2}{2} > \frac{\bar{w} \bar{u}_2}{ 8 \bar{u}_1 + 4 \bar{w} }$. Combining this with \eqref{ac:eqn:ac-example-regional-12}, \eqref{ac:eqn:ac-example-regional-6} and \eqref{ac:eqn:ac-example-regional-11}, we have
\begin{align}
    &\mathbf{E} \mleft [ \mleft \vert \zeta^{\top} \psi(x+W, \alpha(x+W,S,\vartheta) \mright \vert \mright ] \\
    &\geq \max \mleft ( \vert \zeta_2 \vert \frac{\overline{u}_2}{2}, \vert \zeta_1 \vert \frac{\bar{w}}{2} - \mleft \vert \zeta_2 \mright \vert \overline{u}_1  \mright ) \\
    &\geq \min \mleft ( \frac{\bar{w}}{4}, \frac{\bar{w} \bar{u}_2}{ 8 \bar{u}_1 + 4 \bar{w} } \mright ) =: c_{\textnormal{PE1}}.
\end{align}
On the other hand, we have
\begin{align}
    &\mathbf{V}\mathrm{ar} \mleft ( \mleft \vert \zeta^{\top} \psi(x+W, \mathrm{sat}_{\bar{u}_1}(-\vartheta_2^{\dagger}\vartheta_1(x+W)) + S ) \mright \vert \mright ) \\
    &= \mathbf{V}\mathrm{ar} \Big ( \Big \vert \zeta_1(x+W) + \zeta_2 \mathbf{1}_{\{ \tilde{x} \mid \vert \tilde{x} \vert \leq \bar{x} \}}(x + W)   \\
    & \quad \mleft. \mleft. \cdot (\mathrm{sat}_{\bar{u}_1}(-\vartheta_2^{\dagger}\vartheta_1(x+W)) + S) \mright \vert \mright ) \\
    &= \mathbf{V}\mathrm{ar} \mleft ( \mleft \vert \zeta_1(x+W) + \zeta_2 (\mathrm{sat}_{\bar{u}_1}(-\vartheta_2^{\dagger}\vartheta_1(x+W)) + S) \mright \vert \mright ) \\
    &= \mathbf{E} \mleft [ \mleft ( \zeta_1 W + \zeta_2 S + \zeta_2 \mleft ( \mathrm{sat}_{\bar{u}_1}(-\vartheta_2^{\dagger}\vartheta_1(x+W)) \mright. \mright. \mright. \\
    & \quad \mleft. \mleft. \mleft. - \mathbf{E} \mleft[ \mathrm{sat}_{\bar{u}_1}(-\vartheta_2^{\dagger}\vartheta_1(x+W)) \mright ] \mright ) \mright )^2 \mright ] \\
    &\leq 3  \bigg(\mathbf{E} \mleft [ \mleft ( \zeta_1 W \mright )^2 \mright ] + \mathbf{E} \mleft [ \mleft( \zeta_2 S \mright )^2 \mright ] \\
    & \quad + \mathbf{E} \mleft [ \mleft ( \zeta_2 \mleft (\mathrm{sat}_{\bar{u}_1}(-\vartheta_2^{\dagger}\vartheta_1(x+W)) \mright. \mright. \mright. \\
    & \quad \mleft. \mleft. \mleft. - \mathbf{E} \mleft[ \mathrm{sat}_{\bar{u}_1}(-\vartheta_2^{\dagger}\vartheta_1(x+W)) \mright ] \mright ) \mright )^2 \mright ]   \bigg) \label{ac:eqn:ac-exampple-regional-13} \\
    &\leq 3  \max \mleft ( \frac{\bar{w}^2}{3}, \frac{\overline{u}_2^2}{3} + 4\overline{u}_1^2 \mright ) \label{ac:eqn:ac-example-regional-14}
\end{align}
where \eqref{ac:eqn:ac-exampple-regional-13} follows from the QM-AM inequality, and \eqref{ac:eqn:ac-example-regional-14} follows from $\mathbf{E} \mleft [ W^2 \mright ]  \leq \frac{\bar{w}^2}{3}$, $\mathbf{E}\mleft [ S^2 \mright ] \leq \frac{\overline{u}_2^2}{3}$ and $\mathbf{E} \mleft [ \mleft ( \zeta_2 \mleft (\alpha_2(x+W,\vartheta) - \mathbf{E} \mleft[ \alpha_2(x+W,\vartheta) \mright ] \mright ) \mright )^2 \mright ] \leq 4 \bar{u}_1^2$.
Thus, using Lemma~\ref{ac:lemma:regional-excitation-sufficient} we find that $(\psi,\alpha,\mu_s,\mu_w)$ is $(\mathcal{X}_{\textnormal{PE}},c_{\textnormal{PE}},p_{\textnormal{PE}})$\textit{-regionally excited} with 
$c_{\textnormal{PE}} := \frac{1}{4} c_{\textnormal{PE}1}^2$ and $p_{\textnormal{PE}} := \frac{1}{4} \mleft( \frac{ c_{\textnormal{PE}2}}{c_{\textnormal{PE}1}^2} + 1 \mright)^{-1}$, such that Assumption~\ref{ac:assump:regional-excitation} is satisfied.

Next, we verify that Assumption~\ref{ac:assump:rpi} holds.
Before doing so, we note the following lemma holds (as a trivial consequence of Lemma~1 from \cite{siriya2023stability}.
\begin{lemma} \label{ac:lemma:example-local-lipschitz}
    Consider $A \in \mathbb{R}^{n \times n}$ and $B \in \mathbb{R}^{n \times m}$ with $n,m \in \mathbb{N}$. Suppose $A$ is full rank, $\vert A \vert \leq 1$ and $(A,B)$ is $\kappa$-step reachable, that is $\mathrm{rank} \mleft ( \mathcal{R}_{\kappa}(A,B) \mright )=n$ for some $\kappa \in \mathbb{N}$. Let $\bar{u} > 0$ be a constant. Then, there exist $m,C>0$ such that for any $x \in \mathbb{R}^n$ and $\hat{\vartheta}=\begin{bmatrix} \hat{\vartheta}_1 & \hat{\vartheta}_2 \end{bmatrix}^{\top} \in \bar{B}_m(\theta_*)$,
    \begin{align}
    \mleft \vert \mathrm{sat}_{\bar{u}}(\hat{\vartheta}_2^{\dagger}\hat{\vartheta}_1x) - \mathrm{sat}_{\bar{u}}(\mathcal{R}_{\kappa}(A,B)^{\dagger}A^{\kappa}x) \mright \vert \leq C \mleft \vert \hat{\vartheta} - \vartheta_* \mright \vert.
    \end{align}
\end{lemma}
Now, let $\mathcal{X}_{\textnormal{RPI}}:=[-(\bar{x}-\bar{w}-0.1u_{\textnormal{max}}),\bar{x}-\bar{w}-0.1u_{\textnormal{max}}]$. Moreover, let $\bar{\vartheta},C>0$ be such that  $0.1\bar{u}_1 \geq 0.1 (C\bar{\vartheta} + \bar{s}) + \bar{w}$, whose existence is verified via the assumption in \eqref{ac:eqn:ac-example-regional-19} and Lemma~\ref{ac:lemma:example-local-lipschitz}. 
Then, note that $\Gamma(\mathcal{X}_{\textnormal{RPI}}) \subseteq \mathcal{X}_{\textnormal{PE}}$.
Furthermore, for all $x \in \mathcal{X}_{\textnormal{RPI}}$, $u \in \mathbb{U}$, $s\in \mathbb{S}$ and $\hat{\vartheta} = \begin{bmatrix} \hat{\vartheta}_1 & \hat{\vartheta}_2 \end{bmatrix}^{\top}\in\bar{B}_{\bar{\vartheta}}(\theta_*)$, we have
\begin{align}
    &\vert g(x,\alpha(x,s,\hat{\vartheta}_2^{\dagger}\hat{\vartheta}_1),w) \vert\\
    &=\vert x + 0.1 \mathbf{1}_{ \{ \tilde{x} \mid \vert \tilde{x} \vert \leq \bar{x} \} }(x) \cdot (\mathrm{sat}_{\bar{u}_1}(-\hat{\vartheta}_2^{\dagger}\hat{\vartheta}_1x) + s) + w \vert \\
    &=\vert x + 0.1 (\mathrm{sat}_{\bar{u}_1}(-\hat{\vartheta}_2^{\dagger}\hat{\vartheta}_1x) + s) + w \vert \\
    &=\vert x + 0.1 \mathrm{sat}_{\bar{u}_1}(-10 x) \\
    & \quad + 0.1 ( s + \mathrm{sat}_{\bar{u}_1}(-\hat{\vartheta}_2^{\dagger}\hat{\vartheta}_1 x) - \mathrm{sat}_{\bar{u}_1}(-10 x) ) + w \vert\\
    &\leq \vert x + 0.1 \mathrm{sat}_{\bar{u}_1}(-10 x) \vert \\
    & \quad + \vert 0.1 ( s + \mathrm{sat}_{\bar{u}_1}(-\hat{\vartheta}_2^{\dagger}\hat{\vartheta}_1 x) - \mathrm{sat}_{\bar{u}_1}(-10 x) ) + w \vert\\
    &\leq \vert x + 0.1 \mathrm{sat}_{\bar{u}_1}(-10 x) \vert + 0.1( C \bar{\vartheta} + \bar{s} ) + \bar{w} \\
    &\leq \bar{x} - \bar{w} - 0.1u_{\textnormal{max}},
\end{align}
where the final inequality follows from the fact that making use of \eqref{ac:eqn:ac-example-regional-18} and \eqref{ac:eqn:ac-example-regional-19}, we know that
if $\vert x \vert \leq 0.1 \bar{u}_1$ and $\vert x \vert \leq \bar{x} - \bar{w} - 0.1 u_{\textnormal{max}}$, then $\vert x + 0.1 \mathrm{sat}_{\bar{u}_1}(-10 x) \vert + 0.1( C \bar{\vartheta} + \bar{s} ) + \bar{w}=0.1( C\bar{\vartheta} + \bar{s} ) + \bar{w} \leq 0.1 \bar{u}_1 \leq \bar{x} - \bar{w} -0.1 u_{\textnormal{max}}$, 
and if $\vert x \vert > 0.1 \bar{u}_1$ and $\vert x \vert \leq \bar{x} - \bar{w} - 0.1 u_{\textnormal{max}}$, we have
$\vert x + 0.1 \mathrm{sat}_{\bar{u}_1}(-10 x) \vert + 0.1( C \bar{\vartheta} + \bar{s} ) + \bar{w} = \vert x \vert - 0.1 \bar{u}_1 + 0.1( C \bar{\vartheta} + \bar{s} ) + \bar{w} \leq \vert x \vert \leq \bar{x} - \bar{w} - 0.1 u_{\textnormal{max}}$.
This implies $g(x,\alpha(x,s,\hat{\vartheta}),w) \in \mathcal{X}_{\textnormal{RPI}}$.
Therefore, we have established that $(g,\alpha,\mu_w,\mu_s)$ is $(\mathcal{X}_{\textnormal{RPI}},\bar{\vartheta},\mu_w,\mu_s)$-RPI.

We now verify that Assumption~\ref{ac:assump:lyapunov} holds. 
Firstly, note that we already established that $(g,\alpha,\mu_w,\mu_s)$ is $(\mathcal{X}_{\textnormal{RPI}},\bar{\vartheta},\mu_w,\mu_s)$-RPI.
Since we assumed \eqref{ac:eqn:ac-example-regional-19}, it trivially follows that
\begin{align}
    h:=\ln( \mathbf{E}[\exp(0.1 \vert S \vert)]) + \ln(\mathbf{E}[\exp(\vert W \vert)]) < 0.1\bar{u}_1. \label{ac:eqn:example-regional-sat-overpower-noise}
\end{align}
Following similar steps to the proof of Lemma~2 from \cite{siriya2023stability}, we find that for all $x \in \mathcal{X}_{\textnormal{RPI}}$ and  $\hat{\vartheta} = \begin{bmatrix} \hat{\vartheta}_1 & \hat{\vartheta}_2 \end{bmatrix}^{\top} \in \bar{B}_{\bar{\vartheta}}(\theta_*)$,
\begin{align}
    &\mathbf{E} \Big [ \exp\Big (\vert x + 0.1 \mathbf{1}_{\{ \tilde{x} \mid \vert \tilde{x} \vert \leq \bar{x} \}}(x) \\
    & \quad \mleft. \mleft. \cdot \mleft (- \mathrm{sat}_{\bar{u}_1}\mleft (\hat{\vartheta}_2^{\dagger}\hat{\vartheta}_1x\mright )+S\mright ) + W \vert\mright ) - 1 \mright ]\\
    &=\mathbf{E} \mleft [ \exp\mleft (\vert x + 0.1\mleft (- \mathrm{sat}_{\bar{u}_1}\mleft (\hat{\vartheta}_2^{\dagger}\hat{\vartheta}_1x\mright )+S\mright ) + W \vert\mright ) - 1 \mright ]\\
    &\leq \max \mleft ( \exp \mleft (h + 0.1 C\mleft \vert \hat{\vartheta} - \theta_* \mright \vert \mright ), \mright. \\
    & \quad \mleft. \exp\mleft (\vert x \vert - 0.1\bar{u}_1 + h + 0.1 C\mleft \vert \hat{\vartheta} - \theta_* \mright \vert \mright )  \mright ) - 1 \\
    &= \max\mleft (\exp \mleft (h + 0.1 C\mleft \vert \hat{\vartheta} - \theta_* \mright \vert\mright ) - 1, \mright. \\
    & \quad \mleft. \exp\mleft (\vert x \vert\mright ) \exp\mleft ( -0.1 \bar{u}_1 + h + 0.1 C\mleft \vert \hat{\vartheta} - \theta_* \mright \vert \mright ) - 1\mright ) \\
    &\leq \max\mleft (\exp \mleft (h + 0.1 C\mleft \vert \hat{\vartheta} - \theta_* \mright \vert\mright ) - 1, \mleft (\exp\mleft (\vert x \vert \mright )- 1\mright ) \mright. \\ 
    & \quad \mleft. \cdot \exp\mleft ( -0.1\bar{u}_1 + h + 0.1 C\mleft \vert \hat{\vartheta} - \theta_* \mright \vert \mright )\mright ) \label{ac:eqn:ac-example-regional-15} \\
    &\leq \mleft (\exp\mleft (\vert x \vert\mright ) - 1\mright ) \exp\mleft (-0.1\bar{u}_1 + h + 0.1 C\mleft \vert \hat{\vartheta} - \theta_* \mright \vert\mright ) \\
    & \quad + \exp \mleft (h + 0.1 C\mleft \vert \hat{\vartheta} - \theta_* \mright \vert\mright ) - 1 \\
    &= \mleft (\exp\mleft (\vert x \vert\mright ) - 1\mright ) - \Big (1 - \exp\Big (-0.1\bar{u}_1  \\
    & \quad \mleft. \mleft. + h + 0.1 C\mleft \vert \hat{\vartheta} - \theta_* \mright \vert\mright )\mright )\mleft (\exp\mleft (\vert x \vert\mright )-1\mright ) \\
    &\quad + \exp\mleft (h + 0.1 C\mleft \vert \hat{\vartheta} - \theta_* \mright \vert\mright ) - 1  \\
    &\leq \mleft (\exp\mleft (\vert x \vert\mright ) - 1\mright ) - \mleft (1 - \exp\mleft (-0.1 \bar{u}_1 + h + C \bar{\vartheta} \mright )\mright ) \\
    & \quad \cdot \mleft (\exp\mleft (\vert x \vert\mright )-1\mright ) + \exp\mleft (h + 0.1 C\mleft \vert \hat{\vartheta} - \theta_* \mright \vert\mright ) - 1 \label{ac:eqn:ac-example-regional-16} \\
    &\leq \mleft (\exp\mleft (\vert x \vert\mright ) - 1\mright ) - \mleft (1 - \exp\mleft (-0.1 \bar{u}_1 + h + C \bar{\vartheta} \mright )\mright ) \\
    & \quad \cdot \mleft (\exp\mleft (\vert x \vert\mright )-1\mright ) + \exp\mleft (h\mright ) - 1 \\
    & \quad + \exp\mleft ( 0.2 C \mleft \vert \hat{\vartheta} - \theta_* \mright \vert\mright ) - 1, \label{ac:eqn:ac-example-regional-17}
\end{align}
where \eqref{ac:eqn:ac-example-regional-15} follows from \eqref{ac:eqn:example-regional-sat-overpower-noise}, \eqref{ac:eqn:ac-example-regional-16} follows from 
$\mleft \vert \hat{\vartheta} - \theta_* \mright \vert \leq \bar{\vartheta}$, and \eqref{ac:eqn:ac-example-regional-17} follows from the weak triangle inequality. 
 Thus, Assumption~\ref{ac:assump:lyapunov} is satisfied with $V$, $\alpha_1$, $\alpha_2$, $\alpha_3$ chosen as
\begin{align}
    &V\mleft (x\mright )= \exp\mleft (\vert x \vert \mright ) - 1, \quad \alpha_1\mleft (r\mright )=\alpha_2\mleft (r\mright ) = \exp\mleft (r\mright )-1, \\
    &\alpha_3\mleft (r\mright )=\mleft (1 - \exp\mleft (-0.1\bar{u}_1 + h + 0.1 C\mleft \vert \hat{\vartheta} - \theta_* \mright \vert\mright )\mright ) \\
    & \quad \cdot \mleft (\exp\mleft (r\mright ) - 1\mright ), \\
    &\tilde{d}= \exp\mleft (h\mright ) - 1, \quad \sigma_3\mleft (r\mright ) = \exp\mleft (0.1 C r \mright ) - 1,
\end{align}
noting that $\alpha_3 \circ \alpha_2^{-1}\mleft (r\mright ) \geq \mleft ( 1 - \exp\mleft (-0.1\bar{u}_1 + h + C \bar{\vartheta} \mright ) \mright ) r$ for $r \geq 0$ where the RHS of the inequality is clearly convex.

Since we have verified Assumptions~\ref{ac:assump:measurable}-\ref{ac:assump:rpi}, we know via Theorem~\ref{ac:theorem:rls-error-rpi} that if $(\delta,x_0)$ satisfy \eqref{ac:eqn:theorem-rls-error-rpi-condition}, the probabilistic guarantees in \eqref{ac:eqn:theorem-rls-error-rpi-main} hold when applying the described adaptive control strategy to \eqref{ac:eqn:example-1-system}. Moreover, since we also verified Assumption~\ref{ac:assump:lyapunov}, we also know via Theorem~\ref{ac:theorem:regional-stability} that if $(\delta,x_0)$ satisfy \eqref{ac:eqn:theorem-rls-error-rpi-condition-mod}, the probabilistic stability bounds in \eqref{ac:eqn:theorem-regional-stability-main} hold. By keeping all other constants fixed and choosing $\bar{x}$ sufficiently large, both conditions \eqref{ac:eqn:theorem-rls-error-rpi-condition} and \eqref{ac:eqn:theorem-rls-error-rpi-condition-mod} can be verified for the described system and adaptive control algorithm. 

\begin{figure}[ht]
    \centering
    \includegraphics[width=0.48\textwidth]{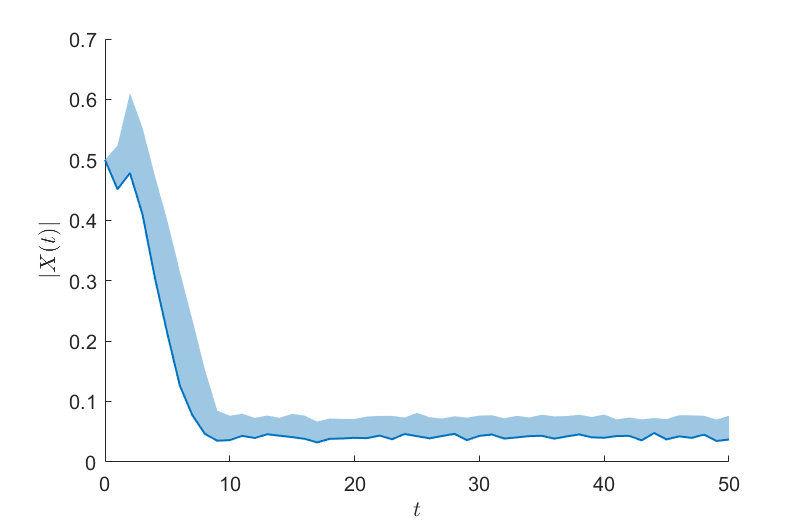}
    \caption{Median and $90$th percentile of $\vert X(t) \vert$ over 100 trials.}
    \label{ac:fig:pwa-success-plot}
\end{figure}

\ssnotebluefour{Finally, we also simulated the closed-loop adaptive PWA system over 100 trials with 
$\overline{x}= 3000$, $\overline{u}_1=0.9$, $\overline{u}_2=0.1$, $\bar{w}=0.07$, $\gamma=0.0001$ and $x_0 = 0.5$. The median and $90$th percentile of the system state over 100 sample paths are shown in Figure~\ref{ac:fig:pwa-success-plot}. The figure highlights the stable behaviour of the adaptive system, supporting Theorem~\ref{ac:theorem:regional-stability}. Note that with these parameters, the conditions in Theorem~\ref{ac:theorem:regional-stability} are satisfied.
However, $T_{\textnormal{converge}}(\delta,x_0)$ and $T_{\textnormal{contained}}(\delta,x_0)$ were found to be significantly larger than the times in Figure~\ref{ac:fig:pwa-success-plot}, reflecting the conservativeness of our analysis.
}

\subsection{Example 2: Input-Constrained Stochastic Linear System} 

We consider stochastic linear systems of the form
\begin{align}
    X(t+1)=AX(t)+BU(t)+W(t+1), \ t \in \mathbb{N}_0, \label{ac:eqn:example-2-system}
\end{align}
with $X(0)=x_0$, where the random sequences $\{X(t)\}_{t \in \mathbb{N}_0}$, $\{U(t)\}_{t \in \mathbb{N}_0}$ and $\{W(t)\}_{t \in \mathbb{N}}$ are the states, controls, and process noise, taking values in $\mathbb{R}^n$, $\mathbb{R}^m$, and $\mathbb{R}^n$ respectively, $x_0 \in \mathbb{R}^n$ is the initial state, and $A \in \mathbb{R}^{n \times n}$ and $B \in \mathbb{R}^{n \times m}$ are the true, unknown, system matrices. We assume that $\{W(t)\}_{t\in\mathbb{N}}$ is sampled i.i.d. from the normal distribution $\mathcal{N}(0,\Sigma_w)$ with $\Sigma_w \succ 0$. Moreover, we assume that $A$ is full rank and satisfies $\vert A \vert \leq 1$, and $(A,B)$ is $\kappa$-step reachable, that is, $\mathrm{rank}(\mathcal{R}_{\kappa}(A,B))=n$, where $\mathcal{R}_{\kappa}(\tilde{A},\tilde{B}) := \begin{bmatrix} \tilde{B} & \tilde{A}\tilde{B} & \hdots & \tilde{A}^{\kappa-1}\tilde{B}  \end{bmatrix}$ is the $\kappa$-step reachability matrix given $\tilde{A}\in\mathbb{R}^{n\times n}$ and $\tilde{B}\in\mathbb{R}^{n\times m}$. The $\kappa$-step sub-sampled system can be written as the linear system
\begin{align} 
    \bar{X}(\tau+1) &= AX(\kappa(\tau+1)-1) + B U(\kappa(\tau+1)-1) \\
    & \quad + W(\kappa(\tau+1)-1)\\
    &=A^{\kappa}\bar{X}(\tau) + \mathcal{R}_*\bar{U}(\tau) + \bar{W}(\tau+1) \label{ac:eqn:example-2-sampled-system}
\end{align}
for $\tau \in \mathbb{N}_0$, with $\mathcal{R}_*:=\mathcal{R}_{\kappa}(A,B)$. Here, $\{\bar{X}(\tau)\}_{t\in\mathbb{N}_0}$ is the sub-sampled state sequence taking values in $\mathbb{R}^n$, $\{\bar{U}(\tau)\}_{\tau \in \mathbb{N}_0}$ is the control sequence for the sub-sampled system defined as
\begin{align}
    \bar{U}(\tau):=\begin{bmatrix} U(\kappa(\tau + 1)-1)^{\top} & \hdots & U(\kappa \tau)^{\top} \end{bmatrix}^{\top},
\end{align}
and $\{\bar{W}(\tau)\}_{\tau \in \mathbb{N}}$ is the process noise for the sub-sampled system defined as
\begin{align}
    \bar{W}(\tau):= \mathcal{R}_{\kappa}(A,I) \begin{bmatrix}W(\kappa(\tau+1)-1)^{\top} & \hdots & W(\tau)^{\top}\end{bmatrix}^{\top}.
\end{align}
Note that $\{\bar{W}(\tau)\}_{\tau \in \mathbb{N}}$ is an i.i.d. random sequence sampled from the distribution $\mathcal{N}(0,\Sigma_{\bar{w}})$, where $\Sigma_{\bar{w}}:=\mathcal{R}_{\kappa}(A,I) \mleft ( \mathbf{I}_{\kappa} \otimes \Sigma_w \mright ) \mathcal{R}_{\kappa}(A,I)^{\top}$ with $\otimes$ denoting the Kronecker product. 

We consider the $\kappa$-step sub-sampled system from \eqref{ac:eqn:example-2-sampled-system} as the dynamical system to be controlled, such that the nominal system $f$, unknown parameter $\theta_*$, and basis functions $\psi$ in \eqref{ac:eqn:system-dynamics} correspond to
\begin{align}
    f(x,u)=0, \quad \theta_* = \begin{bmatrix} A^{\kappa} & \mathcal{R}_* \end{bmatrix}^{\top}, \quad  \psi(x,u)=\begin{bmatrix} x \\ u \end{bmatrix},
\end{align}
for $x \in \mathbb{R}^n$ and $u\in\mathbb{R}^{\kappa m}$. We make this choice because considering the sampled system is a common technique for analyse the stability of input-constrained linear systems with unbounded stochastic disturbances. This is reasonable because the stability of the original system is typically implied by the stability of the sub-sampled system. 
Moreover, $\mu_w$ is the probability measure associated with the distribution $\mathcal{N}(0,\Sigma_{\bar{w}})$.

We now describe the adaptive control strategy based on the framework in Algorithm~\ref{ac:alg:framework}.
Firstly, let $\bar{u}_1 \in (0,u_{\textnormal{max}})$ be a user-chosen parameter that specifies the level of the saturated component of the policy $\alpha$. 
Moreover, let $\bar{u}_2 = \frac{1}{\sqrt{\kappa m}}(u_{\textnormal{max}}-\bar{u}_1)$, and we choose $\mu_s$ as the probability measure associated with the joint distribution of $\kappa m$ independent uniform distributions over $[-\bar{u}_2,\bar{u}_2]$, such that $\mathbb{S} = [-\bar{u}_2,\bar{u}_2]^{\kappa m}$.
The family of control policies $\alpha$ chosen for our framework is then given by
\begin{align}
    \alpha(x,s,\vartheta):=\sat_{\bar{u}_1}(-\vartheta_2^{\dagger} \vartheta_1^{\kappa} x) + s
\end{align}
for $s \in \mathbb{S}$ and $\vartheta = \begin{bmatrix} \vartheta_1 & \vartheta_2 \end{bmatrix}^{\top}$ with $\vartheta_1\in\mathbb{R}^{n \times n}$ and $\vartheta_2 \in \mathbb{R}^{n \times \kappa m}$. 

Next, we verify the assumptions required to establish Corollary~\ref{ac:corollary:global-stability}.
We start by verifying Assumption~\ref{ac:assump:measurable}. We know $f(x,u)=0$ is Borel measurable since constants are Borel measurable, and $\psi(x,u)=\begin{bmatrix} x^{\top} & u^{\top} \end{bmatrix}^{\top}$ is Borel measurable since coordinate functions preserve Borel measurability. Moreover, $\alpha(x,s,\vartheta) = \sat_{\bar{u}_1}(\vartheta_2^{\dagger} \vartheta_1^{\kappa} x) + s$ is Borel measurable since 1) addition, multiplication, composition, and coordinate functions preserve Borel measurability, and 2) both $\sat_{\bar{u}_1}(\cdot)$ and the pseudo-inverse $(\cdot)^{\dagger}$ are Borel measurable functions.

Next, Assumption~\ref{ac:assump:process-noise} holds since $\{\bar{W}(\tau)\}_{\tau \in \mathbb{N}}$ is an i.i.d. sequence by construction, $\{\bar{W}(\tau)\}_{\tau \in \mathbb{N}}$ and $\{S(\tau)\}_{\tau \in \mathbb{N}_0}$ are mutually independent, and $\bar{W}(\tau)$ is zero-mean and $\lambda_{\textnormal{max}}(\Sigma_{\bar{w}})$-sub-Gaussian.

We now verify Assumption~\ref{ac:assump:forward-complete}. 
Firstly, note that for any initial state $\xi \in \mathbb{X}$, sub-sampled time $\tau \in \mathbb{N}_0$, and deterministic input and noise sequences for the sub-sampled system $\{u(i)\}_{i=0}^{\tau-1} \in \mathbb{U}^{\tau}$ and $\{w(i)\}_{i=1}^{\tau} \in \mathbb{W}^{\tau}$, the deterministic trajectory $\phi$ corresponding to the sub-sampled system \eqref{ac:eqn:example-2-sampled-system} satisfies
\begin{align}
    &\vert \phi(\tau,\xi,\{u(i)\}_{i=0}^{\tau-1},\{w(i)\}_{i=1}^\tau) \vert \\
    &\leq \vert A^{\kappa \tau} x_0 + \sum_{i=0}^{\tau - 1} A^{\kappa i} \mathcal{R}_* u(\tau - 1 - i) + \sum_{i=0}^{\tau-1} A^{\kappa i} w(\tau - i) \vert \\
    &\leq \vert A^{\kappa \tau} x_0 \vert + \vert \sum_{i=0}^{\tau - 1} A^{\kappa i} \mathcal{R}_* u(\tau - 1 - i) \vert \\
    & \quad + \vert \sum_{i=0}^{\tau-1} A^{\kappa i} \mathcal{R}_{\kappa}(A,I) w(\tau - i) \vert \\
    &\leq \vert x_0 \vert + \vert \mathcal{R}_* \vert \sum_{i=0}^{\tau - 1} \vert u(i) \vert + \vert \mathcal{R}_{\kappa}(A,I) \vert \sum_{i=1}^{\tau} \vert w(i) \vert.
\end{align}
Thus, Assumption~\ref{ac:assump:forward-complete} is satisfied with $\chi_2(\cdot) = \mathrm{Id}(\cdot)$, $\chi_3(\cdot)=\vert \mathcal{R}_* \vert \mathrm{Id}(\cdot)$, \\$\chi_4(\cdot) =\vert \mathcal{R}_{\kappa}(A,I) \vert \mathrm{Id}(\cdot)$ and $\sigma_1(\cdot)=\sigma_2(\cdot)=\mathrm{Id}(\cdot)$, and with arbitrary $\chi_1 \in \mathcal{K}_{\infty}^{\textnormal{1-SE}}$.

Next, Assumption~\ref{ac:assump:constrained-controls} holds since for all $x \in \mathbb{X}$, $s \in \mathbb{S}$ and $\vartheta = \begin{bmatrix} \vartheta_1 & \vartheta_2 \end{bmatrix}^{\top} \in \mathbb{R}^{d \times n}$,
\begin{align}
    \vert \alpha(x,s,\vartheta) \vert := \vert \sat_{\bar{u}_1}(\vartheta_2^{\dagger} \vartheta_1^{\kappa} x)\vert + \vert s \vert &\leq \bar{u}_1 + \sqrt{\kappa m} \bar{u}_2 \\
    &= u_{\textnormal{max}}
\end{align}

Note that Assumption~\ref{ac:assump:poly-feature-map} is satisfied with $\chi_4(\cdot) = \mathrm{Id}(\cdot)$.

We now verify Assumption~\ref{ac:assump:global-excitation}. 
\ssnotebluefour{The steps we take to verify this follow similarly to Example~2 in \cite{sysidpaper}. In both cases, Lemma~\ref{ac:lemma:regional-excitation-sufficient} is used to establish regional excitation, and linear systems with bounded policies and Gaussian process noise are considered. However, differences arise since we consider multi-dimensional systems, rather than specifically the double integrator which is a $2$-dimensional system.}
Suppose $x \in \mathbb{R}^n$, $\vartheta \in \mathbb{R}^{(n + \kappa m) \times n}$, and $\zeta = [\zeta_1^{\top} \ \zeta_2]^{\top} \in \mathcal{S}^{n + \kappa m - 1}$ (where $\zeta_1 \in \mathbb{R}^n$ and $\zeta_2 \in \mathbb{R}^{\kappa m}$). Then,
\begin{align}
    & \mathbf{E} \mleft [ \mleft \vert \zeta^{\top} \psi(x+W, \alpha(x+W,S,\vartheta) \mright \vert \mright ] \\
    &= \mathbf{E} \mleft [ \mleft \vert \zeta_1^{\top}(x+W) + \zeta_2^{\top} (\mathrm{sat}_{\bar{u}_1}(-\vartheta_2^{\dagger}\vartheta_1^{\kappa}(x+W)) + S) \mright \vert \mright ]. \label{ac:eqn:ac-example-global-12}
\end{align}
Moreover,
\begin{align}
    &\mathbf{E} \mleft [ \mleft \vert \zeta_1^{\top}(x+W) + \zeta_2^{\top} (\mathrm{sat}_{\bar{u}_1}(-\vartheta_2^{\dagger}\vartheta_1^{\kappa}(x+W)) + S) \mright \vert \mright ] \\
    &= \mathbf{E} \mleft [ \mathbf{E} \mleft [ \mleft \vert \zeta_1^{\top}(x+W) \mright. \mright. \mright. \\
    &\quad \mleft. \mleft. \mleft. + \zeta_2^{\top} (\mathrm{sat}_{\bar{u}_1}(-\vartheta_2^{\dagger}\vartheta_1^{\kappa}(x+W)) + S) \mright \vert \mid S \mright ] \mright ] \label{ac:eqn:ac-example-global-4} \\
    & \geq \mathbf{E} \mleft [ \mleft \vert \mathbf{E} \mleft [ \zeta_1^{\top}(x+W) \mright. \mright. \mright. \\
    & \quad \mleft. \mleft. \mleft. + \zeta_2^{\top} (\mathrm{sat}_{\bar{u}_1}(-\vartheta_2^{\dagger}\vartheta_1^{\kappa}(x+W)) + S) \mid S \mright ] \mright \vert\mright ] \\
    &= \mathbf{E} \mleft [ \mleft \vert \zeta_1^{\top} x \mright. \mright. \\
    & \quad \mleft. \mleft. + \zeta_2^{\top} \mathbf{E} \mleft [ \mathrm{sat}_{\bar{u}_1}(-\vartheta_2^{\dagger}\vartheta_1^{\kappa}(x+W)) \mright ] + \zeta_2^{\top} S \mright \vert\mright ] \label{ac:eqn:ac-example-global-5} \\
    &\geq \vert \zeta_2 \vert \mathbf{E}\mleft[ \mleft \vert S \mright \vert \mright] = \vert \zeta_2 \vert \sqrt{\kappa m}\frac{\bar{u}_2}{2} \label{ac:eqn:ac-example-global-6}
\end{align}
where \eqref{ac:eqn:ac-example-global-4} follows from the tower property, and \eqref{ac:eqn:ac-example-global-5} follows from the independence of $W$ and $S$. Moreover, \eqref{ac:eqn:ac-example-global-6} follows from the optimality property of medians and $ \mathbf{E}\mleft[ \mleft \vert S \mright \vert \mright] \geq \frac{1}{\sqrt{\kappa m}} \mathbf{E} \mleft [ \mleft \vert S \mright \vert_1 \mright ] = \frac{1}{\sqrt{\kappa m}} \sum_{i=1}^{\kappa m} \mathbf{E} \mleft [ \mleft \vert S_i \mright \vert \mright ] = \sqrt{\kappa m} \frac{\bar{u}_2}{2} $, where $S_i$ denotes the $i$th coordinate of $S$ and we make use of $\mathbf{E}\mleft [ \mleft \vert S_i \mright \vert \mright ] = \frac{\bar{u}_2}{2}$. Similarly, we have
\begin{align}
    &\mathbf{E} \mleft [ \mleft \vert \zeta_1^{\top}(x+W) + \zeta_2^{\top} (\mathrm{sat}_{\bar{u}_1}(-\vartheta_2^{\dagger}\vartheta_1^{\kappa}(x+W)) + S) \mright \vert \mright ] \\
    &= \mathbf{E} \mleft [ \mathbf{E} \mleft [ \mleft \vert \zeta_1^{\top}(x+W) \mright. \mright. \mright. \\
    & \quad \mleft. \mleft. \mleft. + \zeta_2^{\top} (\mathrm{sat}_{\bar{u}_1}(-\vartheta_2^{\dagger}\vartheta_1^{\kappa}(x+W)) + S) \mright \vert \mid W \mright ] \mright ] \label{ac:eqn:ac-example-global-7} \\
    & \geq \mathbf{E} \Big [ \Big \vert \mathbf{E} \Big [ \zeta_1^{\top}(x+W)  \\
    & \quad \mleft. \mleft. \mleft. + \zeta_2^{\top} (\mathrm{sat}_{\bar{u}_1}(-\vartheta_2^{\dagger}\vartheta_1^{\kappa}(x+W)) + S) \mid W  \mright ] \mright \vert\mright ] \\
    &= \mathbf{E} \mleft [ \mleft \vert \zeta_1^{\top}(x+W) + \zeta_2^{\top} \mathrm{sat}_{\bar{u}_1}(-\vartheta_2^{\dagger}\vartheta_1^{\kappa}(x+W)) \mright \vert\mright ] \label{ac:eqn:ac-example-global-8} \\
    &\geq \mleft \vert \zeta_1 \mright \vert \mathbf{E} \mleft [ \mleft \vert (x+W) \mright \vert\mright ] \\
    & \quad - \mleft \vert \zeta_2 \mright \vert \mathbf{E} \mleft [ \mleft \vert \mathrm{sat}_{\bar{u}_1}(-\vartheta_2^{\dagger}\vartheta_1^{\kappa}(x+W)) \mright \vert\mright ] \label{ac:eqn:ac-example-global-9} \\
    &\geq \mleft \vert \zeta_1 \mright \vert \mathbf{E} \mleft [ \mleft \vert W \mright \vert\mright ] - \mleft \vert \zeta_2 \mright \vert \overline{u}_1 \label{ac:eqn:ac-example-global-10} \\
    &= \vert \zeta_1 \vert \sqrt{\frac{2 \lambda_{\textnormal{min}}(\Sigma_{\bar{w}}) n}{\pi}} - \mleft \vert \zeta_2 \mright \vert \overline{u}_1 \label{ac:eqn:ac-example-global-11}
\end{align}
\sloppy
where \eqref{ac:eqn:ac-example-global-7} follows from the tower property, \eqref{ac:eqn:ac-example-global-8} follows from the independence of $W$ and $S$, \eqref{ac:eqn:ac-example-global-9} follows from the reverse triangle inequality, \eqref{ac:eqn:ac-example-global-10} follows from the optimality property of medians and the fact that $\mathrm{sat}_{\bar{u}_1}$ is upper bounded by $\overline{u}_1$, and \eqref{ac:eqn:ac-example-global-11} follows from $\mathbf{E}\mleft [ \mleft \vert W \mright \vert \mright ] = \mathbf{E} \mleft [ \mleft \vert \Sigma_{\bar{w}}^{1/2} \Sigma_{\bar{w}}^{-1/2} W \mright \vert \mright ] \geq \sqrt{\lambda_{\textnormal{min}}(\Sigma_{\bar{w}})} \mathbf{E} \mleft [ \mleft \vert \Sigma_{\bar{w}}^{-1/2} W \mright \vert \mright ] \geq \sqrt{\frac{\lambda_{\textnormal{min}}(\Sigma_{\bar{w}})}{n}} \mathbf{E}\mleft [ \mleft \vert \Sigma_{\bar{w}}^{-1/2} W \mright \vert_1 \mright] = \sqrt{\frac{\lambda_{\textnormal{min}}(\Sigma_{\bar{w}})}{n}} \sum_{i=1}^n \mathbf{E}\mleft [ \mleft \vert \mleft (\Sigma_{\bar{w}}^{-1/2} W \mright )_i \mright \vert \mright] = \sqrt{\frac{2 \lambda_{\textnormal{min}}(\Sigma_{\bar{w}}) n}{\pi}} $, where $\mleft (\Sigma_{\bar{w}}^{-1/2} W \mright )_i$ denotes the $i$th coordinate of $\Sigma_{\bar{w}}^{-1/2} W$ and we make use of $\mathbf{E}\mleft [ \mleft \vert \mleft (\Sigma_{\bar{w}}^{-1/2} W \mright )_i \mright \vert \mright] = \sqrt{\frac{2}{\pi}}$ since $\Sigma_{\bar{w}}^{-1/2} W$ has distribution $\mathcal{N}(0,I)$.
Note that if $\vert \zeta_2 \vert \leq \frac{\sqrt{\lambda_{\textnormal{min}}(\Sigma_{\bar{w}})n}}{\bar{u}_1 \sqrt{2 \pi} + 2 \sqrt{\lambda_{\textnormal{min}}(\Sigma_{\bar{w}})}n}$, then $\vert \zeta_1 \vert \geq 1 - \frac{\sqrt{\lambda_{\textnormal{min}}(\Sigma_{\bar{w}})n}}{\bar{u}_1 \sqrt{2 \pi} + 2 \sqrt{\lambda_{\textnormal{min}}(\Sigma_{\bar{w}})}n}$, and so $\vert \zeta_1 \vert \sqrt{\frac{2 \lambda_{\textnormal{min}}(\Sigma_{\bar{w}}) n}{\pi}} - \mleft \vert \zeta_2 \mright \vert \overline{u}_1 \geq \mleft ( 1 - \frac{\sqrt{\lambda_{\textnormal{min}}(\Sigma_{\bar{w}})n}}{\bar{u}_1 \sqrt{2 \pi} + 2 \sqrt{\lambda_{\textnormal{min}}(\Sigma_{\bar{w}})}n} \mright ) \sqrt{\frac{2 \lambda_{\textnormal{min}}(\Sigma_{\bar{w}}) n}{\pi}} - \frac{\sqrt{\lambda_{\textnormal{min}}(\Sigma_{\bar{w}})n} \bar{u}_1}{\bar{u}_1 \sqrt{2 \pi} + 2 \sqrt{\lambda_{\textnormal{min}}(\Sigma_{\bar{w}})}n} = \sqrt{\frac{\lambda_{\textnormal{min}}(\Sigma_{\bar{w}})n}{2\pi}}$. 
Moreover, if $\vert \zeta_2 \vert > \frac{\sqrt{\lambda_{\textnormal{min}}(\Sigma_{\bar{w}})n}}{\bar{u}_1 \sqrt{2 \pi} + 2 \sqrt{\lambda_{\textnormal{min}}(\Sigma_{\bar{w}})}n}$, then $\vert \zeta_2 \vert \sqrt{\kappa m} \frac{\overline{u}_2}{2} > \frac{\sqrt{\lambda_{\textnormal{min}}(\Sigma_{\bar{w}})nm\kappa}\bar{u}_2}{2\sqrt{2 \pi} \bar{u}_1 + 4 \sqrt{\lambda_{\textnormal{min}(\Sigma_{\bar{w}})}n}}$. Combining this with \eqref{ac:eqn:ac-example-global-12}, \eqref{ac:eqn:ac-example-global-6} and \eqref{ac:eqn:ac-example-global-11}, we have
\begin{align}
    &\mathbf{E} \mleft [ \mleft \vert \zeta_1^{\top}(x+W) + \zeta_2^{\top} (\mathrm{sat}_{\bar{u}_1}(-\vartheta_2^{\dagger}\vartheta_1^{\kappa}(x+W)) + S) \mright \vert \mright ] \\
    &\geq \max \mleft ( \vert \zeta_2 \vert \sqrt{\kappa m}\frac{\bar{u}_2}{2}, \vert \zeta_1 \vert \sqrt{\frac{2 \lambda_{\textnormal{min}}(\Sigma_{\bar{w}}) n}{\pi}} - \mleft \vert \zeta_2 \mright \vert \overline{u}_1 \mright ) \\
    &\geq \min \mleft ( \frac{\sqrt{\lambda_{\textnormal{min}}(\Sigma_{\bar{w}})nm\kappa}\bar{u}_2}{2\sqrt{2 \pi} \bar{u}_1 + 4 \sqrt{\lambda_{\textnormal{min}(\Sigma_{\bar{w}})}n}}, \sqrt{\frac{\lambda_{\textnormal{min}}(\Sigma_{\bar{w}})n}{2\pi}} \mright ).
\end{align}
On the other hand, we have
\begin{align}
    &\mathbf{V}\mathrm{ar} \mleft ( \mleft \vert \zeta^{\top} \psi(x+W, \mathrm{sat}_{\bar{u}_1}(-\vartheta_2^{\dagger}\vartheta_1^{\kappa}(x+W)) + S ) \mright \vert \mright ) \\
    &\leq \mathbf{V}\mathrm{ar}  \mleft ( \zeta^{\top} \psi(x+W, \mathrm{sat}_{\bar{u}_1}(-\vartheta_2^{\dagger}\vartheta_1^{\kappa}(x+W)) + S ) \mright ) \\
    &= \mathbf{E} \mleft [ \mleft ( \zeta_1^{\top}W + \zeta_2^{\top} S + \zeta_2^{\top} \mleft ( \mathrm{sat}_{\bar{u}_1}(-\vartheta_2^{\dagger}\vartheta_1^{\kappa}(x+W)) \mright. \mright. \mright. \\
    & \quad \mleft. \mleft. \mleft. - \mathbf{E} \mleft[ \mathrm{sat}_{\bar{u}_1}(-\vartheta_2^{\dagger}\vartheta_1^{\kappa}(x+W)) \mright ] \mright ) \mright )^2 \mright ] \\
    &\leq 3 \mleft (\mathbf{E} \mleft [ \mleft ( \zeta_1^{\top} W \mright )^2 \mright ] + \mathbf{E} \mleft [ \mleft( \zeta_2^{\top} S \mright )^2 \mright ] \mright. \\
    & \quad + \mathbf{E} \mleft [ \mleft ( \zeta_2^{\top} \mleft ( \mathrm{sat}_{\bar{u}_1}(-\vartheta_2^{\dagger}\vartheta_1^{\kappa}(x+W)) \mright. \mright. \mright. \\
    & \quad \mleft. \mleft. \mleft. \mleft. - \mathbf{E} \mleft[ \mathrm{sat}_{\bar{u}_1}(-\vartheta_2^{\dagger}\vartheta_1^{\kappa}(x+W)) \mright ] \mright ) \mright )^2 \mright ] \mright ) \label{ac:eqn:ac-example-global-13} \\
    &\leq 3 \mleft ( \max \mleft ( \sigma_w^2, \frac{\overline{u}_2^2}{3} \mright ) + 4\overline{u}_1^2 \mright ), \label{ac:eqn:ac-example-global-14}
\end{align}
where \eqref{ac:eqn:ac-example-global-13} follows from the QM-AM inequality, and \eqref{ac:eqn:ac-example-global-14} follows from $\mathbf{E} \mleft [ \mleft ( \zeta_1^{\top} W \mright )^2 \mright ] = \zeta_1^{\top} \mathbf{E}\mleft[ W W^{\top} \mright ] \zeta_1 \leq \lambda_{\textnormal{max}}(\Sigma_{\bar{w}})$ and $\mathbf{E}\mleft [ ( \zeta_2^{\top} S)^2 \mright ] = \zeta_2^{\top} \mathbf{E}\mleft [ S S^{\top} \mright ] \zeta_2 = \frac{\bar{u}_2^2}{3}$. 
Thus, using Lemma~\ref{ac:lemma:regional-excitation-sufficient} we find that $(\psi,\alpha,\mu_s,\mu_w)$ is $(c_{\textnormal{PE}},p_{\textnormal{PE}})$\textit{-globally excited} with 
$c_{\textnormal{PE}} := \frac{1}{4} c_{\textnormal{PE}1}^2$ and \\$p_{\textnormal{PE}} := \frac{1}{4} \mleft( \frac{ c_{\textnormal{PE}2}}{c_{\textnormal{PE}1}^2} + 1 \mright)^{-1}$, where
$c_{\textnormal{PE1}}:=\min \mleft ( \frac{\sqrt{\lambda_{\textnormal{min}}(\Sigma_{\bar{w}})nm\kappa}\bar{u}_2}{2\sqrt{2 \pi} \bar{u}_1 + 4 \sqrt{\lambda_{\textnormal{min}(\Sigma_{\bar{w}})}n}}, \sqrt{\frac{\lambda_{\textnormal{min}}(\Sigma_{\bar{w}})n}{2\pi}} \mright )$ and $c_{\textnormal{PE2}}:=3 \mleft ( \max \mleft ( \sigma_w^2, \frac{\overline{u}_2^2}{3} \mright ) + 4\overline{u}_1^2 \mright ) $ such that Assumption~\ref{ac:assump:global-excitation} is satisfied.

We now show that Assumption~\ref{ac:assump:global-lyapunov} is satisfied under the additional constraint
\begin{align}
    h:=\ln( \mathbf{E}[\exp(\vert \mathcal{R}_* S \vert)]) + \ln(\mathbf{E}[\exp(\vert W \vert)]) < \sigma_{\textnormal{min}}(\mathcal{R}_*)\bar{u}_1. \label{ac:eqn:example-global-sat-overpower-noise}
\end{align}
To see this, we first note under this constraint and using Lemma~\ref{ac:lemma:example-local-lipschitz}, we know there exist $\bar{\vartheta},C>0$ such that for any $x \in \mathbb{R}^n$ and $\hat{\vartheta}=\begin{bmatrix} \hat{\vartheta}_1 & \hat{\vartheta}_2 \end{bmatrix}^{\top} \in \bar{B}_{\bar{\vartheta}}(\theta_*)$, both $h + \vert \mathcal{R}_* \vert C \bar{\vartheta} < \sigma_{\textnormal{min}}(\mathcal{R}_*) \bar{u}_1$ and $\mleft \vert \mathrm{sat}_{\bar{u}_1}(\hat{\vartheta}_2^{\dagger}\hat{\vartheta}_1x) - \mathrm{sat}_{\bar{u}_1}(\mathcal{R}_*^{\dagger}A^{\kappa}x) \mright \vert \leq C \mleft \vert \hat{\vartheta} - \vartheta_* \mright \vert$ hold.
Now, let $\bar{\vartheta},C$ satisfy this requirement.
Following similar steps to the proof of Lemma~2 in \cite{siriya2023stability}, we find that for all $x \in \mathbb{R}^n$ and  $\hat{\vartheta} = \begin{bmatrix} \hat{\vartheta}_1 & \hat{\vartheta}_2 \end{bmatrix}^{\top} \in \bar{B}_{\bar{\vartheta}}(\theta_*)$,
\begin{align}
    &\mathbf{E} \mleft [ \exp\mleft (\vert A^{\kappa}x + \mathcal{R}_*\mleft (- \sat_{\bar{u}_1}\mleft (\hat{\vartheta}_2^{\dagger}\hat{\vartheta}_1x\mright )+S\mright ) + W \vert\mright ) - 1 \mright ]\\
    &\leq -1 + \max \mleft ( \exp \mleft (h + \mleft \vert \mathcal{R}_* \mright \vert C\mleft \vert \hat{\vartheta} - \theta_* \mright \vert \mright ), \mright. \\
    & \quad \mleft. \exp\mleft (\vert x \vert - \sigma_{\textnormal{min}}\mleft (\mathcal{R}_*\mright )\bar{u}_1 + h + \mleft \vert \mathcal{R}_* \mright \vert C\mleft \vert \hat{\vartheta} - \theta_* \mright \vert \mright )  \mright ) \\
    &= \max\mleft (\exp \mleft (h + \mleft \vert \mathcal{R}_* \mright \vert C\mleft \vert \hat{\vartheta} - \theta_* \mright \vert\mright ) - 1, \exp\mleft (\vert x \vert\mright ) \mright. \\
    & \quad \mleft.  \cdot \exp\mleft ( -\sigma_{\textnormal{min}}\mleft (\mathcal{R}_*\mright )\bar{u}_1 + h + \mleft \vert \mathcal{R}_* \mright \vert C\mleft \vert \hat{\vartheta} - \theta_* \mright \vert \mright ) - 1\mright ) \\
    &\leq \max \Big (\exp \mleft (h + \mleft \vert \mathcal{R}_* \mright \vert C\mleft \vert \hat{\vartheta} - \theta_* \mright \vert\mright ) - 1, \mleft (\exp\mleft (\vert x \vert \mright )- 1\mright )\\
    & \quad \cdot \exp\mleft ( -\sigma_{\textnormal{min}}\mleft (\mathcal{R}_*\mright )\bar{u}_1 + h + \mleft \vert \mathcal{R}_* \mright \vert C\mleft \vert \hat{\vartheta} - \theta_* \mright \vert \mright )\Big ) \label{ac:eqn:ac-example-global-15} \\
    &\leq \exp\mleft (-\sigma_{\textnormal{min}}\mleft (\mathcal{R}_*\mright )\bar{u}_1 + h + \mleft \vert \mathcal{R}_* \mright \vert C\mleft \vert \hat{\vartheta} - \theta_* \mright \vert\mright ) \\
    & \quad \cdot \mleft (\exp\mleft (\vert x \vert\mright ) - 1\mright ) + \exp \mleft (h + \mleft \vert \mathcal{R}_* \mright \vert C\mleft \vert \hat{\vartheta} - \theta_* \mright \vert\mright ) - 1 \\
    &= \mleft (\exp\mleft (\vert x \vert\mright ) - 1\mright ) - \mleft (\exp\mleft (\vert x \vert\mright )-1\mright ) \\
    & \quad \cdot \mleft (1 - \exp\mleft (-\sigma_{\textnormal{min}}\mleft (\mathcal{R}_*\mright )\bar{u}_1 + h + \mleft \vert \mathcal{R}_* \mright \vert C\mleft \vert \hat{\vartheta} - \theta_* \mright \vert\mright )\mright ) \\
    &\quad + \exp\mleft (h + \mleft \vert \mathcal{R}_* \mright \vert C\mleft \vert \hat{\vartheta} - \theta_* \mright \vert\mright ) - 1  \\
    &\leq \mleft (\exp\mleft (\vert x \vert\mright ) - 1\mright ) - \mleft (\exp\mleft (\vert x \vert\mright )-1\mright )\\
    & \quad  \cdot \mleft (1 - \exp\mleft (-\sigma_{\textnormal{min}}\mleft (\mathcal{R}_*\mright )\bar{u}_1 + h + C \bar{\vartheta} \mright )\mright ) \\
    & \quad + \exp\mleft (h + \mleft \vert \mathcal{R}_* \mright \vert C\mleft \vert \hat{\vartheta} - \theta_* \mright \vert\mright ) - 1 \label{ac:eqn:ac-example-global-16} \\
    &\leq \mleft (\exp\mleft (\vert x \vert\mright ) - 1\mright ) - \mleft (\exp\mleft (\vert x \vert\mright )-1\mright ) \\
    & \quad \cdot \mleft (1 - \exp\mleft (-\sigma_{\textnormal{min}}\mleft (\mathcal{R}_*\mright )\bar{u}_1 + h + C \bar{\vartheta} \mright )\mright ) \\
    & \quad + \exp\mleft (\ssredd{2}h\mright ) - 1 + \exp\mleft ( 2 \vert \mathcal{R}_* \vert C \mleft \vert \hat{\vartheta} - \theta_* \mright \vert\mright ) - 1, \label{ac:eqn:ac-example-global-17}
\end{align}
where \eqref{ac:eqn:ac-example-global-15} follows from \eqref{ac:eqn:example-global-sat-overpower-noise}, \eqref{ac:eqn:ac-example-global-16} follows from 
$\mleft \vert \hat{\vartheta} - \theta_* \mright \vert \leq \bar{\vartheta}$, and \eqref{ac:eqn:ac-example-global-17} follows from the weak triangle inequality. Thus, Assumption~\ref{ac:assump:global-lyapunov} is satisfied with $V$, $\alpha_1$, $\alpha_2$, $\alpha_3$ chosen as
\begin{align}
    &V\mleft (x\mright )= \exp\mleft (\vert x \vert \mright ) - 1, \quad \alpha_1\mleft (r\mright )=\alpha_2\mleft (r\mright ) = \exp\mleft (r\mright )-1, \\
    &\alpha_3\mleft (r\mright )= \mleft (\exp\mleft (r\mright ) - 1\mright ) \\
    & \quad \cdot \mleft (1 - \exp\mleft (-\sigma_{\textnormal{min}}\mleft (\mathcal{R}_*\mright )\bar{u}_1 + h + \mleft \vert \mathcal{R}_* \mright \vert C\mleft \vert \hat{\vartheta} - \theta_* \mright \vert\mright )\mright ) , \\
    &\tilde{d}= \exp\mleft (h\mright ) - 1, \quad \sigma_3\mleft (r\mright ) = \exp\mleft (2 \vert \mathcal{R}_* \vert C r \mright ) - 1,
\end{align}
noting that $\alpha_3 \circ \alpha_2^{-1}\mleft (r\mright ) \geq \mleft ( 1 - \exp\mleft (-\sigma_{\textnormal{min}}\mleft (\mathcal{R}_*\mright )\bar{u}_1 + h + C m\mright ) \mright ) r$ for $r \geq 0$ where the RHS of the inequality is clearly convex.

Since, we have shown that Assumptions~\ref{ac:assump:measurable}, \ref{ac:assump:process-noise}, \ref{ac:assump:forward-complete}, \ref{ac:assump:constrained-controls}, \ref{ac:assump:poly-feature-map} are \ref{ac:assump:global-excitation} satisfied, and \ref{ac:assump:global-lyapunov} is also satisfied under the extra assumption \eqref{ac:eqn:example-global-sat-overpower-noise}, the premise of Corollary~\ref{ac:corollary:global-stability} has been verified. Thus, we have successfully established the existence of a high probability stability bounds for the learning-based adaptive control problem in input-constrained linear systems subject to Gaussian disturbances. We point out that high probability stability bounds were established in \cite{siriya2023stability}, but using an analysis specific to the problem setup in \eqref{ac:eqn:example-2-sampled-system}. On the other hand, here we made use of the general analytical framework from Section~\ref{ac:sec:control-method-results} to establish such bounds.
\section{Proofs} \label{ac:sec:control-proofs}

\ssnotebluetwo{We now provide the proofs of all results in this paper. In Section~\ref{ac:sec:proof-error}, we establish the estimation error bound and invariance guarantee result from Theorem~\ref{ac:theorem:rls-error-rpi}. In Section~\ref{ac:sec:proof-stability}, we prove the probabilistic stability bound results in  Theorem~\ref{ac:theorem:regional-stability} and Corollary~\ref{ac:corollary:global-stability}.}

\subsection{Estimation Error Bounds and Invariance Guarantee} \label{ac:sec:proof-error}

\ssnoteblue{Firstly, we provide Lemma~\ref{ac:lemma:random-variables}. It ensures that the states, controls, and parameter estimates, are all random sequences, \ssnotebluetwo{such that all stochastic properties of interest are well-defined. This result is the same as Lemma~2 in \cite{sysidpaper}, which we refer to for the proof.} \ssnotebluetwo{We will not refer to Lemma~\ref{ac:lemma:random-variables} directly in our proofs to simplify the exposition.}}
\begin{lemma}
\label{ac:lemma:random-variables}
    \ssnoteblue{Suppose Assumption~\ref{ac:assump:measurable} holds. Then, $\{ X(t) \}_{t \in \mathbb{N}_0}$, $\{ U(t) \}_{t \in \mathbb{N}_0}$, $\{Z(t)\}_{t \in \mathbb{N}}$ and $\{ \hat{\theta}(t) \}_{t \in \mathbb{N}}$, all satisfy the definition of a random sequence.}
\end{lemma}

\ssnoteblue{Next, we provide Lemma~\ref{ac:lemma:data-dependent-error-bound}, which is a high probability, upper bound on the estimation error $\vert \hat{\theta}(t) - \theta_* \vert$ as a function of $\lambda_{\textnormal{min}}(G(t))$ and $\lambda_{\textnormal{max}}(G(t))$.}
\ssnotebluetwo{This is the same as Lemma~7 in \cite{sysidpaper}, which we refer to for the proof.}
\begin{lemma} \label{ac:lemma:data-dependent-error-bound}
    (Data-Dependent Least-Squares Error Bounds) 
    Suppose Assumptions~\ref{ac:assump:measurable} and \ref{ac:assump:process-noise} are satisfied. Then, for any $\delta \in (0,1)$ and $x_0 \in \mathbb{X}$
    \begin{align}
        &\mathbb{P}  \bigg ( \vert \hat{\theta}(t) - \theta_*  \vert \leq \frac{1}{ \lambda_{\textnormal{min}}  ( G(t)  )^{1/2} } \Big ( \gamma^{1/2}  \vert \theta_*  \vert_F + \\
        & \sigma_w  \sqrt{ 2 n   ( \ln (n / \delta) + (d/2) \ln ( \lambda_{\textnormal{max}}(G(t)) \gamma^{-1} )  ) } \Big ), \ \forall t \in \mathbb{N}  \bigg )\\
        &\geq 1 - \delta.
    \end{align}
\end{lemma}

\ssnotebluetwo{In order to move from the data-dependent bound in Lemma~\ref{ac:lemma:data-dependent-error-bound} to a data-independent bound, we need to obtain a probabilistic upper bound on $\lambda_{\textnormal{max}}(G(t))$ and lower bound on $\lambda_{\textnormal{min}}(G(t))$.
The upper bound can be obtained with the help of Lemma~\ref{ac:lemma:high-prob-state-bound}, which provides high probability upper bounds on $\vert Z(t) \vert$ and $\vert X(t) \vert$ that hold uniformly over all time. This is the same as Lemma~4 in \cite{sysidpaper}, which we refer to for the proof.}
\begin{lemma} \label{ac:lemma:high-prob-state-bound}
    (High Probability State and Regressor Bounds)
    Suppose Assumptions~\ref{ac:assump:measurable}, \ref{ac:assump:process-noise}, \ref{ac:assump:forward-complete}, \ref{ac:assump:constrained-controls} and \ref{ac:assump:poly-feature-map} hold. Then, there exists \ssnoteblue{APB} function $\chi_5 : \mathbb{R}_{\geq 0} \rightarrow \mathbb{R}_{\geq 0}$, such that for any $x_0 \in \mathbb{X}$ and $\delta \in (0,1)$,
    \begin{align}
        &\mathbb{P} \mleft ( \mleft \vert X(t-1) \mright \vert \leq \overline{x}(t-1,\delta,x_0) \mright. \\
        & \quad \mleft. \text{and} \ \mleft \vert Z(t) \mright \vert \leq  \overline{z}(t,\delta,x_0) , \ \forall t \in \mathbb{N} \mright ) \geq 1 - \delta,
    \end{align}
    with $\overline{x}$ and $\overline{z}$ defined in \eqref{ac:def:high-prob-state-bound} and \eqref{ac:def:high-prob-regressor-bound} respectively.
\end{lemma}

\ssnoteblue{Next, we probabilistically lower bound $\lambda_{\textnormal{min}}(\sum_{i=1}^t Z(i) Z(i)^{\top})$ using Lemma~\ref{ac:lemma:regional-pe}. It says that there exists a high probability event $\mathcal{E}_2$, such that for all time steps $t \geq T_{\textnormal{burn-in}}(3\delta,x_0)$, if 1) our sample belongs to this event, 2) the upper bound $\mleft \vert Z(i) \mright \vert \leq  \overline{z}\mleft(i,\delta,x_0\mright)$ holds on the regressor for all $i \leq t$, and 3) the one-step predicted state $\mathbb{E}[X(i) \mid X(i-1), U(i-1) ]$ remains in $\mathcal{X}_{\textnormal{PE}}$ for all $i \leq t-1$, then a linearly increasing lower bound on $\lambda_{\textnormal{min}} \mleft(\sum_{i = 1}^t Z(i) Z(i)^{\top} \mright)$ holds --- i.e. PE is holding. We interpret this as a \textit{regional PE} result due to the requirement that the one-step predicted state remains in $\mathcal{X}_{\textnormal{PE}}$. Note that conditioning on the event $\mathcal{E}_2$ is required due to the stochastic nature of the problem, in the sense that PE also depends on the injected and process noise affecting the system.}
\ssnotebluetwo{This result is the same as Lemma~5 in \cite{sysidpaper}, which we refer to for the full proof.}
\begin{lemma} \label{ac:lemma:regional-pe}
    (Regional Persistency of Excitation)
    Suppose Assumptions~\ssnoteblue{\ref{ac:assump:measurable}}, \ref{ac:assump:process-noise}, \ref{ac:assump:forward-complete}, \ref{ac:assump:constrained-controls}, \ref{ac:assump:poly-feature-map} and \ref{ac:assump:regional-excitation} are satisfied. Moreover, suppose that $(\psi,\alpha,\mu_w,\mu_s)$ is \ssnotebluetwo{$(\mathcal{X}_{\textnormal{PE}},c_{\textnormal{PE}},p_{\textnormal{PE}})$} for some $\mathcal{X}_{\textnormal{PE}} \subseteq \mathbb{X}$, \ssnotebluetwo{$c_{\textnormal{PE}},p_{\textnormal{PE}} > 0$}.
    Then, for any $x_0 \in \mathbb{X}$ and $\delta \in (0,1)$, there exists an event $\mathcal{E}_2 \in \mathcal{F}$ satisfying $\mathbb{P}(\mathcal{E}_2) \geq 1 - \delta$, such that for any $t \geq T_{\textnormal{burn-in}}(\ssnoteblue{3}\delta,x_0)$,
   \begin{align}
        &\lambda_{\textnormal{min}} \mleft(\sum_{i = 1}^t Z(i) Z(i)^{\top} \mright) \geq \frac{\ssnote{\ssnotebluetwo{c_{\textnormal{PE}}} p_{\textnormal{PE}}}}{4}(t-1)
    \end{align}
    on the event\footnote{\ssnotebluetwo{We use ``on the event'' to relate a predicate involving a random variable to a probabilistic event, with the understanding it means that when an outcome belongs to that event, the predicate evaluated on the random variable is true. 
    Specifically, given a collection of random variables $X_1,X_2,\hdots$ taking values in $\mathcal{X}_1, \mathcal{X}_2,\hdots$, a predicate $Q:\mathcal{X}_1 \times \mathcal{X}_2 \times \cdots \rightarrow \{ \mathrm{true}, \mathrm{false} \}$, and an event $E \in \mathcal{F}$, we write ``on the event $E$, $Q(X_1,X_2,\hdots)$'', or equivalently ``$Q(X_1,X_2,\hdots)$ on the event $E$'', if $E \subseteq \{ Q(X_1,X_2,\hdots)\}$.}}
    \begin{align}
        & \mathcal{E}_2 \cap \mleft \{ \mleft \vert Z(i) \mright \vert \leq  \overline{z}\mleft(i,\delta,x_0\mright), \ \forall i \leq t \mright \} \cap \\
        & \quad \{ \mathbb{E}[X(i) \mid X(i-1), U(i-1) ] \in \mathcal{X}_{\textnormal{PE}}, \ \forall i \leq t-1  \}. \label{ac:eqn:lemma-regional-pe-main}
    \end{align}
\end{lemma}

\ssnotebluetwo{We now provide Proposition~\ref{ac:prop:rls-error-rpi-event}. It can be viewed as a result similar to Theorem~\ref{ac:theorem:rls-error-rpi} that provides estimation error bounds and the guarantee of invariance of the state $X(t)$ inside $\mathcal{X}_{\textnormal{RPI}}$, but it also contains the information that this event is a superset of another event $\mathcal{E}_2 \cap \mathcal{E}_3 \cap \mathcal{E}_4$ (defined in Proposition~\ref{ac:prop:rls-error-rpi-event}). We provide this more informative representation since it is useful for the stability analysis later in Section~\ref{ac:sec:proof-stability}.}
\begin{proposition} \label{ac:prop:rls-error-rpi-event}
    \ssnotebluetwo{Suppose Assumptions~\ref{ac:assump:measurable}, \ref{ac:assump:process-noise}, \ref{ac:assump:forward-complete}, \ref{ac:assump:constrained-controls}, \ref{ac:assump:poly-feature-map}, \ref{ac:assump:regional-excitation} and \ref{ac:assump:rpi} are satisfied. 
    Then, for any $x_0 \in \mathcal{X}_{\textnormal{RPI}}$ and $\delta \in (0,1)$ such that \eqref{ac:eqn:theorem-rls-error-rpi-condition} is satisfied, there exists an event $\mathcal{E}_2 \in \mathcal{F}$ satisfying $\mathbb{P}(\mathcal{E}_2) \geq 1 - \delta/3$ such that on the event $\mathcal{E}_2 \cap \mathcal{E}_3 \cap \mathcal{E}_4$, 
    \begin{align}
        &\vert \hat{\theta}(t) - \theta_* \vert \leq e(t,\delta,x_0) \ \text{for all} \ t \geq T_{\textnormal{burn-in}}(\delta,x_0) \\
        & \quad \text{and} \ X(t) \in \mathcal{X}_{\textnormal{RPI}} \ \text{for all} \ t \in \mathbb{N}_0, \label{ac:eqn:prop-rls-error-rpi-event-main}
    \end{align}
    where the events $\mathcal{E}_3,\mathcal{E}_4 \in \mathcal{F}$ are defined as 
    \begin{align}
        &\mathcal{E}_3 :=  \Bigg \{  \vert \hat{\theta}(t) - \theta_*  \vert \leq \frac{1}{ \lambda_{\textnormal{min}}  ( V(t)  )^{1/2} } \bigg( \gamma^{1/2}  \vert \theta_*  \vert_F +  \\
        & \sigma_w \sqrt{ 2 n   \mleft ( \ln \mleft (3n / \delta\mright ) + \mleft (d/2\mright ) \ln \mleft ( \lambda_{\textnormal{max}}\mleft (V\mleft (t\mright )\mright ) \gamma^{-1} \mright )  \mright ) }  \bigg), \ \forall t \in \mathbb{N}  \Bigg \}, \\
        &\mathcal{E}_4 :=  \{  \vert X(t-1)  \vert \leq \overline{x}(t-1,\delta/3,x_0), \\
        & \qquad \quad  \vert Z(t)  \vert \leq  \overline{z}(t,\delta/3,x_0) , \ \forall t \in \mathbb{N}  \}.
    \end{align}
    and moreover $\mathbb{P}(\mathcal{E}_2 \cap \mathcal{E}_3 \cap \mathcal{E}_4) \geq 1- \delta$.}
    \ssnotebluetwo{
    Furthermore, for any $x_0 \in \mathcal{X}_{\textnormal{RPI}}$ and $\delta \in (0,1)$,
    \begin{enumerate}
        \item $\lim_{t \rightarrow \infty} e(t,\delta,x_0) = 0$;
        \item $T_{\textnormal{burn-in}}(\delta,x_0) < \infty$;
        \item $T_{\textnormal{converge}}(\delta,x_0) < \infty$.
    \end{enumerate}
    }
\end{proposition}

\begin{pf}
    \ssnotebluetwo{Throughout this proof, suppose $x_0 \in \mathcal{X}_{\textnormal{RPI}}$, $\delta \in (0,1)$ and \eqref{ac:eqn:theorem-rls-error-rpi-condition} are satisfied.}
    \ssnotebluetwo{We will first prove \eqref{ac:eqn:prop-rls-error-rpi-event-main}, then statements 1)-3) at the end.}

    \ssnotebluetwo{We start by proving \eqref{ac:eqn:prop-rls-error-rpi-event-main}.} Let $\mathcal{E}_2 \in \mathcal{F}$ be an event such that for all $t \geq T_{\textnormal{burn-in}}(\ssnoteblue{\delta},x_0)$,
    \begin{align}
        &\lambda_{\textnormal{min}} \mleft(\sum_{i = 1}^t Z(i) Z(i)^{\top} \mright) \geq \frac{\ssnote{c_{\textnormal{PE}3} p_{\textnormal{PE}}}}{4}(t-1) \quad \text{on the event} \\
        & \ \mathcal{E}_2 \cap \mleft \{ \mleft \vert Z(i) \mright \vert \leq  \overline{z}\mleft(i,\delta/3,x_0\mright), \ \forall i \leq t \mright \} \cap \\
        & \ \{ \mathbb{E}[X(i) \mid X(i-1), U(i-1) ] \in \mathcal{X}_{\textnormal{PE}}, \ \forall i \leq t-1  \}, \label{ac:eqn:prop-rls-error-rpi-exciting-4}
    \end{align}
    and $\mathbb{P}(\mathcal{E}_2) \geq 1 - \delta/3$, where the existence of a satisfactory $\mathcal{E}_2$ is established in Lemma~\ref{ac:lemma:regional-pe}, \ssnotebluetwo{which we can make use of due to Assumptions~\ref{ac:assump:measurable}, \ref{ac:assump:process-noise}, \ref{ac:assump:forward-complete}, \ref{ac:assump:constrained-controls}, \ref{ac:assump:poly-feature-map} and \ref{ac:assump:regional-excitation}}.
    \ssnotebluetwo{Moreover, recall that $\mathcal{E}_3,\mathcal{E}_4 \in \mathcal{F}$ are both defined in the statement of this proposition, and correspond to the event that that the data-dependent estimation error bound from Lemma~\ref{ac:lemma:data-dependent-error-bound} holds, and the event that the state and regressor bound from Lemma~\ref{ac:lemma:high-prob-state-bound} holds, respectively.}
    \ssnotebluetwo{We will establish \eqref{ac:eqn:prop-rls-error-rpi-event-main}} by separately proving that 
    on the event $\mathcal{E}_2 \cap \mathcal{E}_3 \cap \mathcal{E}_4$,
    \begin{align}
        & \vert \hat{\theta}(t) - \theta_* \vert \leq e(t,\delta,x_0) \label{ac:eqn:prop-rls-error-rpi-exciting-2} \\ 
        & \quad \text{for all} \ t \in \{T_{\textnormal{burn-in}}(\delta,x_0), \hdots, \ssnotebluetwo{T_{\textnormal{converge}}(\delta,x_0)} - 1 \}, 
    \end{align}
    and that on the event $\mathcal{E}_2 \cap \mathcal{E}_3 \cap \mathcal{E}_4$, 
    \begin{align}
        &\vert \hat{\theta}(t) - \theta_* \vert \leq e(t,\delta,x_0) \ \text{for all} \ t \geq \ssnotebluetwo{T_{\textnormal{converge}}(\delta,x_0)} \\
        &\quad  \text{and} \ X(t) \in \mathcal{X}_{\textnormal{RPI}} \ \text{for all} \ t \in \mathbb{N}_0. \label{ac:eqn:prop-rls-error-rpi-exciting-3}
    \end{align} 

    \ssnotebluetwo{We first establish \eqref{ac:eqn:prop-rls-error-rpi-exciting-2}.} Note that $\mathcal{E}_4 \subseteq \{ X(i) \in \mathcal{X}_{\textnormal{RPI}} \ \text{for all} \ i \leq \ssnotebluetwo{T_{\textnormal{converge}}(\delta,x_0)}-3 \}$ \ssnotebluetwo{since $\ssnotebluetwo{T_{\textnormal{converge}}(\delta,x_0)} \leq T_{\textnormal{contained}}(\delta,x_0)-1$ due to \eqref{ac:eqn:theorem-rls-error-rpi-condition}.}
    \ssnotebluetwo{This implies that} 
    $\{ X(i) \in \mathcal{X}_{\textnormal{RPI}} \ \text{for all} \ i \leq \ssnotebluetwo{T_{\textnormal{converge}}(\delta,x_0)}-3 \} \subseteq \{ g(X(i),U(i),0) \in \Gamma(\mathcal{X}_{\textnormal{RPI}}) \ \text{for all} \ i \leq \ssnotebluetwo{T_{\textnormal{converge}}(\delta,x_0)}-3 \} \ssnotebluetwo{\subseteq} \{ \mathbb{E}[X(i) \mid X(i-1), U(i-1) ] \in \mathcal{X}_{\textnormal{PE}} \ \text{for all} \ i \leq \ssnotebluetwo{T_{\textnormal{converge}}(\delta,x_0)}-2 \}$, \ssnotebluetwo{where we used the fact that $\mathbb{E}[X(i+1) \mid X(i),U(i)] = g(X(i),U(i),0)$ and $g(X(i),U(i),0) \in \Gamma(\mathcal{X}_{\textnormal{RPI}})$ on the event $\{X(i) \in \mathcal{X}_{\textnormal{RPI}}$ due to Assumption~\ref{ac:assump:constrained-controls}}.
    Making use of \eqref{ac:eqn:prop-rls-error-rpi-exciting-4}, it follows that for all $t \in \{T_{\textnormal{burn-in}}(\delta,x_0), \hdots, \ssnotebluetwo{T_{\textnormal{converge}}(\delta,x_0)}-2 \}$, $\mathcal{E}_2 \cap \mathcal{E}_4 \cap \{ \mathbb{E}[X(i) \mid X(i-1), U(i-1) ] \in \mathcal{X}_{\textnormal{PE}} \ \text{for all} \ i \leq \ssnotebluetwo{T_{\textnormal{converge}}(\delta,x_0)}-2 \} \subseteq \{\lambda_{\textnormal{min}}(G(t)) \geq \frac{\ssnotebluetwo{c_{\textnormal{PE}}}p_{\textnormal{PE}}}{4}(t-1) \}$, and therefore $\mathcal{E}_2 \cap \mathcal{E}_3 \cap \mathcal{E}_4 \subseteq \{ \vert \hat{\theta}(t) - \theta_* \vert \leq e(t,\delta,x_0) \}$, \ssnotebluetwo{which follows by the definition of $\mathcal{E}_3,\mathcal{E}_4$ and $e(t,\delta,x_0)$ from \eqref{ac:eqn:theorem-rls-error-exciting-e}}. \ssnotebluetwo{Then, since} $\mathcal{E}_2 \cap \mathcal{E}_3 \cap \mathcal{E}_4$ is independent of $t$, we subsequently \ssnotebluetwo{have established \eqref{ac:eqn:prop-rls-error-rpi-exciting-2}.}
    
    In order to establish \eqref{ac:eqn:prop-rls-error-rpi-exciting-3}, we first prove using an induction argument that for all $t \geq \ssnotebluetwo{T_{\textnormal{converge}}(\delta,x_0)}$, on the event $\mathcal{E}_2 \cap \mathcal{E}_3 \cap \mathcal{E}_4$,
    \begin{align}
        \vert \hat{\theta}(t) - \theta_* \vert \leq e(t,\delta,x_0) \ \text{and} \ X(i) \in \mathcal{X}_{\textnormal{RPI}} \ \text{for all} \ i \leq t +1. \label{ac:eqn:prop-rls-error-rpi-exciting-1}
    \end{align}
    
    \ssnotebluetwo{\textit{Base case:}} Note that $\mathcal{E}_4 \subseteq \{ X(i) \in \mathcal{X}_{\textnormal{RPI}} \ \text{for all} \ i \leq \ssnotebluetwo{T_{\textnormal{converge}}(\delta,x_0)} + 1 \}$ since \\$\ssnotebluetwo{T_{\textnormal{converge}}(\delta,x_0)} \leq T_{\textnormal{contained}}(\delta,x_0)-1$ due to \eqref{ac:eqn:theorem-rls-error-rpi-condition}. 
    Moreover, we have 
    $\{ X(i) \in \mathcal{X}_{\textnormal{RPI}} \ \text{for all} \ i \leq \ssnotebluetwo{T_{\textnormal{converge}}(\delta,x_0)} + 1 \} \subseteq \{ X(i) \in \mathcal{X}_{\textnormal{RPI}} \ \text{for all} \ i \leq \ssnotebluetwo{T_{\textnormal{converge}}(\delta,x_0)} - 2 \} \subseteq \{ g(X(i),U(i),0) \in \Gamma(\mathcal{X}_{\textnormal{RPI}})  \ i \leq \ssnotebluetwo{T_{\textnormal{converge}}(\delta,x_0)} - 2 \} \subseteq \{ \mathbb{E}[X(i) \mid X(i-1), U(i-1) ] \in \mathcal{X}_{\textnormal{PE}} \ \text{for all} \ i \leq \ssnotebluetwo{T_{\textnormal{converge}}(\delta,x_0)} - 1 \}$.
    Next, note that $\mathcal{E}_2 \cap \mathcal{E}_4 \cap \{ \mathbb{E} [X(i) \mid X(i-1),U(i-1) ] \in \mathcal{X}_{\textnormal{PE}} \ \text{for all} \ i \leq \ssnotebluetwo{T_{\textnormal{converge}}(\delta,x_0)} - 1 \} \subseteq \{ \lambda_{\textnormal{min}}(G(\ssnotebluetwo{T_{\textnormal{converge}}(\delta,x_0)})) \geq \frac{\ssnotebluetwo{c_{\textnormal{PE}}}p_{\textnormal{PE}}}{4}(\ssnotebluetwo{T_{\textnormal{converge}}(\delta,x_0)} - 1) + \gamma \}$ making use of \eqref{ac:eqn:prop-rls-error-rpi-exciting-4}, and also observe that $\mathcal{E}_3 \cap \mathcal{E}_4 \cap \{ \lambda_{\textnormal{min}}(G(\ssnotebluetwo{T_{\textnormal{converge}}(\delta,x_0)})) \geq \frac{\ssnotebluetwo{c_{\textnormal{PE}}}p_{\textnormal{PE}}}{4}(\ssnotebluetwo{T_{\textnormal{converge}}(\delta,x_0)} - 1) + \gamma \} \subseteq \{ \vert \hat{\theta}(\ssnotebluetwo{T_{\textnormal{converge}}(\delta,x_0)}) - \theta_* \vert \leq e(\ssnotebluetwo{T_{\textnormal{converge}}(\delta,x_0)},\delta,x_0) \}$.
    Thus, by combining these set-theoretic inclusions together \ssnotebluetwo{and making use of the definitions of $\mathcal{E}_3,\mathcal{E}_4 \in \mathcal{F}$ and $e(t,\delta,x_0)$ from \eqref{ac:eqn:theorem-rls-error-exciting-e}}, we find that $\mathcal{E}_2 \cap \mathcal{E}_3 \cap \mathcal{E}_4 \subseteq \{ \vert \hat{\theta}(\ssnotebluetwo{T_{\textnormal{converge}}(\delta,x_0)}) - \theta_* \vert \leq $\\$e(\ssnotebluetwo{T_{\textnormal{converge}}(\delta,x_0)},\delta,x_0) \ \text{and} \ X(i) \in \mathcal{X}_{\textnormal{RPI}} \ \text{for all} \ i \leq \ssnotebluetwo{T_{\textnormal{converge}}(\delta,x_0)} + 1\}$, completing the base case.

    \ssnotebluetwo{\textit{Inductive step:}} Suppose $t \geq \ssnotebluetwo{T_{\textnormal{converge}}(\delta,x_0)}$, and that $\mathcal{E}_2 \cap \mathcal{E}_3 \cap \mathcal{E}_4 \subseteq \{ \vert \hat{\theta}(t) - \theta_* \vert \leq e(t,\delta,x_0) \ \text{and} \ X(i) \in \mathcal{X}_{\textnormal{RPI}} \ \text{for all} \ i \leq t + 1 \}$. 
    Note that $\{ X(i) \in \mathcal{X}_{\textnormal{RPI}} \ \text{for all} \ i \leq t + 1 \} \subseteq \{ X(i) \in \mathcal{X}_{\textnormal{RPI}} \ \text{for all} \ i \leq t - 1 \} \subseteq \{ g( X(i), U(i), 0 ) \in \Gamma( \mathcal{X}_{\textnormal{RPI}} )  \ \text{for all} \ i \leq t - 1 \} \subseteq \{ \mathbb{E}[X(i) \mid X(i-1), U(i-1)] \in \mathcal{X}_{\textnormal{PE}} \ \text{for all} \ i \leq t \} $.
    Next, note that $\mathcal{E}_2 \cap \mathcal{E}_4 \cap \{ \mathbb{E} [X(i) \mid X(i-1),U(i-1) ] \in \mathcal{X}_{\textnormal{PE}} \ \text{for all} \ i \leq t \} \subseteq \{ \lambda_{\textnormal{min}}(G(t+1)) \geq \frac{\ssnotebluetwo{c_{\textnormal{PE}}}p_{\textnormal{PE}}}{4}t + \gamma \}$ making use of \eqref{ac:eqn:prop-rls-error-rpi-exciting-4}, and also observe that $\mathcal{E}_3 \cap \mathcal{E}_4 \cap \{ \lambda_{\textnormal{min}}(G(t+1)) \geq \frac{\ssnotebluetwo{c_{\textnormal{PE}}}p_{\textnormal{PE}}}{4}t + \gamma \} \subseteq \{ \vert \hat{\theta}(t) - \theta_* \vert \leq e(t+1,\delta,x_0) \}$.
    Since $t \geq \ssnotebluetwo{T_{\textnormal{converge}}(\delta,x_0)}$, we have $e(t,\delta,x_0) \leq \ssnotebluetwo{\bar{\vartheta}}$ due to \eqref{ac:eqn:def-T-converge} and therefore $\{ \vert \hat{\theta}(t) - \theta_* \vert \leq e(t,\delta,x_0) \} \subseteq \{ \vert \hat{\theta}(t) - \theta_* \vert \leq \ssnotebluetwo{\bar{\vartheta}} \}$ \ssnotebluetwo{making use of the definitions of $\mathcal{E}_3,\mathcal{E}_4 \in \mathcal{F}$ and $e(t,\delta,x_0)$ from \eqref{ac:eqn:theorem-rls-error-exciting-e}}. Next, note from Assumption~\ref{ac:assump:rpi} that $\{ \vert \hat{\theta}(t) - \theta_* \vert \leq \ssnotebluetwo{\bar{\vartheta}} \} \cap \{ X(t+1) \in \mathcal{X}_{\textnormal{RPI}} \} \subseteq \{ X(t+2) = g(X(t+1),\alpha(X(t+1),\hat{\theta}(t),S(t+1)),W(t+2)) \in \mathcal{X}_{\textnormal{RPI}} \}$. Thus, by combining these set-theoretic inclusions together, we find that $\mathcal{E}_2 \cap \mathcal{E}_3 \cap \mathcal{E}_4 \subseteq \{ \vert \hat{\theta}(t+1) - \theta_* \vert \leq e(t+1,\delta,x_0) \ \text{and} \ X(i) \in \mathcal{X}_{\textnormal{RPI}} \ \text{for all} \ i \leq t + 2 \}$, concluding the inductive step.

    Since we have established both the base case and the inductive step, we have proven \eqref{ac:eqn:prop-rls-error-rpi-exciting-1} via induction. Moreover, since $\mathcal{E}_2 \cap \mathcal{E}_3 \cap \mathcal{E}_4$ is independent of $t$ in \eqref{ac:eqn:prop-rls-error-rpi-exciting-1}, 
    \ssnotebluetwo{\eqref{ac:eqn:prop-rls-error-rpi-exciting-3} follows, and \eqref{ac:eqn:prop-rls-error-rpi-event-main} subsequently follows by combining \eqref{ac:eqn:prop-rls-error-rpi-exciting-2} and \eqref{ac:eqn:prop-rls-error-rpi-exciting-3}.}

    \ssnotebluetwo{We also find that $\mathbb{P}( \mathcal{E}_2 \cap \mathcal{E}_3 \cap \mathcal{E}_4 ) \geq 1 - \delta$  by making use of the union bound, and the fact that $\mathbb{P}(\mathcal{E}_3) \geq 1 - \delta/3$ from Lemma~\ref{ac:lemma:data-dependent-error-bound} \ssnotebluetwo{Assumptions~\ref{ac:assump:measurable} and \ref{ac:assump:process-noise}}, $\mathbb{P}(\mathcal{E}_4) \geq 1 - \delta/3$ from Lemma~\ref{ac:lemma:high-prob-state-bound} \ssnotebluetwo{using Assumptions~\ref{ac:assump:measurable}-\ref{ac:assump:poly-feature-map}}, and $\mathbb{P}(\mathcal{E}_2) \geq 1 - \delta/3$ from earlier in this proof.}

    \ssnotebluetwo{We now establish statement 1)-3). The proof steps for 1) are the same as the proof steps of statement~4) from Corollary~1 in \cite{sysidpaper}. Moreover, the proof of 2) is the same as the proof of $T_{\textnormal{burn-in}}(\delta,x_0)<\infty$ in Theorem~1 from \cite{sysidpaper}. Finally, $T_{\textnormal{converge}}(\delta,x_0)$ follows from 1) and 2).}
\end{pf}

Theorem~\ref{ac:theorem:rls-error-rpi} easily follows as a consequence of Proposition~\ref{ac:prop:rls-error-rpi-event}, \ssnotebluetwo{but is more easily interpreted.}
\begin{pf*}{Proof of Theorem \ref{ac:theorem:rls-error-rpi}}

    \ssnotebluetwo{Suppose $x_0 \in \mathbb{X}$, $\delta \in (0,1)$ and \eqref{ac:eqn:theorem-rls-error-rpi-condition} are satisfied. Let $\mathcal{E}_2,\mathcal{E}_3,\mathcal{E}_4 \in \mathcal{F}$ be defined the same as in Proposition~\ref{ac:prop:rls-error-rpi-event}. Then, we establish \eqref{ac:eqn:theorem-rls-error-rpi-main} as follows:
    \begin{align}
        &\mathbb{P} (\vert \hat{\theta}(t) - \theta_* \vert \leq e(t,\delta,x_0) \ \text{for all} \ t \geq T_{\textnormal{burn-in}}(\delta,x_0) \\
        &\quad \text{and} \ X(t) \in \mathcal{X}_{\textnormal{RPI}} \ \text{for all} \ t \in \mathbb{N}_0 ) \\
        &\geq \mathbb{P}( \mathcal{E}_2 \cap \mathcal{E}_3 \cap \mathcal{E}_4 ) \label{ac:eqn:theorem-rls-error-rpi-exciting-5} \\
        &\geq 1 - \delta, \label{ac:eqn:theorem-rls-error-rpi-exciting-6}
    \end{align}
    where \eqref{ac:eqn:theorem-rls-error-rpi-exciting-5} and \eqref{ac:eqn:theorem-rls-error-rpi-exciting-6} both follow from Proposition~\ref{ac:prop:rls-error-rpi-event}.
    }

    \ssnotebluetwo{Statements 1)-3) in Theorem~\ref{ac:theorem:rls-error-rpi} follow directly from statements 1)-3) in Proposition~\ref{ac:prop:rls-error-rpi-event}.}
\end{pf*}

\subsection{Stability Guarantees} \label{ac:sec:proof-stability}

\ssnotebluetwo{We provide Proposition~\ref{ac:prop:agip}. It can be interpreted as a bound on the magnitude of the states $X(t)$ of the closed-loop system \eqref{ac:eqn:system-dynamics} under the adaptive control framework described in Algorithm~\ref{ac:alg:framework}, that depends on the probability that the parameter estimates $\hat{\theta}(t)$ remain in a $\bar{e}(t)$-ball around $\theta_*$ at each time $t \geq t_0$ (for arbitrary $t_0 \in \mathbb{N}_0$ and $\{\bar{e}(t)\}_{t \geq t_0}$).}
\begin{proposition} \label{ac:prop:agip}
    Suppose Assumptions~\ref{ac:assump:measurable}, \ref{ac:assump:process-noise}, and \ref{ac:assump:lyapunov} are satisfied. 
    Then, for all $\delta \in (0,1)$, there exists $\beta_1, \beta_2 \in \mathcal{K} \mathcal{L}$, $\gamma_3 \in \mathcal{K}$, $\eta_2 \in \mathcal{L}$ and $c_2 \geq 0$ such that for any bounded set $\mathcal{X}\subseteq \mathcal{X}_{\textnormal{RPI}}$, 
    \begin{align}
        &\mathbb{P} \bigg( \vert X(t) \vert \leq \max \bigg( \beta_1( \max_{x \in \mathcal{X}}\vert x \vert, t - t_0), \\
        & \qquad \eta_2( t - t_0)  + \beta_2\mleft(  \max_{t_0 - 1 \leq i \leq t_0 + \lfloor (t - t_0)/2 \rfloor - 1} \bar{e}(i), t - t_0 \mright), \\
        & \qquad c_2 + \gamma_3\mleft( \max_{t_0 + \floor{(t-t_0)/2}\leq i \leq t-1} \bar{e}(i) \mright) \bigg) \bigg) \\
        &\geq 1 - \delta - \mathbb{P}( \{ X(t_0) \in \mathcal{X}_{\textnormal{RPI}}\}^{\comp} \\
        & \qquad \cup \big\{ \vert \hat{\theta}(i) - \theta_* \vert \leq \bar{e}(i) \text{ for all } i \geq t_0 - 1 \big\}^{\comp} \\
        & \qquad \cup \{ X(i) \in \mathcal{X}_{\textnormal{RPI}} \text{ for all } i \geq t_0 \}^{\comp} ) \label{ac:eqn:prop-agip-conj-bound}
    \end{align}
    for all $x_0 \in \ssnotebluethree{\mathbb{X}}$, $t_0 \in \mathbb{N}_0$, $t \geq t_0$, and $\{\bar{e}(i)\}_{i \geq t_0} \ssnotebluetwo{\subseteq \mathbb{R}_{\geq 0} }$ satisfying 
    $\overline{e}(i) \leq \ssnotebluetwo{\bar{\vartheta}}$ for all $i \geq t_0 - 1$
\end{proposition}

\begin{pf}
    Let $\mathcal{X}_{\textnormal{RPI}}$, $V: \mathcal{X}_{\textnormal{RPI}} \rightarrow \mathbb{R}_{\geq 0}$, and $\bar{\theta} > 0$, be such that $(g,\alpha,\mu_w,\mu_s)$ is $(V,\mathcal{X}_{\textnormal{RPI}},\bar{\vartheta})$-stochastic Lyapunov (with existence verified via Assumption~\ref{ac:assump:lyapunov}), and let $\alpha_1, \alpha_2, \alpha_3 \in \mathcal{K}_{\infty}$, $\sigma_3 \in \mathcal{K}$, and $\tilde{d} \geq 0$, all satisfy the requirements in Definition~\ref{ac:def:lyapunov} for $(g,\alpha,\mu_w,\mu_s)$ to be $(V,\mathcal{X}_{\textnormal{RPI}},\bar{\vartheta})$-stochastic Lyapunov.

    Throughout this proof, suppose $\mathcal{X} \subseteq \mathcal{X}_{\textnormal{RPI}}$, $x_0 \in \ssnotebluethree{\mathbb{X}}$, $t_0 \in \mathbb{N}_0$, and
    \ssnotebluetwo{$\{\bar{e}(i)\}_{i \geq t_0} \ssnotebluetwo{\subseteq \mathbb{R}_{\geq 0} }$ satisfies 
    $\overline{e}(i) \leq \ssnotebluetwo{\bar{\vartheta}}$ for all $i \geq t_0 - 1$.}
    Moreover, let $\{\tilde{\theta}(t)\}_{t \geq t_0}$ be a random sequence satisfying $\tilde{\theta}(t) = \hat{\theta}(t) - \theta_*$ for all $t \geq t_0 - 1$.
    
    Let $\alpha_v \in \mathcal{K}_{\infty}$ be a convex function satisfying $\alpha_v(r) \leq \alpha_3 \circ \alpha_2^{-1}(r)$ for $r \geq 0$, \ssnotebluetwo{as required} in Definition~\ref{ac:def:lyapunov}. 
    Then, $\alpha_v( V(x) ) \leq \alpha_v \circ \alpha_2 ( \vert x \vert ) \leq \alpha_3( \vert x \vert )$ holds for all $x \in \mathcal{X}_{\text{RPI}}$. Therefore, for all $i \geq t_0$, on the event $\{ \vert \hat{\theta}(i-1)  - \theta_* \vert \leq \bar{e}(i-1)\} \cap \{ X(i) \in \mathcal{X}_{\text{RPI}}\}$,
    \begin{align}
        &\mathbb{E}[V(X(i+1)) \mid X(i), \hat{\theta}(i-1)] - V(X(i)) \\
        &= \Delta V(X(i), \hat{\theta}(i-1)) \label{ac:eqn:prop-agip-1} \\
        &\leq  - \alpha_3(\vert X(i) \vert )+\tilde{d} + \sigma_3(\vert \hat{\theta}(i-1) - \theta_* \vert) \label{ac:eqn:prop-agip-2} \\
        &\leq - \alpha_v(V(X(i)))+\tilde{d} + \sigma_3(\vert \hat{\theta}(i-1) - \theta_* \vert)  \\
        &\leq - \alpha_v(V(X(i)))+\tilde{d} + \sigma_3( \bar{e}(i-1) ) \label{ac:eqn:prop-agip-template-35}, 
    \end{align}
    where \eqref{ac:eqn:prop-agip-1} follows from \eqref{ac:eqn:def-lyapunov-difference} and \eqref{ac:eqn:prop-agip-2} follows from Assumption~\ref{ac:assump:lyapunov}. For convenience, throughout the remainder of this proof, let $d(i) = \tilde{d} + \sigma_3( \bar{e}(i-1) )$ for $i \geq t_0$.

    Next, define $\tilde{\mathcal{E}}(i) :=  \{ X(t_0) \in \mathcal{X} \} \cap \{ \vert \hat{\theta}(j) - \theta_* \vert \leq \bar{e}(j-1) \text{ and } X(j) \in \mathcal{X}_{\text{RPI}} \text{ for all } j \leq i \}$ for $i \geq t_0$. 
    Then, for all $i \geq t_0$, we have
    \begin{align}
        &\mathbb{E} [\bm{1}_{\tilde{\mathcal{E}}(i+1)} V(X(i+1)) ] \leq \mathbb{E} [\bm{1}_{\tilde{\mathcal{E}}(i)} V(X(i+1)) ] \\
        &= \mathbb{E} [\bm{1}_{\tilde{\mathcal{E}}(i)} \mathbb{E} [V(X(i+1)) \mid \{ \hat{\theta}(j)  \}_{j=t_0-1}^{i-1}, \\
        & \qquad \{ X(j) \}_{j=t_0}^i ] ] \label{ac:eqn:prop-agip-template-15} \\
        &= \mathbb{E} [\bm{1}_{\tilde{\mathcal{E}}(i)} \mathbb{E} [V(X(i+1)) \mid \hat{\theta}(i-1), X(i) ] ] \label{ac:eqn:prop-agip-template-16} \\
        &\leq \mathbb{E} [\bm{1}_{\tilde{\mathcal{E}}(i)} ( V(X(i)) - \alpha_v(V(X(i)))+ d(i) ) ] \label{ac:eqn:prop-agip-template-17} \\
        &\leq \mathbb{E} [ \bm{1}_{ \tilde{\mathcal{E}}(i) } V(X(i)) ] - ( \mathbb{E}[ \bm{1}_{\tilde{\mathcal{E}}(i)} \alpha_v ( V(X(i)) ) ] ) + d(i) \label{ac:eqn:prop-agip-template-19} \\
        &\leq \mathbb{E} [ \bm{1}_{\tilde{\mathcal{E}}(i)} V(X(i)) ] - ( \mathbb{E}[ \alpha_v ( \bm{1}_{\tilde{\mathcal{E}}(i)}  V(X(i)) ) ] ) \\
        & \qquad + d(i) \label{ac:eqn:prop-agip-template-20} \\
        &\leq \mathbb{E} [ \bm{1}_{ \{\tilde{\mathcal{E}}(i)\} } V(X(i)) ] - \alpha_v( \mathbb{E}[ \bm{1}_{\tilde{\mathcal{E}}(i)} V(X(i)) ] ) \\
        & \qquad + d(i) \label{ac:eqn:prop-agip-template-21}
    \end{align}
    where \eqref{ac:eqn:prop-agip-template-15} follows from the tower property of conditional expectation and pulling out known factors, \eqref{ac:eqn:prop-agip-template-16} follows from the conditional independence of $X(i+1)$ and $\{\hat{\theta}(j)\}_{j = t_0 - 1}^{i - 2}, \{ X(j) \}_{j = t_0}^{i-1}$ given $\hat{\theta}(i-1),X(i)$, \eqref{ac:eqn:prop-agip-template-17} follows from \eqref{ac:eqn:prop-agip-template-35}, \eqref{ac:eqn:prop-agip-template-20} follows from the convexity of $\alpha_v$ and the fact that $\alpha_v(0)=0$, and \eqref{ac:eqn:prop-agip-template-21} follows via Jensen's inequality.

    Next, define $\tilde{\gamma}(r) := 2 \max( \alpha_v^{-1}( r ) , r )$
    for $r \geq 0$, and note that $\tilde{\gamma} \in \mathcal{K}$.
    We now establish that for all $i \geq t_0$ and $l \leq i$, if $\mathbb{E}[\bm{1}_{\tilde{\mathcal{E}}(i)}V(X(i))] \leq \tilde{\gamma}( \max_{l \leq j \leq i} d(j) )$, then
    \begin{align}
        \mathbb{E}[\bm{1}_{\tilde{\mathcal{E}}(i+1)}V(X(i+1))] \leq \tilde{\gamma}(\max_{l \leq j \leq i} d(j)). \label{ac:eqn:prop-agip-template-36}
    \end{align}
    \ssnotebluetwo{We prove \eqref{ac:eqn:prop-agip-template-36} via two cases.} 
    
    \ssnotebluetwo{\textit{Case 1:} Suppose $\mathbb{E}[V(X(i))\bm{1}_{ \tilde{\mathcal{E}}(i) } ] \leq \tilde{\gamma}(\max_{l \leq j \leq i} d(j)) / 2$. Then,} we have
    \begin{align}
        &\mathbb{E}[\bm{1}_{ \tilde{\mathcal{E}}(i+1) } V(X(i+1)) ] \\
        &\leq \mathbb{E} [ \bm{1}_{ \tilde{\mathcal{E}}(i) } V(X(i)) ] - \alpha_v( \mathbb{E}[ \bm{1}_{ \tilde{\mathcal{E}}(i) } V(X(i)) ] ) + \max_{l \leq j \leq i} d(j) \label{ac:eqn:prop-agip-template-22} \\
        &\leq \mathbb{E} [ \bm{1}_{ \tilde{\mathcal{E}}(i) } V(X(i)) ] + \max_{l \leq j \leq i} d(j) \\
        &\leq \frac{\tilde{\gamma}(\max_{l \leq j \leq i} d(j))}{2} + \max_{l \leq j \leq i} d(j) \label{ac:eqn:prop-agip-template-24} \\
        &\leq \tilde{\gamma}(\max_{l \leq j \leq i} d(j)), \label{ac:eqn:prop-agip-template-25}
    \end{align}
    where \eqref{ac:eqn:prop-agip-template-22} follows from \eqref{ac:eqn:prop-agip-template-21}, \eqref{ac:eqn:prop-agip-template-24} follows since $\mathbb{E}[V(X(i))\bm{1}_{ \tilde{\mathcal{E}}(i) } ] \leq \tilde{\gamma}(\max_{l \leq j \leq i} d(j)) / 2$, and \eqref{ac:eqn:prop-agip-template-25} follows from  $\max_{l \leq j \leq i} d(j) \leq \tilde{\gamma}(\max_{l \leq j \leq i} d(j))/2$ by the definition of $\tilde{\gamma}$. 

    \ssnotebluetwo{\textit{Case 2:} Suppose $\tilde{\gamma}( \max_{l \leq j \leq i} d(j) )/2 \leq \mathbb{E}[V(X(i))\bm{1}_{ \tilde{\mathcal{E}}(i) } ] \leq \tilde{\gamma}(\max_{l \leq j \leq i} d(j))$. Then,}
   \begin{align}
        &\mathbb{E}[\bm{1}_{ \tilde{\mathcal{E}}(i+1) } V(X(i+1)) ] \\
        &\leq \mathbb{E} [ \bm{1}_{ \tilde{\mathcal{E}}(i) } V(X(i)) ] - \alpha_v( \mathbb{E}[ \bm{1}_{ \tilde{\mathcal{E}}(i) } V(X(i)) ] ) \\
        & \quad + \max_{l \leq j \leq i} d(j) \label{ac:eqn:prop-agip-template-27} \\
        &\leq \tilde{\gamma}(\max_{l \leq j \leq i} d(j)) - \alpha_v( \frac{\tilde{\gamma}(\max_{l \leq j \leq i} d(j))}{2} ) \\
        & \quad + \max_{l \leq j \leq i} d(j) \label{ac:eqn:prop-agip-template-28} \\
        &\leq \tilde{\gamma}(\max_{l \leq j \leq i} d(j)) \label{ac:eqn:prop-agip-template-29}
    \end{align}
    where \eqref{ac:eqn:prop-agip-template-27} follows from \eqref{ac:eqn:prop-agip-template-21}, \eqref{ac:eqn:prop-agip-template-28} follows from the assumption that \\$\tilde{\gamma}(\max_{l \leq j \leq i} d(j))/2 \leq \mathbb{E}[V(X(i))\bm{1}_{ \tilde{\mathcal{E}}(i-1) } ] \leq \tilde{\gamma}(\max_{l \leq j \leq i} d(j))$, and \eqref{ac:eqn:prop-agip-template-29} the fact that $-\alpha_v( \frac{\tilde{\gamma}(\max_{l \leq j \leq i} d(j))}{2}) \leq - \max_{l \leq j \leq i} d(j) $ due to the definition of $\tilde{\gamma}$.

    \ssnotebluetwo{Thus, we have  proven \eqref{ac:eqn:prop-agip-template-36} via two cases.}

    \ssnotebluetwo{Next,} define $\lambda_1(s) := s - \alpha_v(s) + \alpha_v(s/2)$, and $\lambda(s):= (1/2)( s + \max_{s' \in [0,1]} \lambda_1(s') )$. Note that $\lambda_1$ is continuous, $\lambda_1(0)=0$, and $\lambda_1(s) < s$ for all $s > 0$, but not necessarily increasing. On the other hand, it can be shown by following similar steps to \cite[Theorem~B.15]{rawlings2017model}, that $\lambda \in \mathcal{K}_{\infty}$ and
    \begin{align}
        \lambda_1(s) \leq \lambda(s) < s \label{ac:eqn:prop-agip-template-37}
    \end{align}
    for all $s > 0$.
    We then have that for all $i \geq t_0$ and $l \leq i$, if $\mathbb{E}[V(X(i)) \bm{1}_{  \tilde{\mathcal{E}}(i) } ] > \tilde{\gamma}(\max_{l \leq j \leq i} d(j))$, then
    \begin{align}
        &\mathbb{E}[ \bm{1}_{\tilde{\mathcal{E}}(i+1)} V(X(i+1) ) ] \\
        &\leq \mathbb{E}[ \bm{1}_{ \tilde{\mathcal{E}}(i) }V(X(i)) ] - \alpha_v( \mathbb{E}[ \bm{1}_{ \tilde{\mathcal{E}}(i) } V(X(i)) ] ) + \max_{l \leq j \leq i} d(j) \label{ac:eqn:prop-agip-template-30} \\
        &\leq \mathbb{E}[ \bm{1}_{ \tilde{\mathcal{E}}(i) }V(X(i)) ] - \alpha_v( \mathbb{E}[ \bm{1}_{ \tilde{\mathcal{E}}(i) } V(X(i)) ] ) \\
        & \quad + \alpha_v \mleft( \frac{\mathbb{E}[V(X(i))\bm{1}_{ \tilde{\mathcal{E}}(i-1) }]}{2} \mright) \label{ac:eqn:prop-agip-template-31} \\
        &\leq \lambda( \mathbb{E}[\bm{1}_{ \tilde{\mathcal{E}}(i) }V(X(i))] ) \label{ac:eqn:prop-agip-template-34}
    \end{align}
    where \eqref{ac:eqn:prop-agip-template-30} follows from \eqref{ac:eqn:prop-agip-template-21}, \eqref{ac:eqn:prop-agip-template-31} follows from $\mathbb{E}[V(X(i)) \bm{1}_{  \tilde{\mathcal{E}}(i) } ] > \tilde{\gamma}(\max_{l \leq j \leq i} d(j))$, and the fact that $\max_{l \leq j \leq i} d(j) \leq \alpha_v( \frac{\tilde{\gamma}(\max_{l \leq j \leq i} d(j))}{2} )$ by definition of $\tilde{\gamma}$, and \eqref{ac:eqn:prop-agip-template-34} follows from \eqref{ac:eqn:prop-agip-template-37}.

    Now, by contraposition of \eqref{ac:eqn:prop-agip-template-36}, for all $i \geq t_0$ and $l \leq i$, if $\mathbb{E}[\bm{1}_{\tilde{\mathcal{E}}(i+1)}V(X(i+1))] > \tilde{\gamma}(\max_{l \leq j \leq i} d(j))$, then
    \begin{align}
        \mathbb{E}[\bm{1}_{ \tilde{\mathcal{E}}(i) }V(X(i))] > \tilde{\gamma}( \max_{l \leq j \leq i} d(j)). \label{ac:eqn:prop-agip-template-38}
    \end{align}

    Let $\lambda^i$ denote the composition of $\lambda$ with itself $i$ times. For all $i \geq t_0$ and $l \in \{ t_0, \hdots, i \}$, if $\mathbb{E}[V(X(i)) \bm{1}_{\tilde{\mathcal{E}}(i) }] > \tilde{\gamma}(\max_{l \leq j \leq i} d(j))$, then it follows that
    \begin{align}
        &\mathbb{E}[V(X(i+1)) \bm{1}_{\tilde{\mathcal{E}}(i+1)}] \leq \lambda(\mathbb{E}[V(X(i)) \bm{1}_{ \tilde{\mathcal{E}}(i) }]) \label{ac:eqn:prop-agip-template-11} \\
        &\leq \lambda^2( \mathbb{E}[V(X_{i-1}) \bm{1}_{ \tilde{\mathcal{E}}(i-1) }] ) \label{ac:eqn:prop-agip-template-14} \\
        &\leq \lambda^{i-l+1} ( \mathbb{E} [V(X_l) \bm{1}_{ \tilde{\mathcal{E}}(l) } ] ) \label{ac:eqn:prop-agip-template-13}
    \end{align}
    where \eqref{ac:eqn:prop-agip-template-11} follows from \eqref{ac:eqn:prop-agip-template-34}, \eqref{ac:eqn:prop-agip-template-14} follows from \eqref{ac:eqn:prop-agip-template-38} and \eqref{ac:eqn:prop-agip-template-34}, and \eqref{ac:eqn:prop-agip-template-13} follows by iteratively repeating this process $i+1$ times.

    Suppose $t \geq t_0$. For all $l \in \{ t_0 , \hdots, t\}$, we have
    \begin{align}
        &\mathbb{E}[V(X(t)) \bm{1}_{\tilde{\mathcal{E}}(t)}] \\
        & \leq \max( \lambda^{t-l} ( \mathbb{E} [V(X_l) \bm{1}_{\tilde{\mathcal{E}}(l)} ] ), \tilde{\gamma}( \max_{l \leq j \leq t - 1} d(j) ) ) \label{ac:eqn:prop-agip-template-47} \\
        & \leq \max( \lambda^{t-l} ( \max( \lambda^{l - t_0} \mathbb{E}[ V(X(t_0)) \mathbf{1}_{\tilde{\mathcal{E}}(t_0)} ], \\
        & \quad \tilde{\gamma}(\max_{t_0 \leq j \leq l-1}d(j)) ) ), \tilde{\gamma}( \max_{l \leq j \leq t - 1}d(i) ) ) \label{ac:eqn:prop-agip-template-48} \\
        & \leq \max( \lambda^{t-l} ( \max( \lambda^{l - t_0} \circ \alpha_2( \max_{x \in \mathcal{X}} \vert x \vert) , \\
        & \quad \tilde{\gamma}(\max_{t_0 \leq j \leq l-1}d(j)) ) ), \tilde{\gamma}( \max_{l \leq j \leq t - 1}d(j) ) ) \\
        &= \max(  \lambda^{t - t_0}  \circ \alpha_2( \max_{x \in \mathcal{X}} \vert x \vert), \\
        & \quad \lambda^{t-l} \circ \tilde{\gamma}(\max_{t_0 \leq j \leq l-1}d(j)) , \tilde{\gamma}( \max_{l \leq j \leq t - 1}d(j) ) ) \label{ac:eqn:prop-agip-template-49}
    \end{align}
    where \eqref{ac:eqn:prop-agip-template-47} follows from \eqref{ac:eqn:prop-agip-template-36} and \eqref{ac:eqn:prop-agip-template-13}, and so does \eqref{ac:eqn:prop-agip-template-48}. By setting $l \leftarrow t_0 + \lfloor (t - t_0)/2 \rfloor$, it follows that
    \begin{align}
        &\mathbb{E}[V(X(t)) \bm{1}_{\tilde{\mathcal{E}}(t)}] \\
        &\leq \max\Bigg(  \lambda^{t - t_0 }  \circ \alpha_2 \mleft ( \max_{x \in \mathcal{X}} \vert x \vert \mright ), \\
        & \quad \lambda^{t-t_0-\lfloor \mleft (t-t_0\mright )/2 \rfloor} \circ \tilde{\gamma}\mleft ( \max_{t_0 \leq j \leq t_0 + \lfloor \mleft (t-t_0\mright )/2 \rfloor - 1 } d\mleft (j\mright ) \mright ) , \\
        & \quad \tilde{\gamma}\mleft ( \max_{t_0 + \lfloor \mleft (t-t_0\mright )/2 \rfloor \leq j \leq t - 1 } d\mleft (j\mright ) \mright ) \Bigg) \label{ac:eqn:prop-agip-template-50} \\
        &= \max(  \lambda^{t-t_0} \circ \alpha_2 ( \max_{x \in \mathcal{X}} \vert x \vert ), \lambda^{t-t_0-\lfloor (t-t_0)/2 \rfloor} \\
        & \quad \circ \tilde{\gamma}( \tilde{d} + \sigma_3(\max_{t_0 - 1 \leq j \leq t_0 + \lfloor (t-t_0)/2\rfloor-1}e(j)) ), \\
        & \qquad \tilde{\gamma}( \tilde{d} + \sigma_3(\max_{t_0 + \lfloor (t-t_0)/2\rfloor \leq j \leq t - 1}e(j))  ) ) \\
        &= \max(  \lambda^{t - t_0} \circ \alpha_2 ( \max_{x \in \mathcal{X}}\vert x \vert ), \lambda^{t_0 + \ceil{(t - t_0)/2}} \\
        & \quad \circ \tilde{\gamma}( \tilde{d} + \sigma_3(\max_{t_0 - 1 \leq j \leq t_0 + \lfloor (t-t_0)/2\rfloor-1}e(j)) ),  \\
        & \quad \tilde{\gamma}( \tilde{d} + \sigma_3(\max_{t_0 + \lfloor (t-t_0)/2\rfloor \leq j \leq t - 1}e(j))  ) )
    \end{align}
    Next, suppose $\delta \in (0,1)$, and let 
    \begin{align}
        E_1 = &\Bigg\{ V(X(t)) \bm{1}_{\tilde{\mathcal{E}}(t)} \leq \frac{1}{\delta} \max \Bigg (  \lambda^{t - t_0} \circ \alpha_2 \mleft ( \max_{x \in \mathcal{X}}\vert x \vert \mright ), \\
        & \quad \lambda^{t_0 + \ceil{\mleft (t - t_0\mright )/2}} \\
        & \quad \circ \tilde{\gamma}\mleft ( \tilde{d} + \sigma_3\mleft (\max_{t_0 - 1 \leq j \leq t_0 + \lfloor \mleft (t-t_0\mright )/2\rfloor-1}e\mleft (j\mright )\mright ) \mright ), \\
        & \quad \tilde{\gamma}\mleft ( \tilde{d} + \sigma_3\mleft (\max_{t_0 + \lfloor \mleft (t-t_0\mright )/2\rfloor \leq j \leq t - 1}e\mleft (j\mright )\mright )  \mright ) \Bigg ) \Bigg\}.
    \end{align}
    From Markov's inequality, we have
    \begin{align}
        \mathbb{P}( E_1 ) \geq 1 - \delta. \label{ac:eqn:prop-agip-template-44}
    \end{align}
    Moreover, on the event $E_1 \cap \tilde{\mathcal{E}}(t)$, we have
    \begin{align}
        &\vert X(t) \vert \\
        &\leq \alpha_1^{-1}( V(X(t)) ) \label{ac:eqn:prop-agip-template-54} \\
        &= \alpha_1^{-1}( V(X(t)) \bm{1}_{ \tilde{\mathcal{E}}(t) } ) \\
        &\leq \alpha_1^{-1} \Bigg( \frac{1}{\delta} \max \Bigg (  \lambda^{t - t_0} \circ \alpha_2 \mleft ( \max_{x \in \mathcal{X}}\vert x \vert \mright ), \\
        & \quad \lambda^{t_0 + \ceil{\mleft (t - t_0\mright )/2}} \\
        & \quad \circ \tilde{\gamma}\mleft ( \tilde{d} + \sigma_3\mleft (\max_{t_0 - 1 \leq j \leq t_0 + \lfloor \mleft (t-t_0\mright )/2\rfloor-1}e\mleft (j\mright )\mright ) \mright ), \\
        & \quad \tilde{\gamma}\mleft ( \tilde{d} + \sigma_3\mleft (\max_{t_0 + \lfloor \mleft (t-t_0\mright )/2\rfloor \leq j \leq t - 1}e\mleft (j\mright )\mright )  \mright ) \Bigg ) \Bigg) \\
        &\leq  \max \Bigg ( \alpha_1^{-1} \circ \frac{1}{\delta} \lambda^{t - t_0} \circ \alpha_2 \mleft ( \max_{x \in \mathcal{X}}\vert x \vert \mright ), \\
        & \quad \alpha_1^{-1} \circ \frac{1}{\delta}\lambda^{t_0 + \ceil{\mleft (t - t_0\mright )/2}} \\
        & \quad \circ \tilde{\gamma}\mleft ( \tilde{d} + \sigma_3\mleft (\max_{t_0 - 1 \leq j \leq t_0 + \lfloor \mleft (t-t_0\mright )/2\rfloor-1}e\mleft (j\mright )\mright ) \mright ), \\
        & \quad \alpha_1^{-1} \circ \frac{1}{\delta}\tilde{\gamma}\mleft ( \tilde{d} + \sigma_3\mleft (\max_{t_0 + \lfloor \mleft (t-t_0\mright )/2\rfloor \leq j \leq t - 1}e\mleft (j\mright )\mright )  \mright ) \Bigg ) \\
        &\leq \max( \alpha_1^{-1} \circ \frac{1}{\delta} \lambda^{t - t_0} \circ \alpha_2 ( \max_{x \in \mathcal{X}}\vert x \vert ), \\
        & \quad \alpha_1^{-1} \circ \frac{1}{\delta} \lambda^{t_0 + \ceil{(t - t_0)/2}} \circ \tilde{\gamma}( 2\tilde{d} ) + \alpha_1^{-1} \circ \frac{1}{\delta} \lambda^{t_0 + \ceil{(t - t_0)/2}} \\
        & \quad \circ \tilde{\gamma}(2\sigma_3(\max_{t_0 - 1 \leq j \leq t_0 + \lfloor (t-t_0)/2\rfloor-1}e(j)) ), \\
        & \quad \alpha_1^{-1} \circ \frac{1}{\delta}\tilde{\gamma}( 2\tilde{d} ) + \alpha_1^{-1} \\
        & \quad \circ \frac{1}{\delta}\tilde{\gamma}( 2\sigma_3(\max_{t_0 + \lfloor (t-t_0)/2\rfloor \leq j \leq t - 1}e(j))  ) )\label{ac:eqn:prop-agip-template-56} \\
        &\leq \max( \beta_1(\max_{x \in \mathcal{X}} \vert x \vert, t - t_0), \eta_2(t - t_0) \\
        & \quad + \beta_2(\max_{t_0 - 1 \leq j \leq t_0 + \lfloor (t-t_0)/2\rfloor-1}e(j),t-t_0), \\
        & \qquad c_2 + \gamma_3(\max_{t_0 + \lfloor (t-t_0)/2\rfloor \leq j \leq t - 1}e(j))  ) \label{ac:eqn:prop-agip-template-57}
    \end{align}
    where \eqref{ac:eqn:prop-agip-template-54} follows from Definition~\ref{ac:def:lyapunov}, \eqref{ac:eqn:prop-agip-template-56} follows from the weak triangle inequality for $\mathcal{K}$ functions in \cite[Equation~6]{jiang1994small}. Line \eqref{ac:eqn:prop-agip-template-57} follows by setting 
    $\beta_1(r,k) = \alpha_1^{-1} \circ \frac{1}{\delta} \lambda^k \circ \alpha_2(r)$, 
    $c_2=\alpha_1^{-1} \circ \frac{1}{\delta}\tilde{\gamma}( 2\tilde{d} )$, 
    and 
    $\gamma_3(r) = \alpha_1^{-1} \circ \frac{1}{\delta}\tilde{\gamma}( 2\sigma_3( r  ) )$, where $\beta_1$ and $\gamma_3$ are clearly of class $\mathcal{K}\mathcal{L}$ and $\mathcal{K}$ respectively, and moreover by letting \ssnotebluetwo{$\eta_2$} and $\beta_2$ be class $\mathcal{L}$ and $\mathcal{K}\mathcal{L}$ functions respectively that satisfy
    $\eta_2(k)=\alpha_1^{-1} \circ \frac{1}{\delta} \lambda^{t_0 + \ceil{k/2}} \circ \tilde{\gamma}( 2\tilde{d} )$, 
    $\beta_2(r,k)=\alpha_1^{-1} \circ \frac{1}{\delta} \lambda^{t_0 + \ceil{k/2}} \circ \tilde{\gamma}(2\sigma_3(r) )$.

    Therefore, we find that
    \begin{align}
        &\mathbb{P}\Bigg( \vert X(t) \vert \leq \max\Bigg( \beta_1\mleft (\max_{x \in \mathcal{X}} \vert x \vert, t - t_0\mright ),  \\
        & \quad \eta_2\mleft (t - t_0\mright ) + \beta_2\mleft (\max_{t_0 - 1 \leq j \leq t_0 + \lfloor \mleft (t-t_0\mright )/2\rfloor-1}e\mleft (j\mright ),t-t_0\mright ), \\
        & \quad c_2 + \gamma_3\mleft (\max_{t_0 + \lfloor \mleft (t-t_0\mright )/2\rfloor \leq j \leq t - 1}e\mleft (j\mright )\mright )  \Bigg) \Bigg) \\
        & \geq \mathbb{P}(E_1 \cap \tilde{\mathcal{E}}(t)) \label{ac:eqn:prop-agip-template-58} \\
        &\geq 1 - \mathbb{P}(E_1^{\comp}) - \mathbb{P}( \tilde{\mathcal{E}}^{\comp}(t)) \label{ac:eqn:prop-agip-template-41} \\
        &\geq 1 - \delta - \mathbb{P}(\tilde{\mathcal{E}}^{\comp}(t)). \label{ac:eqn:prop-agip-template-42}
    \end{align}
    where \eqref{ac:eqn:prop-agip-template-58} follows from \eqref{ac:eqn:prop-agip-template-57}, \eqref{ac:eqn:prop-agip-template-41} follows from the union bound, and \eqref{ac:eqn:prop-agip-template-42} follows from \eqref{ac:eqn:prop-agip-template-44}. 
\end{pf}

\ssnotebluetwo{We now prove Theorem~\ref{ac:theorem:regional-stability}.}
\begin{pf*}{Proof of Theorem~\ref{ac:theorem:regional-stability}}
    Suppose \ssnotebluethree{$\delta \in (0,1)$, and let $\tilde{\delta} = \delta/2$}.
    Let $\bar{\vartheta}$ and $V:\mathcal{X}_{\textnormal{RPI}} \rightarrow \mathbb{R}_{\geq 0}$ be such that $(g,\alpha,\mu_w,\mu_s)$ is $(V,\mathcal{X}_{\textnormal{RPI}},\bar{\vartheta})$-stochastic Lyapunov. 
    Let $\beta_1, \beta_2 \in \mathcal{K} \mathcal{L}$, $\gamma_3 \in \mathcal{K}$, $\eta_2 \in \mathcal{L}$ and $c_2 \geq 0$ satisfy \eqref{ac:eqn:prop-agip-conj-bound}, where the existence of these objects is ensured since under our assumptions, Proposition~\ref{ac:prop:agip} is satisfied. Now, suppose $x_0 \in \mathcal{X}_{\textnormal{RPI}}$, and $(\ssnotebluethree{\tilde{\delta}},x_0)$ satisfies condition \eqref{ac:eqn:theorem-rls-error-rpi-condition} \ssnotebluethree{equivalent to \eqref{ac:eqn:theorem-rls-error-rpi-condition-mod} being satisfied)}. Then, using Proposition~\ref{ac:prop:rls-error-rpi-event}, we know there exists an event $\mathcal{E}_2 \in \mathcal{F}$ satisfying \eqref{ac:eqn:prop-rls-error-rpi-event-main}, with $\mathcal{E}_3,\mathcal{E}_4 \in \mathcal{F}$ defined in Proposition~\ref{ac:prop:rls-error-rpi-event}. Next, note that on the event $\mathcal{E}_2 \cap \mathcal{E}_3 \cap \mathcal{E}_4$, 
    \begin{align}
        &\vert \hat{\theta}(t) - \theta_* \vert \leq e(t,\ssnotebluethree{\tilde{\delta}},x_0) \ \text{for all} \ t \geq T_{\textnormal{burn-in}}(\ssnotebluethree{\tilde{\delta}},x_0) \\
        &\ \text{and} \ X(t) \in \mathcal{X}_{\textnormal{RPI}} \ \text{for all} \ t \in \mathbb{N}_0.
    \end{align}
    \ssnotebluetwo{For convenience of notation, let} $T_0 = \ssnotebluetwo{T_{\textnormal{converge}}(\tilde{\delta},x_0)}$.
    Using this result, and the fact that $e(t,\ssnotebluethree{\tilde{\delta}},x_0) \leq \bar{\vartheta}$ for all $t \geq T_0$ by the definition of $T_{\textnormal{converged}}(\ssnotebluethree{\tilde{\delta}},x_0)$ in \eqref{ac:eqn:def-T-converge}, we apply Proposition~\ref{ac:prop:agip} to find that for all $t \geq T_0 + 1$,
    \begin{align}
        &\mathbb{P} \bigg( \vert X(t) \vert \leq \max \bigg( \beta_1( \overline{x}(T_0+1,\ssnotebluethree{\tilde{\delta}}/3,x_0), t - (T_0+1)), \\
        & \quad \beta_2\mleft(  \max_{(T_0+1)-1 \leq i \leq (T_0 + 1) + \lfloor (t - (T_0 + 1))/2 \rfloor - 1} e(i,\ssnotebluethree{\tilde{\delta}},x_0), \mright. \\
        &\quad \mleft. t - (T_0+1) \mright) + \eta_2( t - (T_0+1))  , \\
        & \quad c_2 + \gamma_3\mleft( \max_{(T_0+1) + \floor{(t-(T_0+1))/2}\leq i \leq t-1} e(i,\ssnotebluethree{\tilde{\delta}},x_0) \mright) \bigg) \bigg ) \\
        &\geq 1 - \ssnotebluethree{\tilde{\delta}} - \mathbb{P}( \{ \vert X(T_0 + 1) \vert \leq \overline{x}(T_0 + 1, \ssnotebluethree{\tilde{\delta}}/3, x_0) \}^{\comp} \\
        & \quad \cup \big\{ \vert \hat{\theta}(i) - \theta_* \vert \leq e(i,\ssnotebluethree{\tilde{\delta}},x_0) \text{ for all } i \geq (T_0+1) - 1 \big\}^{\comp} \\
        & \quad \cup \{ X(i) \in \mathcal{X}_{\text{RPI}} \text{ for all } i \geq (T_0+1) \}^{\comp} ) \\
        & \geq 1 - \ssnotebluethree{\tilde{\delta}} - \mathbb{P}((\mathcal{E}_2 \cap \mathcal{E}_3 \cap \mathcal{E}_4 )^{\comp}) \geq 1 - 2\ssnotebluethree{\tilde{\delta}} \label{ac:eqn:theorem-regional-stability-1}
    \end{align}
    where the last inequalities follows from $\mathcal{E}_2 \cap \mathcal{E}_3 \cap \mathcal{E}_4 \subseteq \{ \vert X(T_0 + 1) \vert \leq \overline{x}(T_0 + 1, \ssnotebluethree{\tilde{\delta}}/3, x_0) \} \cap \big\{ \vert \hat{\theta}(i) - \theta_* \vert \leq e(i,\ssnotebluethree{\tilde{\delta}},x_0) \text{ for all } i \geq (T_0+1) - 1 \big\} \cap \{ X(i) \in \mathcal{X}_{\text{RPI}} \text{ for all } i \geq (T_0+1) \}$, and the fact that $\mathbb{P}(\mathcal{E}_2 \cap \mathcal{E}_3 \cap \mathcal{E}_4) \geq 1 - \ssnotebluethree{\tilde{\delta}}$ from Proposition~\ref{ac:prop:rls-error-rpi-event}.
    
    Next, define $\eta$ as a function from $\mathbb{N}_0$ to $\mathbb{R}_{\geq 0}$ satisfying $\eta(t) := \max_{t'\geq t}\tilde{\eta}(t')$ for $t \in \mathbb{N}_0$, with $\tilde{\eta}(t)$ defined as
    \begin{align}
    &\tilde{\eta}(t) \\
    &= \begin{cases} \overline{x}(T_0+1,\delta/3,x_0), \quad 0 \leq t \leq T_0, \\
        \max \bigg( \beta_1\mleft ( \overline{x}\mleft (T_0+1,\delta/3,x_0\mright ), t - \mleft (T_0+1\mright )\mright ), \\
        \quad  \beta_2(  \max_{T_0 \leq i \leq T_0 + \lfloor \mleft (t - \mleft (T_0 + 1\mright )\mright )/2 \rfloor} e\mleft (i,\delta,x_0\mright ), \\
        \quad t - (T_0+1) ) + \eta_2\mleft ( t - \mleft (T_0+1\mright )\mright ), \\
        \quad \gamma_3\mleft ( \max_{T_0+1 + \floor{\mleft (t-\mleft (T_0+1\mright )\mright )/2}\leq i \leq t-1} e\mleft (i,\delta,x_0\mright ) \mright ) \bigg), \\
        \qquad t \geq T_0 + 1.
        \end{cases} \label{ac:eqn:theorem-regional-stability-2}
    \end{align}
    for all $t \in \mathbb{N}_0$, where \ssnotebluefour{clearly $\eta \in \mathcal{L}$ since} since $\lim_{t \rightarrow \infty}\tilde{\gamma}(t) = 0$ due to the fact that $\lim_{t \rightarrow \infty} e(t,\ssnotebluethree{\tilde{\delta}},x_0) = 0$ from Theorem~\ref{ac:theorem:rls-error-rpi}. The conclusion follows by combining \eqref{ac:eqn:theorem-regional-stability-1} with \eqref{ac:eqn:theorem-regional-stability-2}\ssnotebluethree{, and substituting $\tilde{\delta}$ with $\delta/2$.}
\end{pf*}

\ssnotebluetwo{We now provide the proof of Corollary~\ref{ac:corollary:global-stability}, which follows from Theorem~\ref{ac:theorem:regional-stability}.}
\begin{pf*}{Proof of Corollary~\ref{ac:corollary:global-stability}}
    \ssnotebluetwo{Since $(\psi,\alpha,\mu_w,\mu_s)$ is $(c_{\textnormal{PE}},p_{\textnormal{PE}})$\textit{-globally excited} by assumption, it follows that $(\psi,\alpha,\mu_w,\mu_s)$ is $(\mathcal{X}_{\textnormal{PE}},c_{\textnormal{PE}},p_{\textnormal{PE}})$\textit{-regionally excited} with $\mathcal{X}_{\textnormal{PE}}=\mathbb{X}$, satisfying Assumption~\ref{ac:assump:regional-excitation}. Moreover, note that Assumption~\ref{ac:assump:rpi} is trivially satisfied with $\mathcal{X}_{\textnormal{RPI}}=\mathbb{X}$. It follows trivially that $T_{\textnormal{contained}}(\ssnotebluethree{\delta}/2,x_0) = \sup \mleft \{ T \in \mathbb{N} \mid B_{\overline{x}(\ssnotebluetwo{T},\ssnotebluethree{\delta}/6,x_0) }(0) \cap \mathbb{X} \subseteq \mathcal{X}_{\textnormal{RPI}} \mright \}=\infty$. Moreover, from Proposition~\ref{ac:prop:rls-error-rpi-event}, we know that $T_{\textnormal{converge}}(\ssnotebluethree{\delta}/2,x_0) < \infty$.}
    \ssnotebluetwo{Next, note that $(g,\alpha,\mu_w,\mu_s)$ is $(V,\bar{\vartheta})$-global stochastic Lyapunov by Assumption~\ref{ac:assump:global-lyapunov}, so $(g,\alpha,\mu_w,\mu_s)$ is \textit{$(V,\mathbb{X},\bar{\vartheta})$-stochastic Lyapunov}, satisfying Assumption~\ref{ac:assump:lyapunov}.
    Since $T_{\textnormal{converge}}(\ssnotebluethree{\delta}/2,x_0)<T_{\textnormal{contained}} = \infty$, we have verified condition \eqref{ac:eqn:theorem-rls-error-rpi-condition} and hence the premise of Theorem~\ref{ac:theorem:regional-stability}, and therefore can establish \eqref{ac:eqn:theorem-regional-stability-main}, concluding the proof.}
\end{pf*}
\section{Conclusion} \label{ac:sec:control-conclusion}

In this work, we have provided a framework for AC in linearly parameterised stochastic systems, that combines a parameterised stabilising policy with regularised least squares (RLS) for parameter estimation. We derived non-asymptotic error bounds for the parameter estimate that holds for sufficiently large time with positive probability under some assumptions, in particular, requiring that the policy will render the states of the system positively invariant in a regionally exciting set when the parameter estimate is close to the true parameter. By additionally assuming the existence of a stochastic Lyapunov function over the invariant set, probabilistic stability bounds were then derived. These bounds were shown to exist for a regionally controllable PWA system example. Then, under the stricter assumption of global excitation and the existence of a global stochastic Lyapunov function with small parameter estimation error, high probability stability bounds were derived. The usefulness of this result was showcased on an input-constrained linear system example.

\bibliographystyle{plain}        
\bibliography{docs/references}

\appendix

\end{document}